\documentclass[twocolumn,10pt,pra,aps]{revtex4-1}

\def\by#1{#1,}
\def\and{and }
\def\yr#1{{(#1)}}
\def\paper#1{#1}
\def\jour#1{{\it #1}}

\def\vol#1{{\bf #1},}
\def\issue#1{}
\def\pages#1{\hbox{#1},}

\usepackage{amsmath}
\usepackage{amsfonts}
\usepackage{amssymb}
\usepackage{graphicx}
\usepackage{appendix}
\usepackage{placeins}

\begin{document}

\title{\large Capillary transport in low saturated sands: superfast non-linear diffusion model versus direct experimental observations.}

\author{Alex V. Lukyanov$^{\dag}$, Vladimir Mitkin$^{\ddag}$, Theo G. Theofanous$^{\S}$ and Mike Baines$^{\dag}$}

\affiliation{$^{\dag}$School of Mathematical and Physical Sciences, University of Reading, Reading, RG6 6AX, UK}
\affiliation{$^{\ddag}$Aerospace Research Laboratory, University of Virginia, Charlottesville, VA 22903, USA}
\affiliation{$^{\S}$University of California, Santa Barbara, CA 93106, USA}

\begin{abstract}
We have established previously, in a pilot study, that the spreading of liquids in granular porous materials at low levels of saturation, typically less than 10\% of the available void space, has very distinctive features in comparison to that at higher saturation levels. In particular, it has been shown, on theoretical grounds, that the spreading is controlled by a special type of diffusional process, that its physics can be captured by an equation of the super-fast diffusion class, and that these findings were supported by first-of-a-kind experiments. In this paper, we take these findings to the next level including deeper examination and exposition of the theory, an expanded set of experiments to address scaling properties, and systematic evaluations of the predictive performance against these experimental data, keeping in mind also potential practical applications.
\end{abstract}

\maketitle
 
\section{Introduction}

Even a small amount of a liquid added to a dry granular material may dramatically change its structural properties due to the appearance of a strong capillary cohesion force between the particles~\cite{Herminghaus-2005, Herminghaus-2008, Herminghaus-2008-2, Hornbaker1997, Halsey1998, Melnikov2015, Melnikov2016}. The strong capillary force, of the order of $F\sim 2\pi R\gamma\cos\theta_c$,  is due to the liquid bridges (pendular rings) formed at the point of particle contact~\cite{Herminghaus-2005, Herminghaus-2008, Herminghaus-2008-2, Hornbaker1997, Halsey1998, Orr-Scriven-1975, Willett-2000}. Here, $R$ is the average particle radius, $\gamma$ is the surface tension coefficient and $\theta_c$ is the static contact angle of the liquid formed at the three-phase contact line on the flat surface of the solid. A simple estimate for water at room temperature ($\gamma=72\, \mbox{mN}/\mbox{m}$) and sand particles ($\theta_c=30^{\circ}$) of $400\,\mu\mbox{m}$ in diameter results in $F\approx 8\times 10^{-5}\,\mbox{N}$, which is much larger than the gravity force acting on each particle $\approx 8\times 10^{-8}\,\mbox{N}$. It is interesting to note, that the cohesive force is practically independent of the liquid content, that is the value of saturation, as long as the liquid morphology consists of isolated pendular rings. 

\begin{figure}[ht!]
\begin{center}
\includegraphics[trim=-1.5cm 3.cm 1cm -0.5cm,width=\columnwidth]{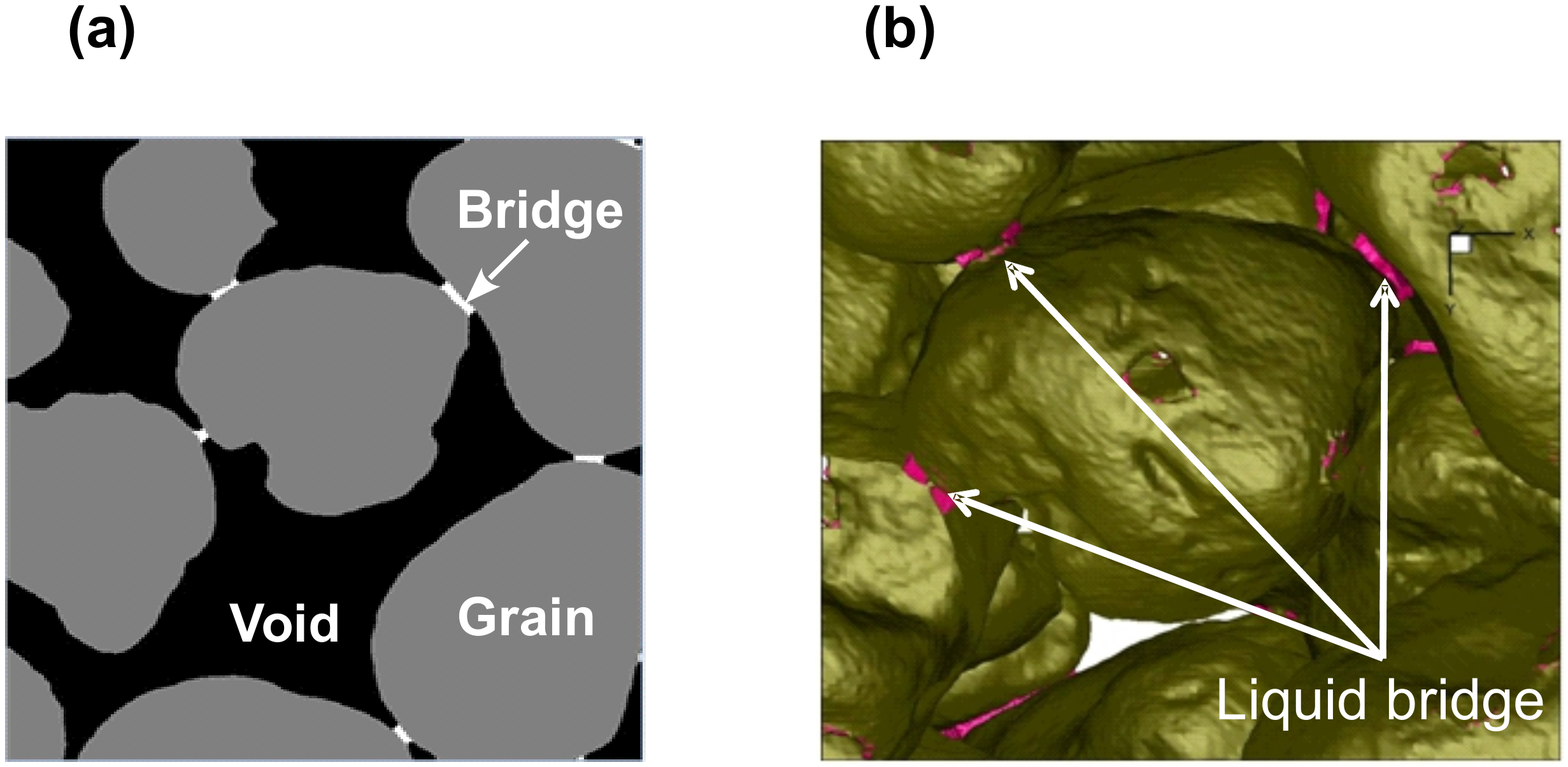}
\end{center}
\caption{Illustration of isolated bridges at low levels of saturation. (a) Micro-x-ray computer tomography (MicroXCT) image, typical from our experiments. (b) 3D
image reconstruction of MicroXCT data. The liquid within the grain roughness is invisible to MicroXCT, since resolution is limited to a few micrometres.} 
\label{Pendular-Rings}
\end{figure}

The formation of isolated liquid bridges is the main characteristic feature of the  pendular regime of wetting in porous materials, when liquid volumes inside the porous matrix are only connected via liquid layers developed on rough surfaces of the particles, Fig. \ref{Pendular-Rings}. The pendular regime of wetting is observed in experiments and computer simulations in a range of saturations $0.2\% \le s\le 10\%$, where the saturation $s$ is defined as the ratio of the liquid volume $V_L$ within a sample volume element $V$ to the available void space $s=\frac{V_L}{V_E}$~\cite{Herminghaus-2005, Herminghaus-2008, Herminghaus-2008-2, Denoth-1999, Lukyanov2012, Melnikov2015, Melnikov2016}. 

\begin{figure}[ht!]
\begin{center}
\includegraphics[trim=1cm 1.cm 1cm -0.5cm,width=\columnwidth]{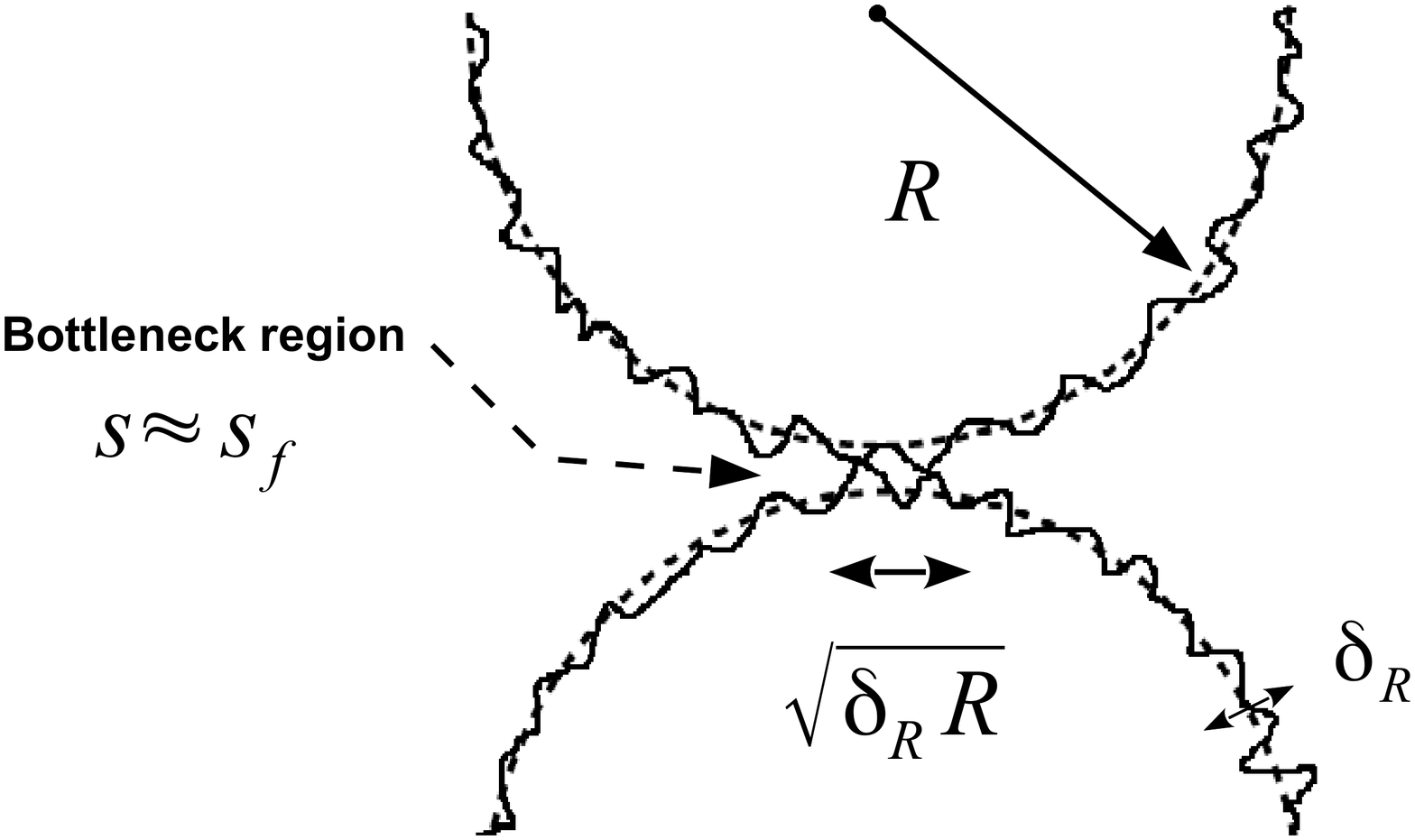}
\end{center}
\caption{The contact zone between two rough spherical particles. The size of the contact zone is $\sqrt{\delta_R R}$~\cite{Halsey1998}.} 
\label{Pendular-Rings-roughness}
\end{figure}

The minimal saturation level is observed when the liquid bridges start to disappear, and when the porous network starts to lose its cohesive and transport properties~\cite{Herminghaus-2005, Herminghaus-2008, Herminghaus-2008-2, Lukyanov2012}. At this level of saturation, the bridges are predominantly formed between asperities on the grains, as is illustrated in Fig. \ref{Pendular-Rings-roughness}, leading to the formation of bottleneck regions at the points of particle contacts, so that permeability of the entire porous network is bound to be greatly reduced when the saturation is approaching this critical level~\cite{Halsey1998, He2001, Herminghaus-2005, Herminghaus-2008, Herminghaus-2008-2, Lukyanov2012}; at this point essentially the whole quantity of the liquid resides in liquid layers formed within the surface roughness of the grains. We will later discuss this scenario in relation to our experimental observations and the formulation of our theoretical model. Here, we note, that as a consequence, we will further distinguish two critical quantities $s_0$ and $s_f$ associated with the minimal saturation level. The first quantity $s_0$ corresponds to the critical saturation level, which would be obtained if we considered the liquid content only residing within the surface roughness of the grains, basically excluding the liquid in the bridges from the consideration. While, the second quantity $s_f$ designates critical saturation level due to the total liquid content in the porous matrix, including the liquid bridges. Apparently, by the definition, $s_f>s_0$, if liquid bridges do not cease to exist completely in the domain of consideration, which is assumed to be always the case in our study. Also, as we will see further, $s_f\approx s_0$. The latter may be intuitively obvious, since the bottleneck regions occur when the bridge liquid content is lower than the potential surface roughness capacity. 

The value of $s_0$, according to its definition, can be parametrized by the non-dimensional quantity $\frac{\delta_L}{R}$, where parameter $\delta_L$ has the dimension of length and can be interpreted as the characteristic average thickness of the liquid layer in the surface roughness; the amplitude of the surface roughness is designated by $\delta_R$. Apparently, two parameters should be consistent, that is $\delta_L \le \max(\delta_R)$. For example, a threshold value $s_f\approx s_0 \approx  0.2\%$ has been observed in experiments using spherical particles, average radius $R=187.5\,\mu\mbox{m}$, with the maximum surface roughness amplitude of $\max(\delta_R)\approx 500\,\mbox{nm}$ as determined by scanning force microscopy~\cite{Herminghaus-2005}. At the same time, in our experiments with  Ottawa sands of average grain radius $R\approx 250 \,\mu\mbox{m}$ a minimal value of $s_f \approx s_0\approx 0.6\%$ was observed.  In Ottawa sands, the surface roughness amplitude $\delta_R$ is distributed between $\min(\delta_R)\approx 250\,\mbox{nm}$ and $\max(\delta_R)\approx 3\,\mu\mbox{m}$ with the mean value found in the range $0.7\,\mu\mbox{m} \le \bar{\delta}_R\le 1\,\mu\mbox{m}$ depending on the average particle radius~\cite{Alshibli2004}. One can see then that the lower is the surface roughness on average, the lower are the critical values $s_0$ and $s_f$. 

If we now consider spherical (or nearly spherical) grains with identical, on average, surface area $4\pi R^2$ and volume $V_0=\frac{4}{3}\pi R^3$ and take into account that only some part of the grain surface volume $4 \pi R^2 \delta_L$ is available for the liquid during the spreading, then the value of saturation due to the liquid distributed on the rough surface of the grains is 
\begin{equation}
\label{Surface-saturation}
s_0=3\alpha_R\frac{1-\phi}{\phi} \, \frac{\delta_L}{R},
\end{equation}
where parameter $\alpha_R$ is the fraction of the surface (roughness) volume occupied by the liquid and  $\phi$  is the porosity. Indeed, if we consider a sample volume element $V$ containing $N\gg 1$ solid particles of volume $V_0$, then following the definition of the saturation
$$
s_0=\frac{V_L}{V_E}=\frac{4 \pi R^2 \alpha_R \delta_L N}{\phi V} 
$$
and
$$
(1-\phi)V \approx N V_0.
$$
The result (\ref{Surface-saturation}) then follows. The quantity $\alpha_R$ is a phenomenological parameter of the model defined by the properties of the surface roughness~\cite{Tokunaga-1997,Tuller-2000, Tuller-2005, deGennes-1985}. 

In our experiments, as we will show, parameter $\alpha_R$ is found to be $\alpha_R\approx 0.3$ at equilibrium. We can thus estimate, using (\ref{Surface-saturation}), that to get $s_0=0.6\%$ at $R=250\,\mu\mbox{m}$ and $\phi=30\%$, one needs to have $\delta_L\approx 0.7\,\mu\mbox{m}$, which is well in the range of the surface roughness amplitudes in the sands used in the experiments. At the same time, to get $s_0=0.2\%$ at $R=187.5\,\mu\mbox{m}$ and $\phi=30\%$, one needs to have $\delta_L\approx 180\,\mbox{nm}$, which is also below the maximum value of the surface roughness observed $500\,\mbox{nm}$.  

\begin{figure}[ht!]
\begin{center}
\includegraphics[trim=-1.5cm 3.cm 1cm -0.5cm,width=1\columnwidth]{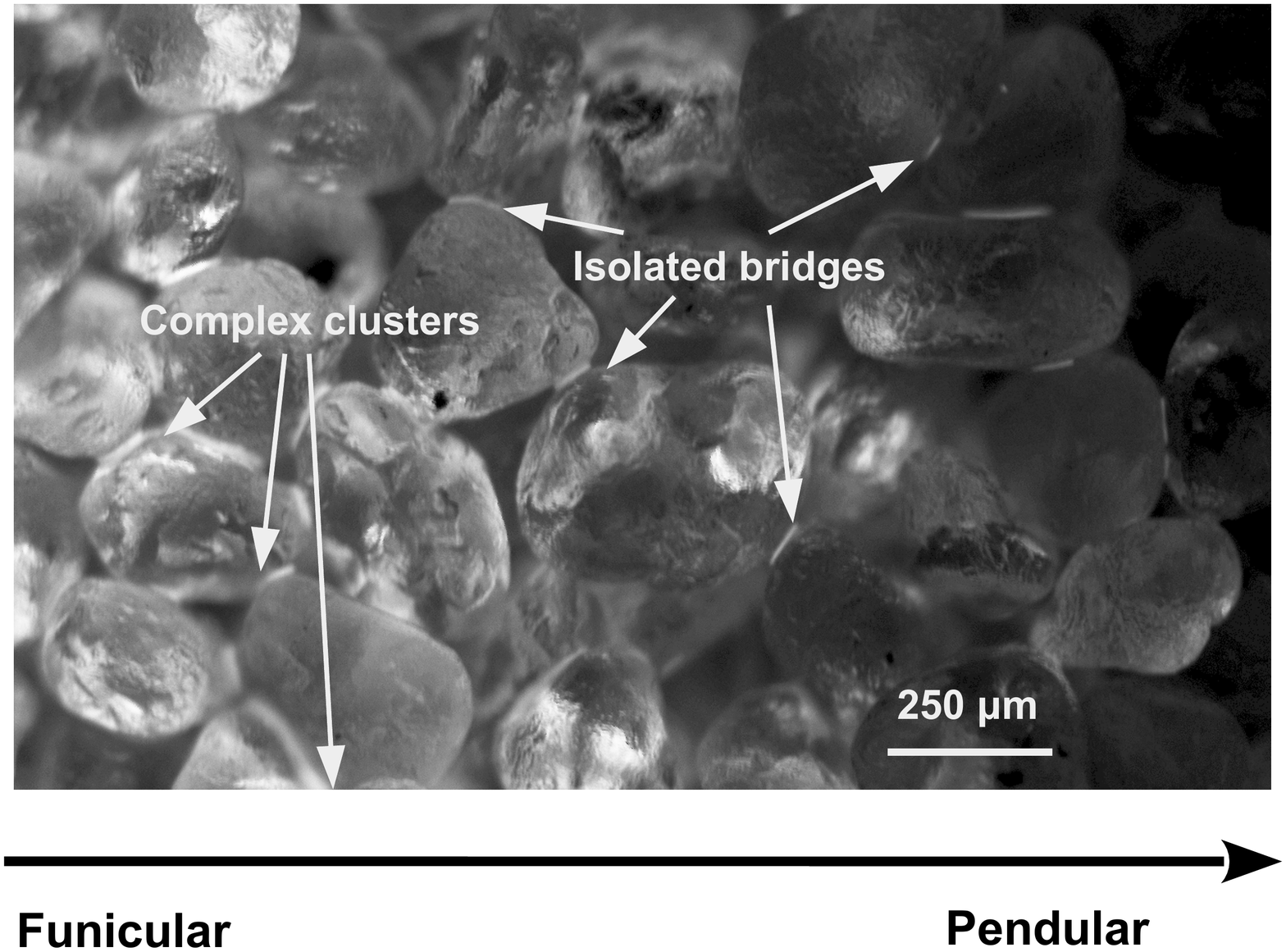}
\end{center}
\caption{An UV fluorescence image of the liquid (TEHP) distribution in sands in the transition from the pendular to the funicular regimes of wetting, at $s>10\%$.} 
\label{Clusters}
\end{figure}

Above  $s_c\approx 10\%$, liquid bridges coalesce into more complex structures, like trimers and pentamers,  and the pendular wetting state gradually transforms into the so called funicular  regime, Fig. \ref{Clusters}, while the global connectivity of the liquid volumes is still absent~\cite{Herminghaus-2005, Herminghaus-2008, Herminghaus-2008-2, Melnikov2015, Melnikov2016}. Finally, at $s\approx 30\%$ a percolation transition occurs when the largest clusters contain about $90\%$ of the available liquid.

Our prime concern here is liquid transport in the range of saturations corresponding to the pendular regime of wetting, which is important for accurate representation of soil-liquid characteristic curves at the lower end of saturations to study biological processes, such as plant water uptake and microbial activity, and spreading of persistent (non-volatile) liquids in arid environments and dry industrial installations~\cite{Tuller-2005, Lukyanov2012}. 

The peculiar character of the diffusion processes in the pendular regime of wetting, when porous network connectivity is conditioned by thin liquid films, has been recognized previously~\cite{deGennes-1985, Bacri1985, Scriven1989, Scriven1993, Tokunaga-1997,Tuller-2000, Tuller-2005, Lukyanov2012}. It has been shown that specific features of liquid transport at low levels of saturation could lead to a special class of mathematical problems, when effective coefficient of non-linear diffusion $D(s)$ diverges at the lower end of saturation, that is in the limit $\displaystyle \lim_{s\to 0}D(s)=s^{\lambda}$ with $\lambda<0$. 

For the first time, the diverging behaviour of the diffusion coefficient $D(s)$, named {\it hyperdispersion}, was predicted in the analysis of spreading in porous networks driven by the disjoining pressure $\Pi(h)$ of nanoscale (thickness $h \sim 1-100\,\mbox{nm}$) wetting films~\cite{Scriven1989, Scriven1993}. A range of admissible $\lambda$ has been predicted depending on the behaviour of the disjoining pressure $\Pi(h)$ as a function of the film thickness $h$, including hyperdispersive exponents $\lambda<0$. Evidence of hyperdispersive behaviour has been observed in two-phase fluid flows with the exponent $\lambda\approx -1$~\cite{Bacri1985}. One needs to note, though, that the values of the effective diffusion coefficient measured in~\cite{Bacri1985} were two-three orders of
magnitude higher than those predicted in~\cite{Scriven1989}. On the other hand, studies of persistent liquids spreading in sands have revealed another mechanism leading to the formulation of a {\it super-fast} non-linear diffusion model~\cite{Lukyanov2012}. The driving force in this model is due to the macroscopic capillary pressure developed on a scale of the surface roughness $\delta_R$ with $D(s)\propto (s-s_0)^{-3/2}$, formally diverging (in the model $s>s_0$ is always the case) at much higher values of saturation $s=s_0\approx 0.6\%$ than that anticipated in \cite{Scriven1989, Scriven1993} and with a different exponent value $\lambda=-3/2$. A comparison between a lead-in theoretical model of superfast diffusion and experimental observations has shown quite good agreement~\cite{Lukyanov2012}. In this study, we further pursue this work, aiming for enhanced definition of the theoretical approach and a more detailed comparison with experiments, including new ones designed to explore scaling properties of the process.    

\section{Experimental observations}

Our experiments have been conducted, as in our previous work, by carefully placing  small liquid drops of a controlled volume, $3\,\mbox{mm}^3 \le V_D \le 12\,\mbox{mm}^3$, on  naturally packed sand beds (slightly shaken to level out) with porosity levels of $\phi\approx 0.3$. To obtain the desired low-dispersion samples, we processed from the standard Ottawa Sand (EMD Chemicals, product SX0075) using a mini-sieves set (Bel-Art Products). The average radii obtained were $R=0.32, 0.26, 0.25, 0.2$ and $0.14\, \mbox{mm}$ with the standard deviations,  $w_R$, as is presented in Table \ref{Table1}. The surface roughness amplitude according to a previous study was distributed in the range $\min(\delta_R)\approx 0.25\,\mu\mbox{m}\le \delta_R\le \max(\delta_R)\approx 3\,\mu\mbox{m}$ with the mean value found in the range $ 0.7\,\mu\mbox{m} \le \bar{\delta}_R\le 1\,\mu\mbox{m}$ depending on the average grain size~\cite{Alshibli2004}. For liquids, we have used several  low-volatility (organophosphate) liquids of varying viscosity and surface tension: tributyl phosphate (TBP, molar weight $266.32\,\mbox{g/mol}$), CAS 126-73-8; Tris(2-ethylhexyl) phosphate (TEHP, molar weight $434.63\,\mbox{g/mol}$), CAS 78-42-2 and tricresyl phosphate (TCP, molar weight $368.37\,\mbox{g/mol}$), CAS 1330-78-5 (Sigma-Aldrich); the details can be found in Table \ref{Table1}.  The contact angles $\theta_c$ of TBP, TEHP and TCP measured on smooth/rough flat glass surfaces in our laboratory at $20\,\mbox{C}^{\circ}$ were found to be at $10^{\circ}/0^{\circ}$, $10^{\circ}/0^{\circ}$ and $30^{\circ}/20^{\circ}$ respectively, though no detailed characterization of the surface roughness was made, therefore later on in the analysis we will use those numbers as the range of the contact angles variations.  

\begin{figure}[ht!]
\begin{center}
\includegraphics[trim=-1.5cm 1.cm 1cm -0.5cm,width=0.9\columnwidth]{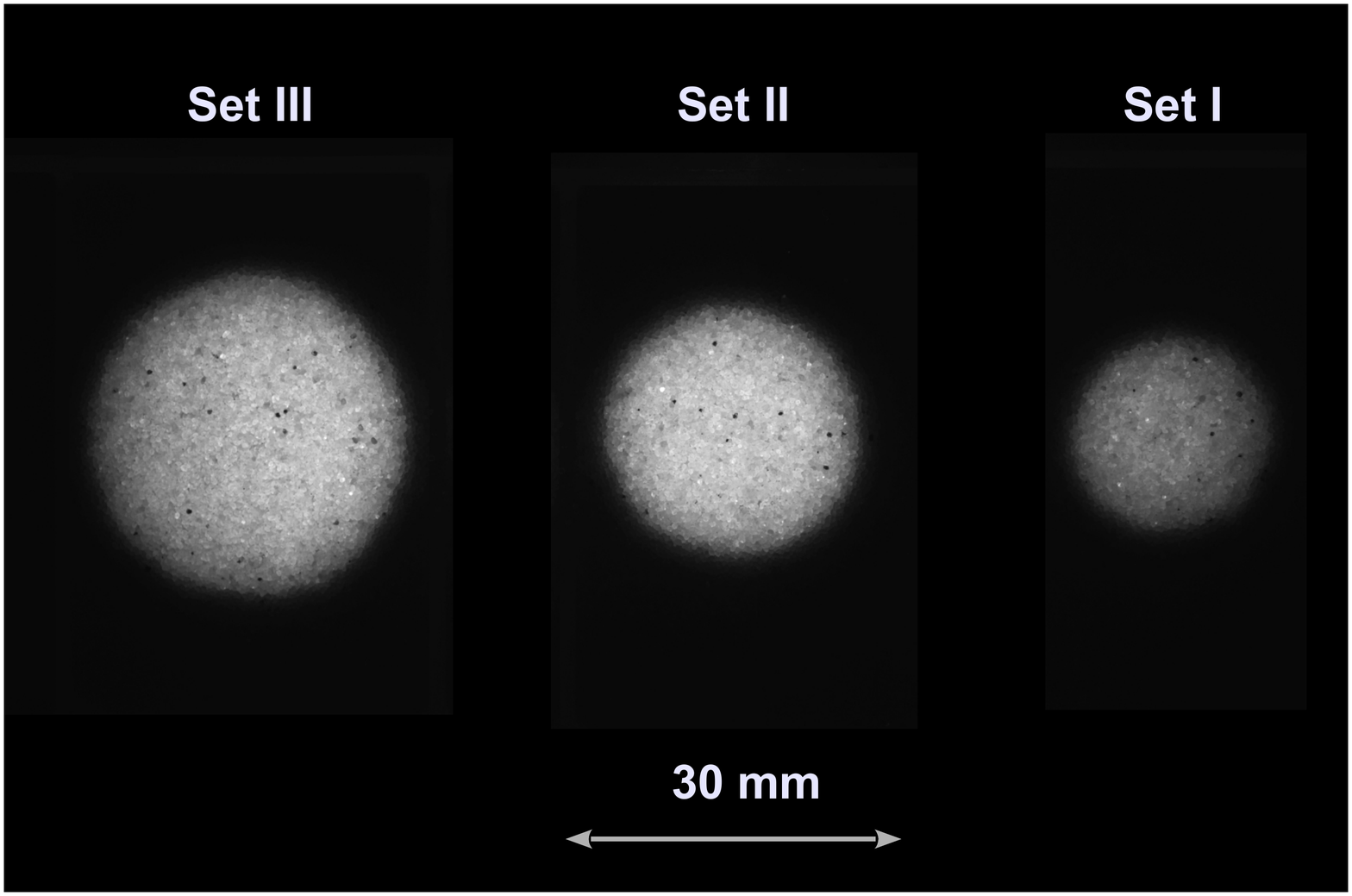}
\end{center}
\caption{UV fluorescence images of wet areas taken after  $\approx 45 000\,\mbox{min}$ of spreading into a sand bed prepared using $R\approx 0.26\,\mbox{mm}$ particles. From left to right the images are from runs III, II and I, as seen in Table \ref{Table1}.} 
\label{Wet-spot}
\end{figure}

The spreading process has been monitored by time-lapse photography using UV-excited fluorescence of the liquid obtained by adding a small amount ($1\%$ by weight) of Coumarin 503 dye. We have verified that the liquid properties were unaffected by the presence of the dye. The photographs, Fig. \ref{Wet-spot}, were taken by $10.7$ MPixel digital cameras (Lumenera Corporation) equipped with a macro-lens and focused to resolve individual grains. The lens was covered by long-pass glass filters to cut off scattered excitation light. No significant background signal could be detected in the absence of the dyed liquid in the range of exposures used in the experiments. 

\begin{table*}[ht!]
\resizebox{2.\columnwidth}{!}{
    \begin{tabular}{ | c | c | c | c | c | c | c | c | c | c | c | c | c | c | c |}
      \hline 
      Run & Liquid & $\mu$ ($\,\mbox{mPa}\cdot\mbox{s}$)  & $\gamma$ (\mbox{mN/m}) & $P_{ve}$ (Pa) & $V_D$ ($\mbox{mm}^3$)  & $R$ ($\mbox{mm}$)  & $w_R$ ($\mbox{mm}$)  & $s_f$ (\%)  & $s_f-s_0^e$ (\%) & $p_f$ ($10^4$\, Pa) & $\bar{\delta}_R (\mu\mbox{m})$ & $\delta_L (\mu\mbox{m})$  & $D_f$ ($10^{-14} \,\mbox{m}^2/\mbox{s}$) & $D_0^e$ ($10^{-14} \,\mbox{m}^2/\mbox{s}$)\\  
      \hline
			
	     I	&	TCP & $20$  & $42.5$ & $8\times 10^{-5}$ &  $3$  & $0.26$ & $0.06$ & $0.61$  & $0.043$ & $-3.6$ & $0.8$ & $0.7$ & $9.7\pm 1.9$ & $9.7\pm 4$ \\  
      \hline
			
			II	&	TCP & $20$  & $42.5$ & $8\times 10^{-5}$ & $6$  & $0.26$ & $0.06$ & $0.61$  & $0.043$ & $-3.6$ & $0.8$ & $0.7$ & $9.7 \pm 1.9$ &  $9.7 \pm 4$ \\  
      \hline

			III	&	TCP & $20$  & $42.5$ & $8\times 10^{-5}$ & $12$  & $0.26$ & $0.06$ & $0.61$  & $0.043$ & $-3.6$ & $0.8$ & $0.7$ & $9.7 \pm 1.9$ &  $9.7\pm 4$ \\  
      \hline
																
			IV	&	TCP & $20$  & $42.5$ & $8\times 10^{-5}$ & $6$  & $0.32$ & $0.08$ & $0.49$   & $0.028$ & $-3.6$ & $0.8$ & $0.7$ & $7.5 \pm 1 $ & $6.3 \pm 3$\\  
      \hline
																			
			V 	&	TCP & $20$  & $42.5$ & $8\times 10^{-5}$ & $6$  & $0.2$ & $0.06$  & $0.73$   & $0.073$ & $-3.6$ & $0.7$ & $0.6$ &  $9.2\pm 1.2$ & $17 \pm 10$\\  
      \hline	
																					
			VI	&	TCP & $20$  & $42.5$ & $8\times 10^{-5}$ & $6$  & $0.14$ & $0.04$ & $1.16$  & $0.15$ & $-3.6$ & $0.7$ & $0.6$ & $7.6\pm 2$  & $34 \pm 19$ \\  
      \hline		
																						
			VII	&	TBP & $3.9$  & $28$  & $1.5\times 10^{-1}$ & $6$  & $0.25$ & $0.08$ & $0.68$  & $0.057$ & $-2.4$ & $0.8$ & $0.8$ &  $164\pm 32$  & $150 \pm 96$\\  
      \hline																							
              
		 VIII &	TEHP & $15$  & $29$  & $1.1\times 10^{-5}$  & $6$  & $0.25$ & $0.08$ & $0.68$  & $0.057$ & $-2.5$ & $0.8$ & $0.8$ & $59\pm 6$  &  $41 \pm 25 $ \\  
      \hline

    \end{tabular} }
    \caption{Parameters of the drop spreading experiments: liquid viscosity $\mu$ at $20^{\circ}\,C$ , surface tension $\gamma$ at $25^{\circ}\,C$, equilibrium vapour pressure $P_{ev}$ at $20^{\circ}\,C$~\cite{Skene1995, Handbook}, drop volume $V_D$, average grain radius $R$, standard deviation around the average grain radius $w_R$, steady state saturation level $s_f$, the model parameter $s_f-s_0^e$ calculated at $B_f=29\, \mu\mbox{m}^{2}$ on the basis of (\ref{scaling-sf-s0}), capillary pressure at the moving front $p_f$, the average surface roughness amplitude $\bar{\delta}_R$, the average liquid layer thickness $ \delta_L$ calculated using (\ref{dl-scaling}), coefficient of diffusion $D_f$ obtained in the comparison with experimental data, coefficient of diffusion $D_0^e$ calculated on the basis of (\ref{Permeability-contact-angle}), (\ref{D0E}) and parameters of the liquids and the sands at $\theta_c=30^{\circ}$ in the case of TCP and $\theta_c=10^{\circ}$ in the case of TBP and TEHP, and $\xi_{f}=0.038$.}
    \vspace{5pt}
    \label{Table1}
\end{table*}

It has been demonstrated previously that after several minutes following the drop contact with the porous bed, the wet region in the sand had the shape of a hemisphere~\cite{Lukyanov2012}. This implies that the roles of gravity and evaporation are negligible. The effects of evaporation can be also seen directly by observing a decrease in the fluorescence intensity. We have noticed that evaporation begins to come into play late in the TBP runs after about six days of exposure, near achieving the steady state, at which point the measurements were terminated. This agrees with calculations (the vapour pressures are given in Table \ref{Table1}), which also agree with experiments that for the other two liquids evaporation was utterly negligible. 

\begin{figure}[ht!]
\begin{center}
\includegraphics[trim=-1.5cm 2.cm 1cm -0.5cm,width=\columnwidth]{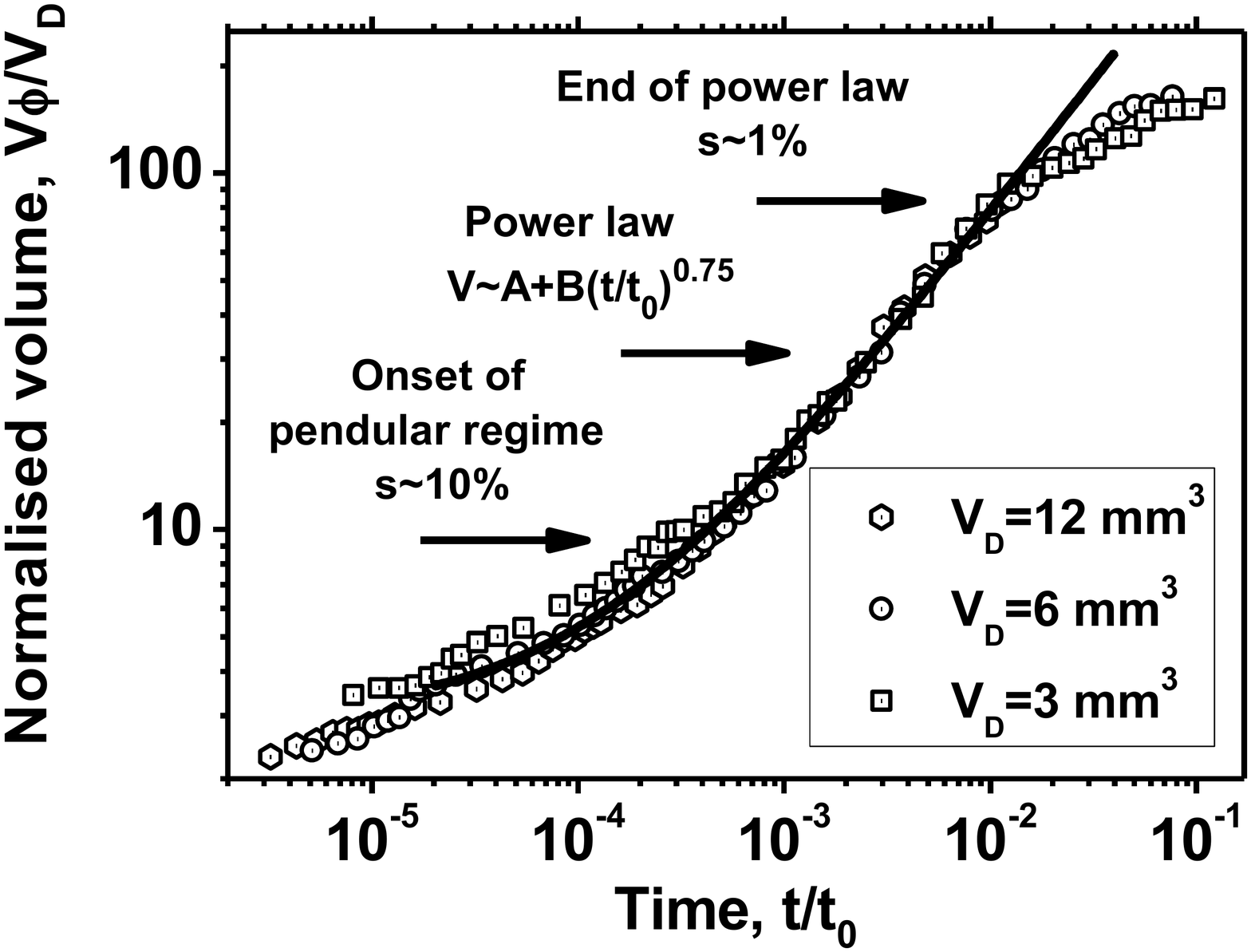}
\end{center}
\caption{Spreading of TCP liquid drops of different volumes $V_D=3, 6$ and $12\,\mbox{mm}^3$ in $R\approx 0.26\,\mbox{mm}$ sand in a three-dimensional case. Normalized volume $V\phi/V_D$ (inverse average saturation $\bar{s}^{-1}$) as a function of reduced time $t/t_0$, where $t_0=\frac{V_D^{2/3}}{D_f}$, $D_f=9.7\times 10^{-14}\,\mbox{m}^2/\mbox{s}$ ($t_0\approx 3.4\times 10^7\, \mbox{s}$ at $V_D=6\,\mbox{mm}^3$), Table \ref{Table1}. The solid line is the fit $V\phi/V_D=A + B (t/t_0)^{0.75}$ at $A=2.9$ and $B=2400$.} 
\label{Fig1}
\end{figure}

The externally visible wet spot diameter can be directly converted into the wet volume $V$. The wet volume, in turn, can be converted into average saturation $\bar{s}=\frac{V_D}{\phi V}$. Typical evolution dynamics of the wet regions obtained by depositing TCP liquid drops of different volumes ($V_D=3, 6$ and $12\,\mbox{mm}^3$) is shown in Fig. \ref{Fig1}. One can see that the wet volume monotonically increases with time eventually saturating at $\bar{s}=s_f \approx s_0$, with parameter $s_f$ apparently being independent of the amount of the liquid deposited, $V_D$ (Table \ref{Table1}). 

At $\bar{s}\approx 10\%$ the increase of the wet volume with time becomes a power law $V(t)\propto t^{\lambda_3}$ with $\lambda_3\approx 0.75$, that is the wetting front radius in this three-dimensional case behaves as $X_3(t)\propto t^{0.25}$. This power law has been previously identified to be universal for the pendular regime in the case of three dimensional geometry of wetting volumes~\cite{Lukyanov2012}. One may notice that using reduced time $t/t_0$ with the scaling dictated by a diffusion law, that is $t_0=\frac{V_D^{2/3}}{D_f}$, one can bring the evolution curves corresponding to different drop volumes $V_D$ into a master curve. Here, $D_f$ is the coefficient of diffusion obtained from comparison with experiments, Table \ref{Table1}. The result indicates that macroscopically the process of spreading can be described by a diffusion-like model, which will be explored in the next parts. Notice, that there is some memory effect in early scaled times, but the data collapse to a single master curve over the whole duration of the pendular regime ($0.006<\bar{s}<0.10$). Also notice, that this poorly-scaled portion of the evolution is less than 10$\%$ of the total duration of the spreading process. This characteristic behaviour was observed in all our experiments conducted using different liquids (TCP, TEHP and TBP) and sands with different grain radii $R$, see further discussions. 

The steady state has been reached usually after about two weeks of spreading. We continued to monitor the wet spots for another month (in some test runs up to three months) without observing any changes in the position of the wetting front within the accuracy of our measurements, Fig. \ref{Fig1}. To ensure that we actually observe a steady state, which is supposed to be independent of the sensitivity of our measurements, we varied the UV-light intensity tenfold and observed no changes in the visible position of the wetting front. This implies that the position of the wetting front was well-defined, in particular that there was no some small quantities of liquid running ahead of the brightly visible front.    

The steady state at a particular value of saturation $s=s_f$ can be, we argue, for two reasons. First, due to the small, but essentially non-zero static contact angles of the liquid-solid combinations used in our experiments, such that the spreading parameter $\gamma_{SV}-\gamma_{SL}-\gamma<0$ was always negative, where $\gamma_{SV}$, $\gamma_{SL}$ and $\gamma$ are the solid-vapour, solid-liquid and liquid-gas surface tensions respectively. In this case, thin liquid films observed in complete wetting case can not be formed, and the minimal liquid layer thickness should be controlled by the available minimal surface roughness length scales~\cite{Popescu2012}. We note, though, that even in our case of incomplete wetting, the observed liquid layer thickness $\delta_L\approx 0.7\,\mu\mbox{m}$ was found to be above the minimal length scale of the grain surface roughness $\min(\delta_R)=250\,\mbox{nm}$~\cite{Alshibli2004}, so that there should have been an additional factor leading to the observed steady state behaviour. This is, as we argue, the formation of bottleneck regions due to the surface roughness at the points of particle contacts, Fig. \ref{Pendular-Rings-roughness}. At sufficiently low saturation levels, the remaining contact area would be between asperities on the rough surface, so that the permeability is expected to be greatly reduced, by at least two orders of magnitude~\cite{Halsey1998, He2001, Herminghaus-2005, Herminghaus-2008, Herminghaus-2008-2}.

This specific feature of the phenomenon, the existence of a minimal saturation level $s_f$ was essentially used in the developing of the theoretical model. It implies that the main driving force could be only the capillary pressure developed on the average length scale of the surface roughness, so that even the lower end of the roughness length scale distribution should be practically cut off from participating in the spreading of the liquid.   

Should the initial spreading parameter be positive (equilibrium spreading parameter is equal to zero), the liquid dispersion may lead to formation of very thin liquid layers on the molecular length scale with disjoining pressure playing a significant role, as we know from experiments and theoretical studies on dynamic wetting phenomena~\cite{DeGennes1985, Cazabat1989, Cazabat1990, Herminghaus1996, Popescu2012}. This can potentially lead to different types of non-linearity in the effective coefficient of dispersion in the system, as it has been discussed in~\cite{deGennes-1985, Bacri1985, Scriven1989, Scriven1993, Tuller-2005}. At the same time, we argue, that due to the formation of the bottleneck regions at the point of particle contacts at extremely low saturations levels, even in this case the equilibration period after the flow was inhibited by the bottlenecks is expected to be extremely long following the dramatic reduction of the permeability of the contact area. For example, the equilibration period observed in our experiments, when the flow domain was connected by the liquid bridges of the size above the threshold value, was about $15$ days for $V_D=6\,\mbox{mm}^3$ TCP drops, Fig. \ref{Fig1}. Then, following an expected two-order-of-magnitude fall in the permeability of the contact area, the equilibration period would be prohibitively long, about $4$ years, so that the steady state would be still controlled by the average surface roughness length scale. Should, on the other hand, the bottleneck regions be absent, then the effects of disjoining pressure shall manifest themselves to the fullest extent. We note in this respect, though this is not part of this study, that liquid spreading in porous paper materials, we observed, where bottleneck regions are absent, has not demonstrated such clear steady state behaviour.

\begin{figure}[ht!]
\begin{center}
\includegraphics[trim=-.5cm 4.cm 1cm -0.5cm,width=\columnwidth]{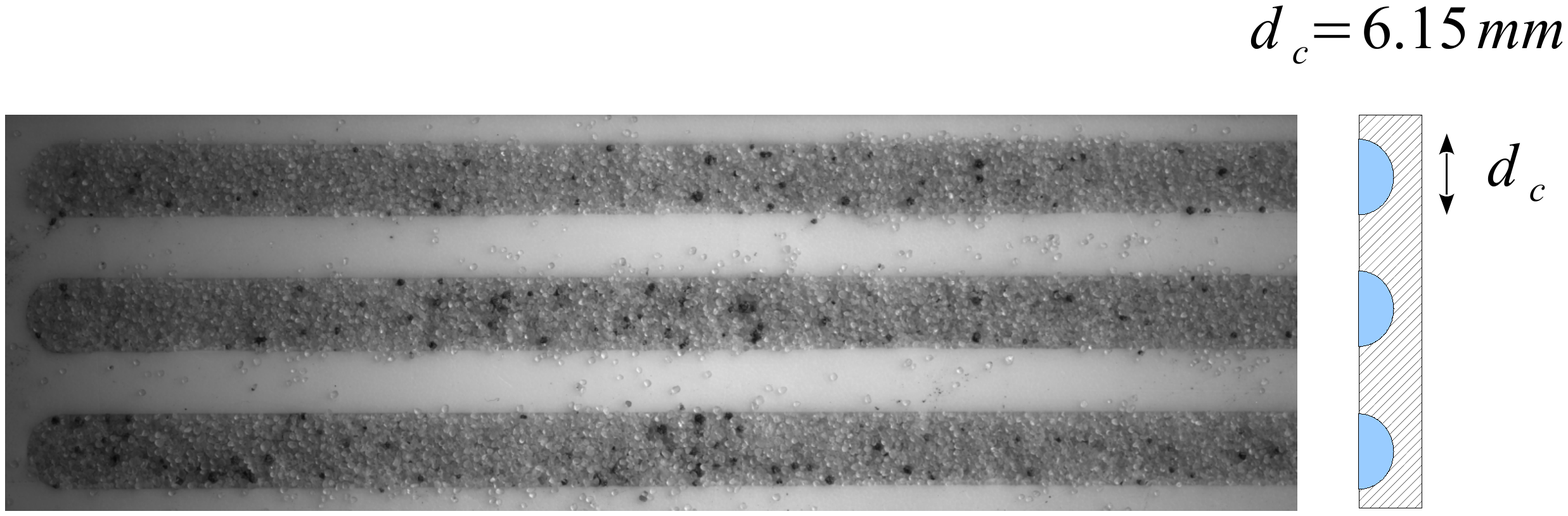}
\end{center}
\caption{Channels in Teflon (diameter of the hemicylinder $d_c=6.15\,\mbox{mm}$) filled in by the standard Ottawa sand ($R\approx 0.25\,\mbox{mm}$) before depositing $V_D=3\,\mbox{mm}^3$ liquid drops.} 
\label{Wet-channel}
\end{figure}

\begin{figure}[ht!]
\begin{center}
\includegraphics[trim=.5cm 2.cm 1cm -0.5cm,width=1\columnwidth]{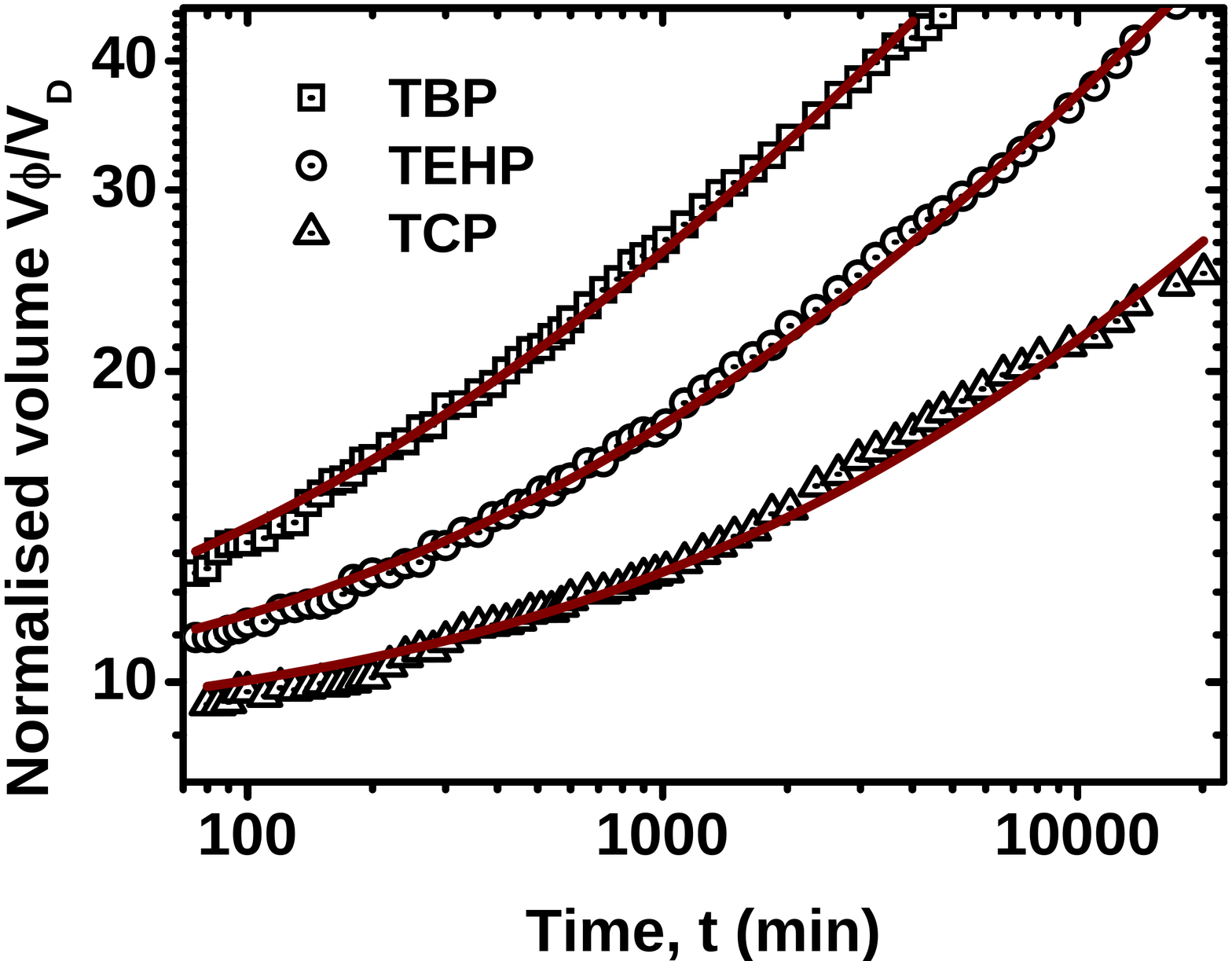}
\end{center}
\caption{Spreading of TCP, TEHP and TBP liquid drops ($V_D=3\,\mbox{mm}^3$) in $R\approx 0.25\,\mbox{mm}$ sand in the channels, as in Fig. \ref{Wet-channel}. Normalised wet volume $V\phi/V_D$ (inverse average saturation $\bar{s}^{-1}$) as a function of time. The experimental data are shown by symbols and the solid lines (brown) are the fits $V\phi/V_D=A + B t^{0.5}$.} 
\label{Fig-1-D}
\end{figure}

In another set of experiments, we studied liquid spreading in essentially one-dimensional geometry, Fig. \ref{Wet-channel}. As in the three dimensional geometry, the behaviour is characterized by an initial phase of liquid spreading and a power law corresponding to the main phase of the pendular regime, Fig. \ref{Fig-1-D}. As expected, the spreading is faster for the less viscous, well-wetting TBP liquid and slower for more viscous TCP liquid with a larger contact angle. 

The power law observed in the evolution of the moving front in the one-dimensional geometry, $X_1(t)\propto t^{0.5}$, and in the three-dimensional case, $X_3(t)\propto t^{0.25}$, suggests that in general there should be universal behaviour $X_n(t)\propto t^{1/(n+1)}$, where $n$ designates the dimension of the experimental setup. In what follows, we examine these data on theoretical grounds.  

\begin{figure}[ht!]
\begin{center}
\includegraphics[trim=.5cm 2.cm 1cm -0.5cm,width=1\columnwidth]{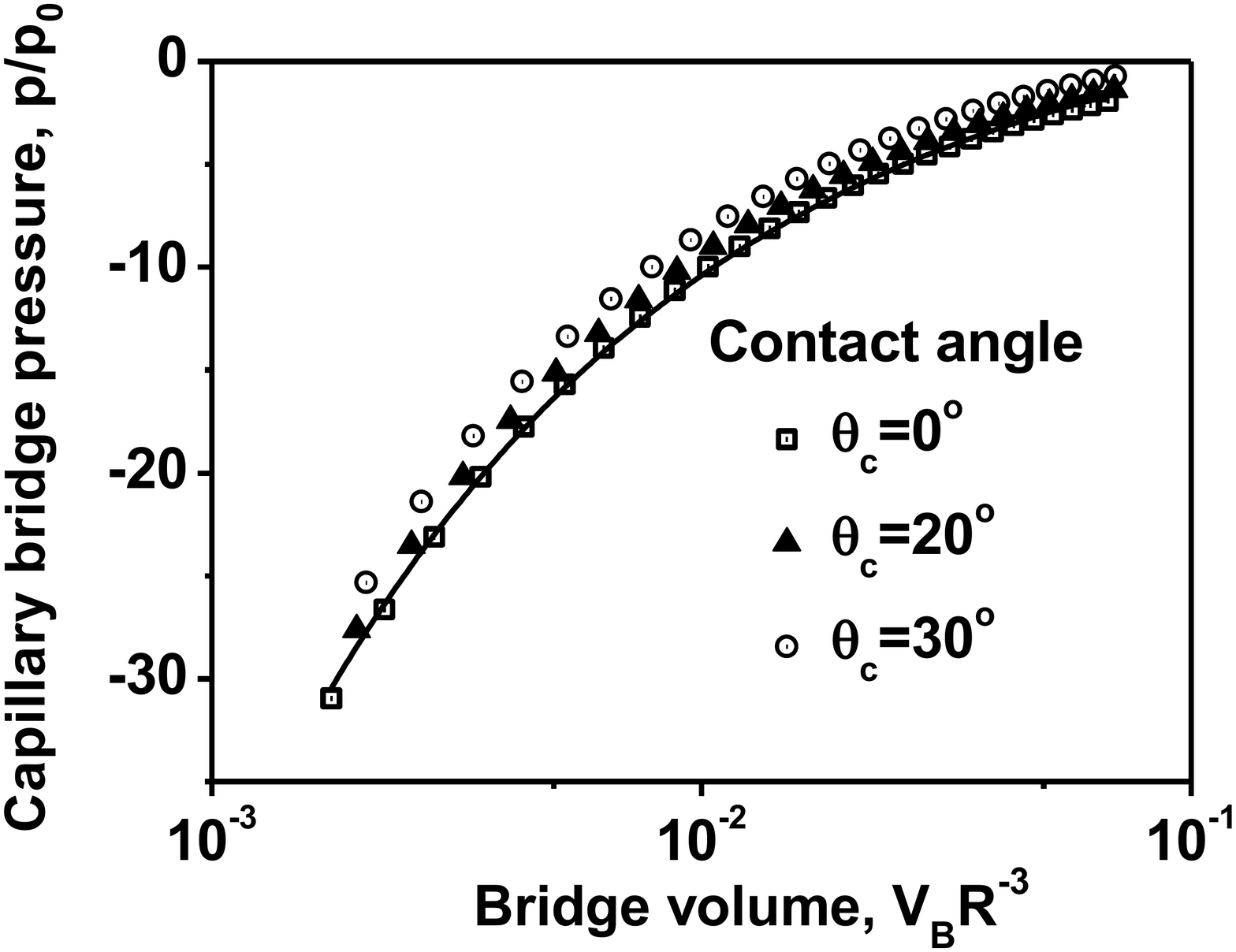}
\end{center}
\caption{Reduced capillary bridge pressure $p/p_0$ in the case of two identical solid spheres in contact (zero separation distance) as a function of the reduced bridge volume $V_B R^{-3}$ at different contact angles $\theta_c$. Symbols indicate exact solutions from~\cite{Orr-Scriven-1975}  and the solid line is the fit $p/p_0=C_0-C_1 (V_B\, R^{-3})^{-1/2}$ at $C_0=3.7, C_1=1.3$.} 
\label{Fig-Bridge}
\end{figure}

\begin{figure}[ht!]
\begin{center}
\includegraphics[trim=0.5cm 2.cm 1cm -0.5cm,width=0.85\columnwidth]{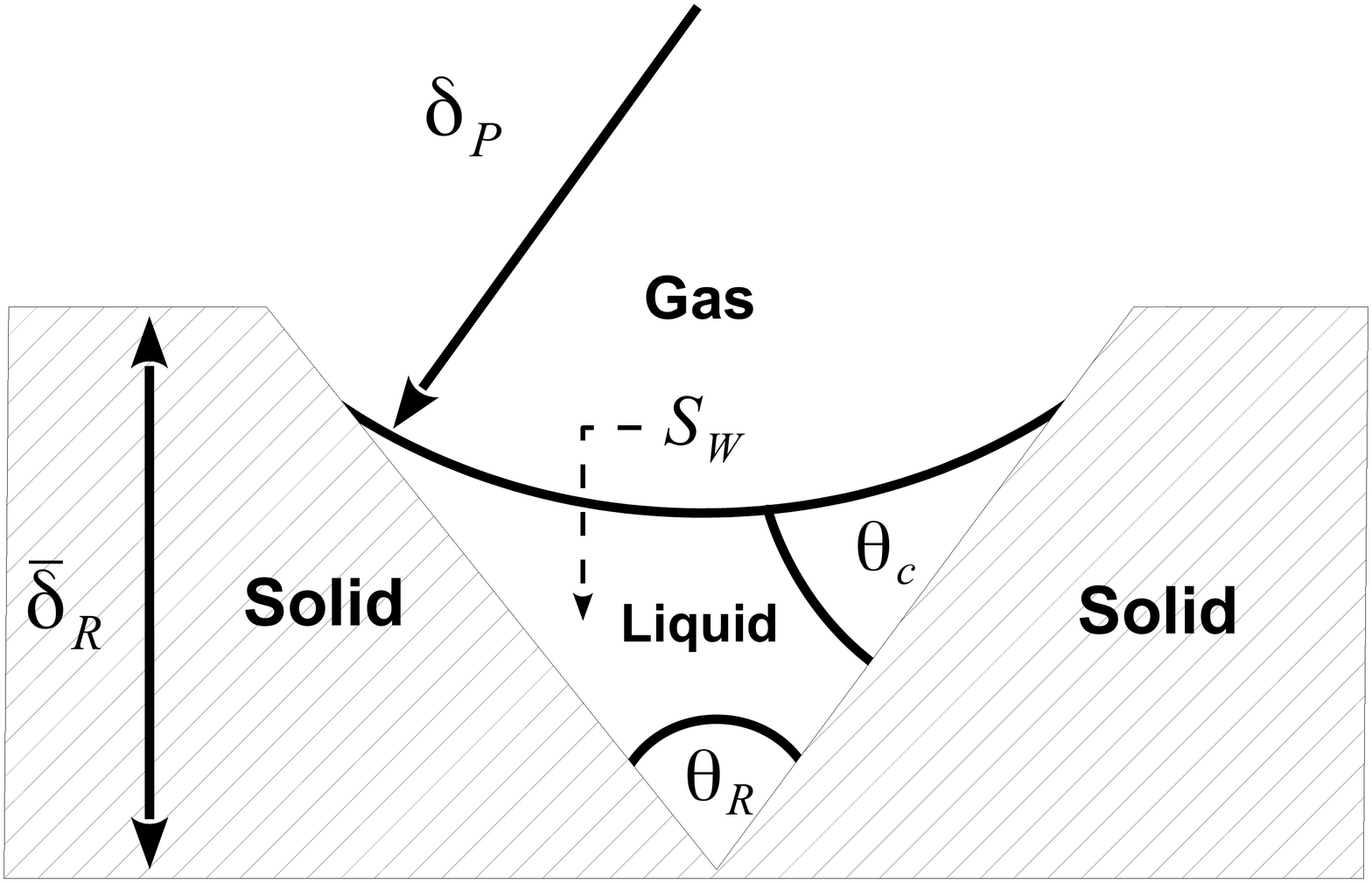}
\end{center}
\caption{Illustration of the model groove geometry with an opening angle $\theta_R$ and a contact angle $\theta_c$ used in the analysis of $\kappa_0$ and $s_0$. In the illustration, the liquid filament cross-section area $S_W$ is shown with the free surface at the capillary pressure $p=-\gamma/\delta_P$.} 
\label{Fig-groove}
\end{figure}

\section{Macroscopic model}
Consider the pendular regime of wetting, when the liquid bridges are completely isolated (that is when more complex clusters like trimmers, for example, are practically absent) and only connected via liquid films (with the thickness on the roughness length scale $\delta_R$) on the particle surfaces. The morphology of the porous media is assumed to be in static conditions, that is the particles are not moving in the process of the liquid spreading and, therefore, macroscopic properties of the porous matrix such as porosity, for example, are not functions of time. 

To obtain governing equations in the continuum limit, we consider a sample volume element $V$ in the flow domain containing many particles. On the microscopic, grain size length scale, the liquid flow in the domain, and in each sample element, takes place on the surface of particles, in the surface roughness, and through the liquid bridges connecting the flow between the particles. In the setting relevant to our experiments, the main driving force of the flow, which creates the gradient of pressure, is wetting of the dry rough solid areas ahead of the moving front. Depending on the wetting conditions (incomplete or complete wetting), the microscopic capillary pressure on the surface of grain particles forming the moving front could be either generated on the scale of the surface roughness available (incomplete wetting) or on the scale of the wetting films. While the driving pressure can reach very high levels, $\sim 10^{5}\,\mbox{Pa}$ in the case of incomplete wetting according to the roughness length scales available ($\min(\delta_R)\approx 250\,\mbox{nm}$) or $\sim 10^{7}\,\mbox{Pa}$ in the case of complete wetting, the central physics to the flow dynamics is the interplay between the capillary pressure and the size or the volume of the liquid bridges. 

 This is because while the liquid bridges do not play any active role in driving the flow, they serve as connecting elements, so that upon a substantial decrease in their volumes, the flow will be inhibited. It has been established previously, that at very high (negative) capillary pressures, the liquid bridges can only exist at the point of contacts of asperities on the rough surface areas of the grains, Fig. \ref{Pendular-Rings-roughness},~\cite{Halsey1998, He2001, Herminghaus-2005}. At the lower levels of the (negative) capillary pressure, the gap between the asperities is filled in with the liquid providing a limited, but still sufficiently large contact area. 
 
 When, on the other hand, that contact area is reduced with increase of the negative capillary pressure, the permeability of the contact, which is proportional to the square of the contact area length scale, can be dramatically diminished, so that when this critical negative pressure level is attained in the whole flow domain, the spreading should slow down dramatically, practically it should stop. 
 
 To estimate an order of magnitude of the reduction, consider typical distribution of roughness in Ottawa sands with $\min(\delta_R)\approx 250\,\mbox{nm}$ and $\max(\delta_R)\approx 3\,\mu\mbox{m}$~\cite{Alshibli2004}. The length scale of the asperities at the contact is defined by the maximal level of the surface roughness available, while the maximum capillary pressure is defined by the lower end of the roughness length scales (or by the disjoining pressure in the films). So the size of the filled-in contact area, Fig. \ref{Pendular-Rings-roughness}, is expected to be $\sqrt{\max(\delta_R) R}\sim 30\,\mu\mbox{m}$ at $R=250\,\mu\mbox{m}$~\cite{Halsey1998}, while the size of the tip contact area $\sim 1\,\mu\mbox{m}$, so that there should be an order of magnitude reduction in the size of the area and then two orders of magnitude reduction in the permeability. 

As we have already discussed, we have observed this scenario very clearly. The existence of the maximal level of the negative capillary pressure will be used in our macroscopic model. At the moving front, we should have a gradual transition between the contact zones of both kinds, with reduced contact areas above (the gap between the asperities is filled in) and below (the liquid bridges are formed only between the tips of the asperities) the critical level corresponding to the maximal level of the negative capillary pressure.  The critical pressure at the moving front should be defined by the mean values of roughness length scale distribution $\bar{\delta}_R$. The implication of this scenario (taken from the experimental observations) for our model is that, we will presume that the capillary pressure at the front is constant. This should be equivalent to some constant level of saturation, which will be defined from experimental observations.

To obtain a relationship between capillary pressure  and the liquid content, that is saturation, consider pendular rings first. In any element $V$, the averaged value of pressure in the pendular rings should be the same as the average pressure in the macroscopic liquid films in the creeping flow conditions. Basically, in the macroscopic limit, there should be no strong variations of pressure in any part of the liquid in the volume element at all, otherwise, the macroscopic description may be inadequate. 

In each individual pendular ring, the liquid content is a function of pressure, unlike liquid content in the surface roughness grooves, which should start to vary only when local (negative) capillary pressure is on the level or larger than $\gamma/\delta_R$. For a liquid bridge formed between two identical spheres of radius $R$, there is an analytical solution relating bridge free surface shape, and hence the liquid volume contained in the ring, to the capillary pressure~\cite{Orr-Scriven-1975}. The analytical expressions are quite lengthy involving, implicitly, a chain of elliptic integrals, but, for small contact angles $\theta_c\ll 1$ (between $0^{\circ}$ and approximately $30^{\circ}$) we have shown numerically, Fig. \ref{Fig-Bridge}, that the approximate relationship 
\begin{equation}
\label{Pressure-law-main}
p\approx p_0\left\{ C_0-C_1\left(\frac{R^3}{V_B} \right)^{1/2} \right\}
\end{equation}
is quite adequate;  $p_0=\frac{2\gamma}{R} \, \cos\theta_c $, $C_0=3.7$, $C_1=1.3$ and $V_B$ is the bridge volume per particle (that is a half of the actual bridge volume)~\cite{Lukyanov2012}. One can see that as the liquid content increases the capillary pressure decreases and ultimately tends to a constant value (independent of saturation). This trend was observed in both spherical grains and real sieved sands~\cite{Herminghaus-2008}. Since in the pendular regime  $s\ll 1$, that is $V_B\, R^{-3}\ll 1$, equation (\ref{Pressure-law-main}) can be further simplified to 
\begin{equation}
\label{Pressure-law}
p\approx -p_0\, C_1 \left(\frac{R^3}{V_B} \right)^{1/2}.
\end{equation}

To parametrize in terms of saturation, we split average liquid content in a sample volume $V$ containing $N\gg 1$ identical grains (neglecting dispersion of the grain particles) into two parts: the liquid contained on the rough surface of particles of volume $V_r=4\pi \alpha_R R^2 \delta_L N$ and the liquid contained in the capillary bridges $V_c=\langle V_B\rangle N_c\, N$. Here, $\langle V_B \rangle$ is the average bridge volume in $V$ and parameter $N_c$ is the coordination number, that is the average number of bridges per a particle. In our experiments the value of $N_c$ was found to be around $N_c\approx 7$, which is further assumed to be constant $N_c=const$.   

Combining both contributions, saturation $s$ can be presented as
\begin{equation}
\label{Saturation}
s=\frac{V_c+V_r}{\phi V}=\langle V_B\rangle  R^{-3}A_s  + s_0,
\end{equation}
where
$$
A_s=\frac{3}{4}\, \frac{1-\phi}{\phi} \frac{N_c}{\pi}
$$
and $s_0$ is given by (\ref{Surface-saturation}).

Treating the bridge volume $V_B$ as an average, using (\ref{Pressure-law}) and (\ref{Saturation}), the average capillary bridge pressure $P=\langle p\rangle ^l$ in the volume element $V$ can be presented as  
\begin{equation}
\label{Pressure-saturation-pr}
P=-p_0 \frac{A_c}{(s-s_0)^{1/2}}, \quad A_c=C_1 A_s^{1/2},
\end{equation}
where
$\langle ...\rangle ^l=V_l^{-1}\int_{V_l} d^3 x$ is intrinsic liquid averaging, and $V_l$ is liquid volume within the sample volume $V$. We would like to emphasize that so far in obtaining the non-linear pressure-saturation relationship, we had made no assumptions about possible dependence of $s_0$ on the capillary pressure itself. The obtained result solely takes into account the fact how the average bridge volume is reacting to variable average capillary pressure. The potential effects of variations of $s_0$ with capillary pressure are discussed in the next section.

We note, that the singularity in (\ref{Pressure-saturation-pr}) as saturation $s$ tends to the critical value $s_0$ is formal. In a similar way, capillary pressure in a drop formally diverges as its radius $R$ vanishes, $p\propto 1/R$. We presume that within the macroscopic domain, including the boundary, where the liquid flow takes place, the average bridge volume is always non-zero, so that it is always the case that $s\ge s_f > s_0$. Also, relationship (\ref{Pressure-saturation-pr}) is only valid, strictly speaking, when the bridge is formed between two particles, where the characteristic length scale is the particle radius $R$, but not between the tips of the surface roughness asperities, where the characteristic length scale is the size of the asperity, so that when the bridge volume actually vanishes, the relationship should be corrected to take into account the change in the solid surface curvature, if this would be necessary. In the present formulation, we do not analyse the asperity regime in details assuming that the system comes into equilibrium just when this transition occurs.  

In the remaining of this section, in part A, we discuss variations of parameter $s_0$ with the capillary pressure, in part B, consider local flow in the grooves and their permeability, and in part C, using the pressure-saturation relationship (\ref{Pressure-saturation-pr}) obtain macroscopic governing equations. To enhance the accuracy of the model predictions, we estimate surface permeability of spherical particles in part D to obtain a correction to the effective coefficient of dispersion. In part E, before turning into a comparison of the model with experimental data, we analyse and discuss similarity properties of the main governing equation with a set of boundary conditions to understand potential asymptotic behaviour, which might be expected from this kind of mathematical problems.

\subsection{Surface liquid content at variable capillary pressure}

We have assumed previously that parameter $s_0$ is constant, that is independent of the capillary pressure~\cite{Lukyanov2012}. This is a good approximation over a range of capillary pressures, but could be possibly violated at small values of $s\approx  s_f \approx s_0$, when the absolute value of (negative) capillary pressure is at its maximum. Here, we test the accuracy of this assumption on the basis of a model one-dimensional surface groove geometry shown in Fig. \ref{Fig-groove}. 

Liquid steady states and surface flows in that kind of geometry have been studied previously in detail both experimentally and theoretically~\cite{Concus1969, Ransohoff1988, Yost-1996, Yost-1998, Tuller-2000, Herminghaus2011}. The first thing to mention here is that liquid morphology in such V-shaped grooves can be either a liquid drop or a filament depending on the groove opening angle $\theta_R$~\cite{Concus1969, Herminghaus2011}. Clearly, imbibition into the groove is only possible when the liquid volume has a shape of a filament. The liquid morphology changes from a filament to a drop, when the opening angle $\theta_R$ obtains a critical value $\theta_R^m= \pi - 2\theta_c$ from below at a given contact angle $\theta_c$. So that in our analysis we assume that the opening groove angle $\theta_R$ is always smaller than the critical value $\theta_R^m$. Given the range of contact angles in our case, that is $0\le \theta_c\le \pi/6$, condition $\theta_R\le \theta_R^m$ does not impose significant restrictions.

Depending on the capillary pressure, the meniscus radius of curvature in the groove could be much larger, about equal, or smaller than the characteristic length scale of the groove. For example, at  $s=s_c=10\%$, $R=250\,\mu\mbox{m}$ ($s_c$ is the saturation at the onset of the pendular regime of wetting), $\theta_c=0$ and $s_0=0.6\%$ from (\ref{Pressure-saturation-pr}) the radius of curvature of the meniscus would be around $15\,\mu\mbox{m}$, which is much larger than the maximum surface roughness amplitude in our sand grains, $\max(\delta_R)\approx 3\,\mu\mbox{m}$. But as $s$ tends to $s_f$, for example at $s=0.65\%$, the radius of curvature would be already only $1\,\mu\mbox{m}$, which is comparable with the characteristic length scale of the groove, as is shown in Fig. \ref{Fig-groove}. 

When the radius of curvature is much larger than the groove dimensions, the meniscus contact line points are pinned to the groove edges, the meniscus shape is almost flat and the groove is completely filled in with the liquid~\cite{Tuller-2000}. In the one dimensional geometry shown in Fig. \ref{Fig-groove}, liquid content in this state could be approximately (neglecting small curvature of the free surface) characterised by the cross-sectional area of the fully filled-in groove $S_F$, which is obviously
$$
S_F(\theta_R)=\bar{\delta}_R^2\frac{\sin(\theta_R/2)}{\cos(\theta_R/2)}.
$$

At the same time, when the radius of curvature is smaller than the groove dimensions, the liquid, simply by geometrical considerations at a given contact angle $\theta_c$, would only partially fill in the groove available volume, as is shown in Fig. \ref{Fig-groove}. The cross-sectional area in this case at a given capillary pressure $p=-\gamma/\delta_P$ can be represented as 
\begin{equation}
\label{GA}
S_W(\theta_R, \theta_c,p^2)=\frac{\gamma^2} {p^2}\, F_c(\theta_R, \theta_c),
\end{equation}
where 
$$
F_c(\theta_R, \theta_c)=\left\{\frac{\cos(\theta_R/2)}{\sin(\theta_R/2)}\cos^2(\theta_c+\theta_R/2)-\right.
$$
$$
\left. \frac{\pi}{2}+\theta_c + \frac{\theta_R}{2} +\cos(\theta_c+\theta_R/2) \sin(\theta_c+\theta_R/2)\right\}.
$$
One can formally notice, that when the opening angle $\theta_R$ is attending the critical value $\theta_R^m$ from below, the surface area tends to zero $S_W(\theta_R^m)=0$, as one can see from (\ref{GA}). This is another manifestation of the liquid morphology change at $\theta_R=\theta_R^m$. Obviously, grooves with $\theta_R\ge \theta_R^m$ are unlikely to be filled during natural spreading.  

Based on the groove geometry, it is not difficult to discern that the contact line would be at the groove edge, when the following condition is satisfied
$$
\frac{\cos(\theta_R^c/2)}{\sin(\theta_R^c/2)}\cos(\theta_c + \theta_R^c/2) = \frac{\bar{\delta}_R |p|}{\gamma},
$$
which defines a critical angle $\theta_R=\theta_R^c$ at given capillary pressure $p$ and the groove size $\bar{\delta}_R$. When the opening corner angle is attending the critical value $\theta_R^c$ from above, the contact line moves to the groove edge and remains there, due to the contact line pinning to the edges, for any further reduction of $\theta_R$ and the absolute value of the capillary pressure. Using characteristic front pressure $\displaystyle p_f\approx - 3.6\times 10^{4}\,\mbox{Pa}$ for TCP and $\bar{\delta}_R=0.8\,\mu\mbox{m}$, Table \ref{Table1}, one can estimate that $\theta_R^c\approx \frac{3}{8}\pi$ at $\theta_c=\pi/6$.

In general, surface roughness, even in a simplified case, should be represented by some distribution of grooves having different parameters, such as opening angle $\theta_R$ and the groove depth. Following a statistical approach applied in the similar kind of the groove geometry~\cite{Tuller-2000}, we obtain averaged microscopic properties using several simplifying assumptions. In particular, we apply averaging over the opening corner angle $\theta_R$  assuming constant groove depth $\bar{\delta}_R$ and a uniform angular distribution in a range $\theta_R\in [0,\pi/2]$, where the upper limit was chosen to avoid large, obtuse opening angles, which are rarely observed~\cite{Alshibli2004}. For simplicity, we presume that there are only two states of the groove filling separated by the critical value $\theta_R^c$; the grooves with the free surface pinned to the groove edges are assumed to be fully filled in, their liquid content is constant and is characterized by the cross-sectional area $S_F$. Otherwise, the grooves are assumed to be partially filled in, and their cross-sectional area is characterized by $S_W$ given by (\ref{GA}). 

From this simple geometric and statistical considerations, the saturation level $s_0$ being an average quantity is expected to be inversely proportional the square of the capillary pressure and decrease with $\theta_c$ increasing due to the presence of partially filled in grooves. Apparently, as the pressure amplitude decreases, all grooves in the range would be eventually filled in and parameter $s_0$ would attain a constant value. It is not difficult to estimate that for TCP, for example, at $p\approx - 1.2\times 10^{4}\,\mbox{Pa}$ ($s\approx 1\%$), the critical angle $\theta_R^c>\pi/2$. This implies that the saturation due to the liquid residing  in the surface grooves would vary in a range $s_0\in[s_0^e,s_0^m]$, when saturation $s$ changes in $s\in[s_f,s_c]$ ($s_c\approx 10\%$), where the maximum value $s_0^m$ should be solely defined in the one-dimensional geometry by the averaged cross-section surface area of fully filled-in grooves $s_0^m\propto 2/\pi \int_0^{\pi/2}\, S_F\, d\theta_R$, while the minimal value $s_0^e$ should also reflect a contribution from the surface area of partially filled-in grooves $s_0^e\propto  2/\pi \left(\int_{0}^{\theta_R^c}\, S_F\, d\theta_R +  \int_{\theta_R^c}^{\pi/2}\, S_W\, d\theta_R\right)$. In particular, one can obtain two important averaged parameters, the ratio
\begin{equation}
\label{s0-scaling}
\frac{ s_0^m}{ s_0^e} =\frac{\int_{0}^{\pi/2} S_F\,d\theta_R }{\int_{0}^{\theta_R^c} S_F\,d\theta_R + \int_{\theta_R^c}^{\pi/2} S_W\,d\theta_R }
\end{equation}
and the characteristic length scale of the liquid layer at equilibrium
\begin{equation}
\label{dl-scaling}
\frac{\delta_L}{\bar{\delta}_R} = \sqrt{ \frac{\int_{0}^{\theta_R^c} S_F\,d\theta_R + \int_{\theta_R^c}^{\pi/2} S_W\,d\theta_R } {\int_{0}^{\pi/2} S_F\,d\theta_R } }.
\end{equation}

If we now take the minimum value of $s_0$ from our experiments with TCP, $s_0^e \approx 0.6\%$, and fix the capillary pressure at the characteristic value $\displaystyle p_f\approx - 3.6\times 10^{4}\,\mbox{Pa}$ and the surface roughness amplitude at $\bar{\delta}_R=0.8\,\mu\mbox{m}$, Table \ref{Table1}, then the average maximum value $s_0^m$ can be estimated using (\ref{s0-scaling}) at $ s_0^m \approx 0.9\%$ at $\theta_c=\pi/6$ giving a range of variations of $s_0$. At the same time, the averaged depth of the liquid layer $\delta_L$ can be estimated using (\ref{dl-scaling}) at $\delta_L\approx 0.7\,\mu\mbox{m}$, Table \ref{Table1}.

Considering that $s_0^m$ was found to be close to $s_0^e$ and $s_0^m\ll s_c$, one can set parameter $s_0$ without loss of accuracy at its equilibrium value $s_0^e$, so that the pressure-saturation relationship becomes  
\begin{equation}
\label{Pressure-saturation}
P\approx -p_0 \frac{A_c}{(s-s_0^e)^{1/2}}.
\end{equation}

Further, in part B, we consider grooves permeability on the basis of the simplified model groove geometry, Fig. \ref{Fig-groove}, and the statistical approach, which have been implemented in this section. While we have established that variations of the surface liquid content with capillary pressure can be in principle neglected in our problem, the same variation of the capillary pressure can have much stronger effect on the surface permeability. This is due to the fact that only $10\%$ of the surface grooves are actually fully connected and can conduct the flow on the particle surface~\cite{Tuller-2000}. As we will show, if the properly connected grooves are those that experience partial filling, the effect is expected to be much stronger, exactly as we observed in our experiments.

\subsection{Surface conductivity and the groove geometry}

Consider now the local transport on the surface of particles, which is described by the average surface flux density ${\bf q}$. The quantity is defined by averaging the volumetric flux over a sample cross-section area containing many grooves and including areas of both solid and liquid. According to a study of liquid spreading on rough surfaces made of microscopic grooves of various shapes and dimensions~\cite{Ransohoff1988, Yost-1996, Yost-1998, Tuller-2000},  the flow on average obeys a Darcy-like law
\begin{equation}
\label{Micro-Darcy-1}
{\bf q}=-\frac{\kappa_m}{\mu} \nabla \psi,
\end{equation} 
where $\mu$ is liquid viscosity, $\psi$ is the averaged pressure within the surface roughness and $\kappa_m$ is the effective coefficient of permeability. 

We consider a non-dimensional quantity $\kappa_0$, which is defined by $\kappa_m=\kappa_0 \bar{\delta}_R^2$. To understand its parametric dependencies, we again consider the surface grooves of a simplified geometry, as is shown in Fig. \ref{Fig-groove}, and use the example as a guide. 

We note, that even in this simplified one-dimensional case, there are no closed form analytical solutions available to describe the flow, and a numerical treatment should be applied~\cite{Ransohoff1988}. The results of numerical analysis of corner flows performed in one-dimensional geometry assuming a fully developed rectilinear Hagen-Poiseuille flow in the open channel, Fig. \ref{Fig-groove}, at different opening and contact angles, $\theta_R$ and $\theta_c$ respectively, can be represented in terms of a non-dimensional coefficient of flow resistance 
$$
\beta = -\frac{\bar{\delta}_R^2}{\mu \hat{q}}\frac{d\psi}{dz},
$$
where the $z$-axis is along the groove, $\psi$ is average pressure in the grooves and $\hat{q}$ is the average volumetric flux density (the average liquid velocity) inside the groove~\cite{Ransohoff1988}. Since the averaging in (\ref{Micro-Darcy-1}) included areas of solid, the two quantities $q$ and $\hat{q}$ are related in the one-dimensional case through parameter $\alpha_R$ (the fraction of the surface roughness volume occupied by the liquid ), that is $\alpha_R \hat{q} = q$.  

As a result, $\beta$ can be interpreted as the inverse non-dimensional permeability of the surface grooves $\beta=\alpha_R/\kappa_0$.  In the setting, Fig. \ref{Fig-groove}, parameter $\beta$ is a function of the contact and opening angles, $\theta_c$ and $\theta_R$, which has been tabulated using numerical simulations~\cite{Ransohoff1988, Tuller-2000}. In a particular case of complete wetting $\theta_c=0$ and capillary pressure $p=-\gamma/\sqrt{2}\bar{\delta}_R$, $\beta$ can be parametrized as~\cite{Tuller-2000} 
\begin{equation}
\label{beta-exact}
\beta(\theta_R)=\frac{1}{2}\exp\left(\frac{2.124+0.4486\, \theta_R}{1-0.2377\, \theta_R}\right).
\end{equation}

\begin{figure}[ht!]
\begin{center}
\includegraphics[trim=-1.5cm 2.cm 1cm -0.5cm,width=0.9\columnwidth]{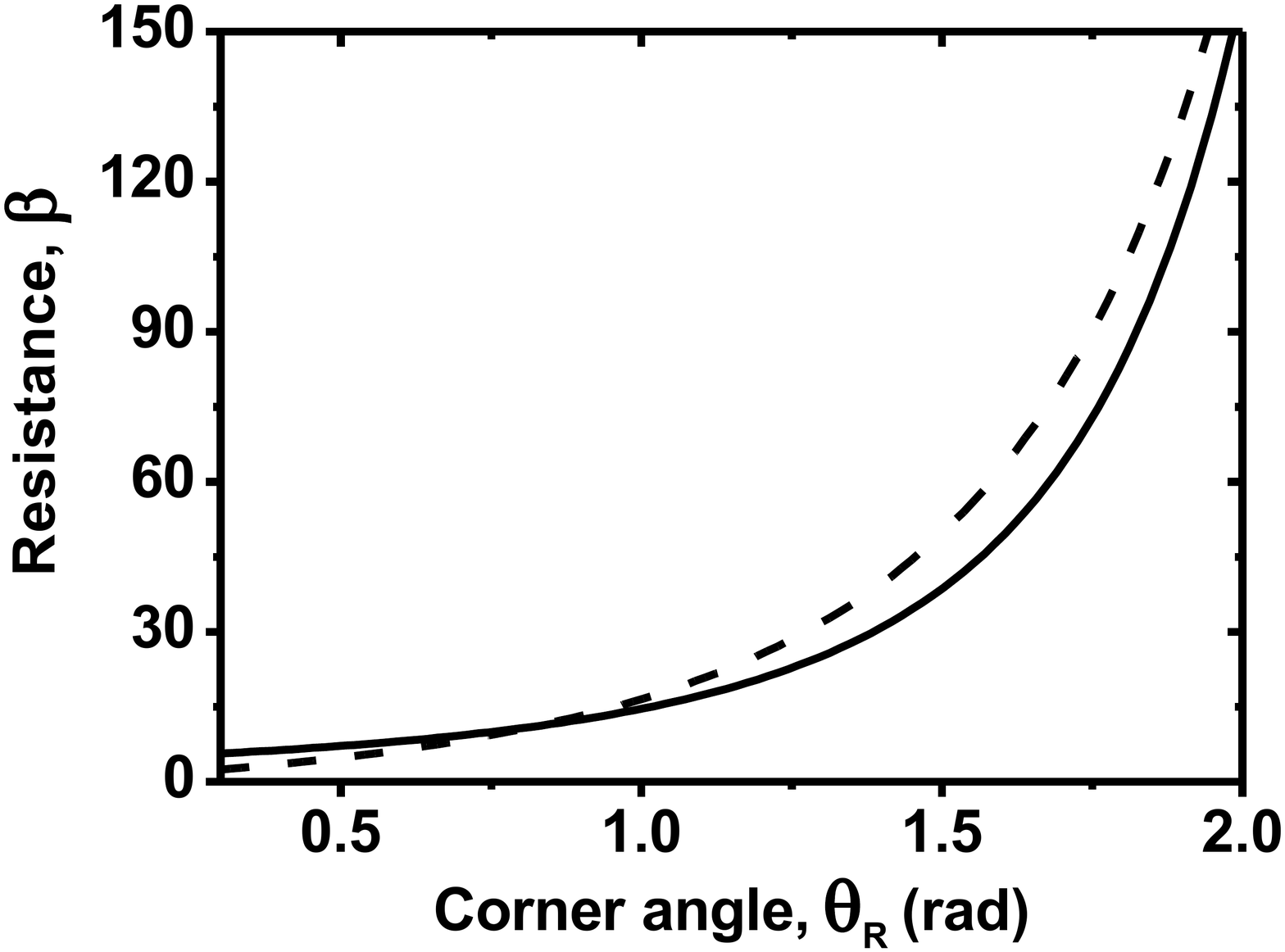}
\end{center}
\caption{Comparison between the hydraulic approximation and the exact solution. Non-dimensional flow resistance $\beta$ as a function of the groove corner angle $\theta_R$ at $\theta_c=0$ and $p=-\gamma/\sqrt{2}\bar{\delta}_R$. The solid line is exact numerical solution (\ref{beta-exact}) and the dashed line is the result in the hydraulic approximation (\ref{Hydraulic}).} 
\label{Resistance}
\end{figure}

This would be instructive to compare the exact result obtained in the one-dimensional flow configuration with the approximation of the hydraulic diameter, when 
\begin{equation}
\label{Hydraulic}
\beta\approx \frac{8\pi \bar{\delta}_R^2}{S},
\end{equation} 
where $S$ is the liquid cross-section area. The approximation (\ref{Hydraulic}) is exact for circular cross-section $S=\pi\bar{\delta}_R^2$, and it provides a reasonable approximation to calculate $\beta$ in the corner flow configurations, Fig. \ref{Resistance}, given that the actual surface flows, we approximate, take place over complex two-dimensional landscape, where the flow conditions are complicated by effects of tortuosity~\cite{Matyka-2008, Tortuosity-Review}. Therefore, in what follows, we use the hydraulic approximation. 

So far, we effectively assumed that all grooves, where the liquid is residing, can conduct the flow. This is not the case according to the analysis done in~\cite{Tuller-2000}, where the connectivity factor accounting for the portion of the grooves contributing to the surface flow was found to be as low as $0.1$. So that, we also introduce a phenomenological parameter $\xi_{f}$ to account for the portion of the grooves, which is able to conduct the flow. The parameter will be defined by comparison with experiments. Obviously, it should also incorporate parameter $\alpha_R$ and the effects of the complex three-dimensional groove geometry, such as tortuosity. That is, $\xi_{f}\propto \alpha_R$, and given $\alpha_R\approx 0.3$ found from our comparison with the experiments, it is expected that $\xi_{f} \ll 0.3$.

To introduce the effects of the contact angle into the model, we use averaging over $\theta_R$ and approximate using a linear relationship
$$
\kappa_0(s) = \frac{\kappa_0^{(2)} - \kappa_0^{(1)}}{s_c-s_f} \left ( s-s_f \right) +\kappa_0^{(1)}, \quad s\in [s_f, s_c],
$$
where $s_c\approx 10\%$, and parameters $\kappa_0^{(1)}$ and $\kappa_0^{(2)}$ are the average quantities corresponding to configurations with and without partially filled-in grooves respectively. The maximum value of the permeability $\kappa_0^{(2)}$ is fully defined by the average cross-section area $$\langle S_F \rangle =\frac{1}{\delta \theta_R}\int_{\min(\theta_R)}^{\pi/2}S_F\,d \theta_R, \quad\delta \theta_R=\frac{\pi}{2}-\min(\theta_R).$$ That is
$$
\kappa_0^{(2)}=\frac{\xi_{f} \langle S_F \rangle}{8\pi\bar{\delta}_R^2}.
$$ 
While the minimum value $\kappa_0^{(1)}$ is also defined by a contribution of parameter $S_W$ at the capillary pressure given by (\ref{Pressure-saturation}) at $s=s_f$. That is
\begin{equation}
\kappa_0^{(1)}= \frac{\xi_{f}}{8\pi \bar{\delta}_R^2\, \delta\theta_R}\left\{\int_{\min(\theta_R)}^{\theta_R^c} S_F\, d\theta_R + \right.
\label{Permeability-contact-angle}
\end{equation}
$$
\left. \int_{\theta_R^c}^{\pi/2} S_W\, d\theta_R \right\}.
$$

The minimum value of $\theta_R$ used in the averaging of the permeability coefficient was set to the critical value $\theta_R^c=\frac{3}{8}\pi$ found in the case of TCP at $\theta_c=\pi/6$ and the characteristic pressure at the moving front $p_f$. The choice is to maximize the effect of the contact angle on the permeability of the surface grooves, and is dictated by our experimental observations of spreading of different liquids. Indeed, as one can see from Table \ref{Table1}, the factors $\gamma\cos\theta_c/\mu$ contributing into the effective coefficient of diffusion $D_0$ in (\ref{Superfast-2}) for TCP and TEHP liquids are practically identical, while, as we discuss in detail later, the propagation curves shown in Fig. \ref{Numerical-solution-3} suggest that the coefficient of diffusion should be at least a factor of six different, see Table \ref{Table1}, indicating that the difference should come from $\kappa_0(s)$. The grooves with sharp angles below $\theta_R^c=\frac{3}{8}\pi$ can be assumed to be fully filled in with the liquid for both TCP and TEHP, so that their permeability would be the same. So, the inclusion of the grooves with sharp angles in the averaging procedure would reduce the effect of the contact angle on the surface permeability. 

Apparently, the choice of the minimal angle $\theta_R$ should not depend on the liquids considered, since it reflects the connectivity properties of the surface roughness.  Whether or not such a choice is fully justified should be seen in further experimental and theoretical studies of the surface flows using more realistic models of the surface roughness. Here, we use (\ref{Permeability-contact-angle}) as a guide to understand, if it is feasible within the model to unify all experimental observations with sufficient accuracy. We also note that the value of $\xi_f=0.038$ found later in the comparison with the experiments indicates that in the chosen partition of the groove opening angles, the connectivity factor was about $0.25$ given that $\alpha_R\approx 0.3$ and the tortuosity effects would reduce permeability at least twofold. The larger value of the connectivity factor than that found in~\cite{Tuller-2000} indicates that indeed the grooves with sharp opening angle are very poorly connected.   

\subsection{Macroscopic governing equation of the super-fast diffusion model}

According to the spatial averaging theorem~\cite{Whitaker-1969}, applying intrinsic liquid averaging $\langle ...\rangle ^l$
\begin{equation}
\label{Macroscopic-Darcy-1}
-\frac{\kappa_m}{\mu} \left\{ \nabla \langle \psi\rangle ^l + V_l^{-1}\, \int_{S_l} \psi\, {\bf n}\, dS \right\}=\langle {\bf q}\rangle ^l, 
\end{equation} 
where $S_l$ is the area of the liquid interface confined inside the volume element $V$ and with normal vector $\bf n$. The surface integral in the creeping flow conditions, when the pressure variations across the liquid layer are insignificant, can be neglected $V_l^{-1}\, \int_{S_l}\, \psi\, {\bf n}\, dS\approx 0$ and 
\begin{equation}
\label{Macroscopic-Darcy-2}
-\frac{\kappa_m}{\mu}  \nabla \langle \psi\rangle ^l =\langle {\bf q}\rangle ^l. 
\end{equation} 
Thus, one can cast the continuity equation,
$$
\frac{\partial \phi s}{\partial t} + \nabla\cdot {\bf Q}=0
$$
into
\begin{equation}
\label{Gov-1}
\frac{\partial \phi s}{\partial t}=\nabla\cdot \left\{ \frac{K}{\mu}  \nabla P\right\}.
\end{equation}
Here, 
\begin{equation}
\label{Def-SE}
{\bf Q}=\frac{S_e}{S}\langle {\bf q}\rangle ^l,
\end{equation} 
$S$ is the surface area of the sample volume $V$ with the effective area of entrances and exits $S_e$ and coefficient $K=\kappa_m \frac{S_e}{S}$. It is assumed that in creeping flow conditions $P= \langle p\rangle ^l \approx \langle \psi\rangle ^l$. Note, that the ratio $S_e/S$ is not strictly speaking just a geometric factor. It is an average quantity defined by (\ref{Def-SE}), which incorporates connectivity and the shape of the surface elements. 

To estimate effects of gravity, we first notice that the capillary pressure is assumed to be generated on a length scale $\bar{\delta}_R\sim 0.8\,\mu\mbox{m}$. If we now compare the capillary length $l_c=\sqrt{\gamma/\rho g}\sim 2\,\mbox{mm}$, where $\rho$ is liquid density and $g$ is the gravity constant, with the length scale associated with the gradient of capillary pressure $\sqrt{\bar{\delta}_R L_0}$, where $L_0\sim 10\,\mbox{mm}$ is the characteristic length scale of the wetting area in our experiments, then $l_c\gg \sqrt{\bar{\delta}_R L_0} \sim 0.1\,\mbox{mm}$. This implies that the gravity effects can be ignored. At the same time, the length scale associated with the gradient of capillary pressure $\sqrt{R L_0}$ in the funicular regime may be comparable with $l_c$  so that the accuracy of our approximation may be reduced.

Assuming further that porosity $\phi$ is constant and using expression (\ref{Pressure-saturation}) for the average pressure, one can transform the governing equation (\ref{Gov-1}) into a non-linear diffusion equation for the saturation $s({\bf x},t)$
\begin{equation}
\label{Superfast-1}
\frac{\partial s}{\partial t}= \nabla\cdot  \left\{  \frac{D_s \, \nabla s}{(s-s_0^e)^{3/2}}\right\}, 
\end{equation}
where
$$
D_s=\frac{1}{2}\frac{K}{\mu}\frac{p_0 A_c}{\phi }. 
$$

To address a moving boundary value problem set in an open domain with a smooth boundary $\partial \Omega$ moving with velocity ${\bf v}$, the governing equation (\ref{Superfast-1}) is complemented with the boundary conditions
\begin{equation}
\label{BC1}
\left. s\right|_{\partial \Omega}=s_f, \quad s_f>s_0^e
\end{equation}
and
\begin{equation}
\label{BC2}
\left. {\bf v\cdot n}\right|_{\partial \Omega}=\left. v_n\right|_{\partial \Omega}=-D_s\frac{{\bf n}\cdot\nabla s}{s_f(s_f-s_0^e)^{3/2}},
\end{equation}
where $\bf n$ is the normal vector to the boundary $\partial \Omega$. The boundary value of the saturation $s_f$ is defined by the capillary pressure developed at the moving front. To be precise, the inverse of the reduced capillary pressure (capillary pressure normalized by $2\gamma/R$) is related with the difference of two parameters $s_f-s_0^e$. So that the first boundary condition at the moving front is set by the assumption of the maximum capillary pressure, which is presumed to be constant in the model. At the same time, parameter $s_f$ defines a steady state saturation level, when the network connectivity is reduced but not broken. The second boundary condition sets the velocity of the moving front in the assumption that the front is moving into a dry area. We would like to point out that in the study, we treat parameters $s_f$ and $s_0^e$ as phenomenological, and determine them from the observations. 

To get an estimate of the typical values of the boundary pressure and the saturation, we assume that the pressure is generated by capillaries with a characteristic size of the order of $\bar{\delta}_R$. Then, for example for TCP, taking  surface tension $\gamma=42.5\,\mbox{mN}/\mbox{m}$ at $25^{\circ}\, \mbox{C}$, the capillary pressure $\displaystyle |P|=\frac{\gamma}{\bar{\delta}_R} \approx 5.3 \times 10^{4}\,\mbox{Pa}$ at $\bar{\delta}_R=0.8\,\mu\mbox{m}$. As a result, from (\ref{Pressure-saturation}), taking typical parameter values $R = 250\,\mu\mbox{m}$ and $\phi=0.3$, parameter $s_f-s_0^e\approx 4\times 10^{-4}$, which is close to the values found in the previous analysis of experimental data, ~\cite{Lukyanov2012}.  Note that  $\frac{s_f-s_0^e}{s_0^e}\ll 1$, considering that $s_0^e\approx 0.006$. 

In general, using (\ref{Pressure-saturation}), one can obtain the following scaling of $s_f-s_0^e$ with the grain size $R$
\begin{equation}
\label{Scaling-P-1}
s_f-s_0^e=4 \frac{A_c^2 \gamma^2 \cos^2\theta_c}{p_f^2 R^2},
\end{equation}
where $p_f$ is the capillary pressure at the front.
That is, taking into account (\ref{Surface-saturation}), a similar scaling for $s_f$ is given by
\begin{equation}
\label{Scaling-P-2}
s_f=3\alpha_R  \, \frac{1-\phi}{\phi} \frac{\delta_L}{R} + 4 \frac{A_c^2 \gamma^2 \cos^2\theta_c}{p_f^2 R^2}.
\end{equation}
These relationships will be further used in the analysis of experimental data to estimate the main non-dimensional model parameters $s_f$ and $s_f-s_0^e$.

\subsection{Global surface permeability of a system of spherical particles.}

To simulate liquid spreading with the help of (\ref{Superfast-1}), the coefficient of permeability $K$ and hence the parameter $S_e/S$ need to be determined somehow.

\begin{figure}[ht!]
\begin{center}
\includegraphics[trim=-1.5cm 3.cm 1cm -0.5cm,width=0.9\columnwidth]{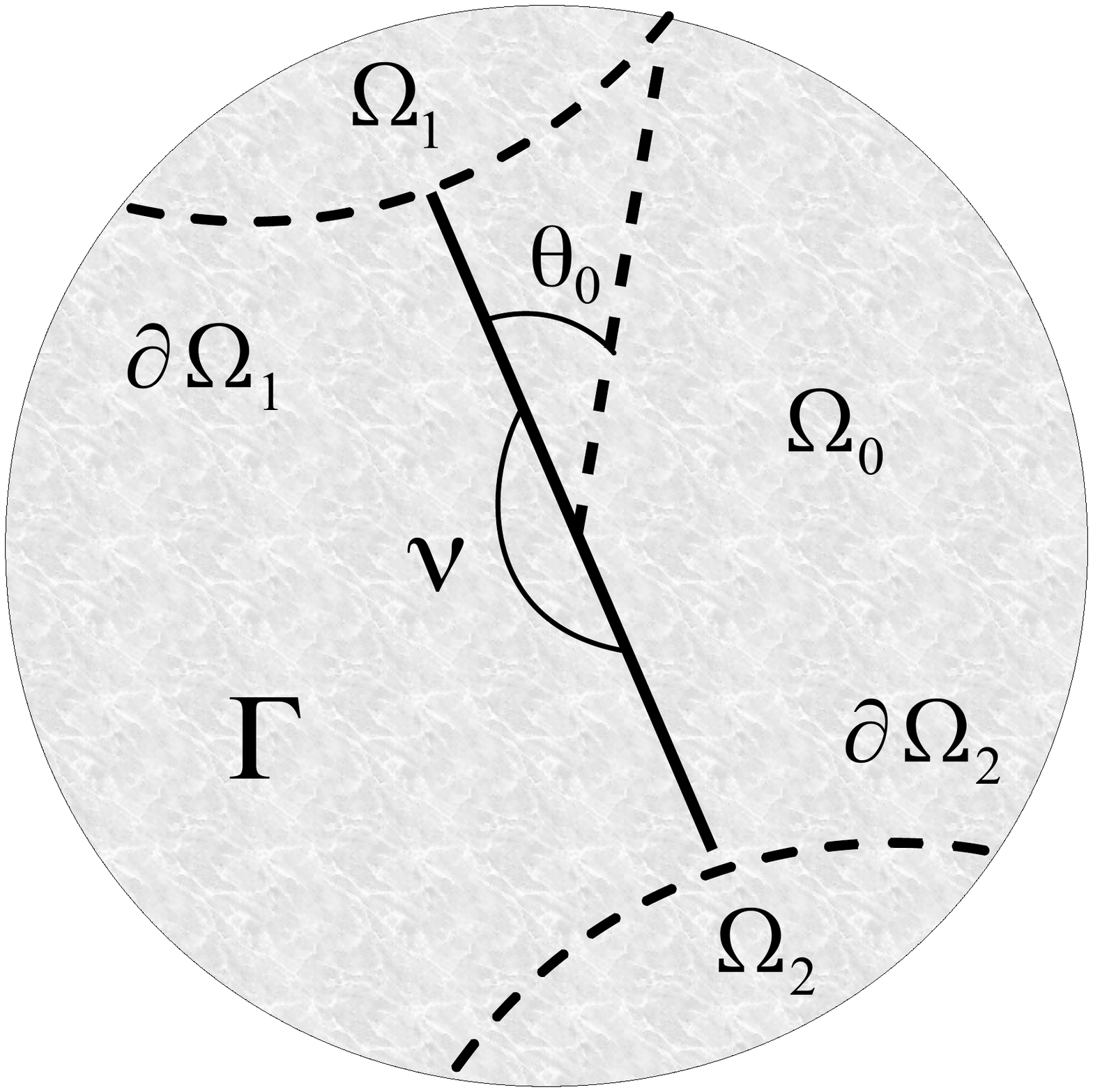}
\end{center}
\caption{Illustration of the surface diffusion domains.} 
\label{Fig-Beltrami-sphere}
\end{figure}

To obtain an estimate of these parameters, we consider surface flow in steady state conditions over just one single particle with a closed surface $\Gamma$, as is shown in Fig. \ref{Fig-Beltrami-sphere}. The particle surface is split into three sub-domains $\Omega_0$, $\Omega_1$ and $\Omega_2$ with surface boundaries between them $\partial \Omega_1$ and $\partial \Omega_2$, Fig. \ref{Fig-Beltrami-sphere}, whose positions are fixed in the steady state. The sub-domains $\Omega_1$ and $\Omega_2$ correspond to the area covered by the liquid in the bridges, while the surface flow, described by (\ref{Micro-Darcy-1}), takes place in $\Omega_0$. 

The transport process in the surface layer of the granular elements is described by a Darcy's like law (\ref{Micro-Darcy-1}) relating average liquid pressure $\psi$ with averaged volumetric flux density $\bf q$. The capillary pressure variations on the scale of one grain particle are assumed to be small enough, $\delta\psi \ll \gamma/\bar{\delta}_R$, so that the groove filling and, hence, the local coefficient of permeability $\kappa_m$ can be considered constant. Then, due to incompressibility of the liquid $\nabla\cdot \bf q=0$, and from (\ref{Micro-Darcy-1}), the problem can be reduced to a boundary-value problem for the Laplace-Beltrami equation
\begin{equation}
\label{Laplace-Beltrami}
\Delta_{\Omega_0}\psi =0
\end{equation}
defined on the surface element $\Omega_0$ of the particle. 

At the same time, liquid pressure variation in the bridges is negligible in slow creeping flow conditions in comparison to that in $\Omega_0$, so that in steady state one can assume that
\begin{equation}
\label{Laplace-Beltrami-BCS}
\left. \psi\right|_{\partial \Omega_1}=\psi_1=const, \quad \left. \psi\right|_{\partial \Omega_2}=\psi_2=const.
\end{equation}
As one can see, physically, the problem formulation (\ref{Laplace-Beltrami})-(\ref{Laplace-Beltrami-BCS}) is equivalent to calculation of the surface flow in $\Omega_0$, which is driven by the constant pressure difference $\psi_2-\psi_1$ applied to the boundaries of the surface element $\Omega_0$.

The boundary-value problem (\ref{Laplace-Beltrami})-(\ref{Laplace-Beltrami-BCS}) has a unique solution, which, if it is found, allows to calculate the total flux $Q_T$ through any contour $\partial \Omega$ on $\Omega_0$, which can not be contracted to a point
$$
Q_T=\delta_L\frac{\kappa_m}{\mu}\int_{\partial \Omega} {\bf n}\cdot \nabla\psi \, dl,
$$
where $\bf n$ is the normal vector to the contour $\partial \Omega$ on the surface, $\delta_L$ is the average width of the surface layer conducting the liquid flux. In particular, due to conservation of the liquid mass and in steady state
$$
Q_T=\delta_L\frac{\kappa_m}{\mu}\int_{\partial \Omega_1} {\bf n}\cdot \nabla\psi \, dl =  -\delta_L\frac{\kappa_m}{\mu} \int_{\partial \Omega_2} {\bf n}\cdot \nabla\psi  \, dl.
$$ 
If the surface flux $Q_T$ is found, then given constant pressure difference $\psi_2-\psi_1$, permeability of the surface element $\Omega_0$ can be defined and deduced. 

To obtain analytical results, we restrict ourselves to the case of a spherical particle of radius $R$. In this case, domain boundaries $\partial \Omega_1$ and $\partial \Omega_2$ will be circular cross sections of the spherical surface $\Gamma$,  Fig. \ref{Fig-Beltrami-sphere}, where we used a spherical coordinate system with the polar angle $\theta$ counted from the axis of symmetry of $\partial \Omega_1$. The location of the sub-domains $\Omega_1$ and $\Omega_2$ with respect to each other on the surface is fixed by an angle $\nu$.

We consider an azimuthally symmetric case, $\nu=\pi$, with equal in size (radius of curvature) domain boundaries  $\partial \Omega_1$ and $\partial \Omega_2$, as is shown in Fig. \ref{Fig-Beltrami-sphere}. The setting implies that the liquid bridges are formed between identical particles in contact. The size of the boundary contours, that is their radius $R\sin\theta_0$, will be characterized by the polar angle $\theta_0$ counted from the axis of symmetry of each contour and the particle radius $R$. Then, due to the nature of the boundary conditions (\ref{Laplace-Beltrami-BCS}), the problem (\ref{Laplace-Beltrami})-(\ref{Laplace-Beltrami-BCS}) is equivalent to 
\begin{equation}
\label{Laplace-Beltrami-p}
\frac{1}{\sin\theta}\frac{\partial }{\partial \theta}\left( \sin\theta \frac{\partial \psi}{\partial \theta}\right)=0, \quad \theta_0\le \theta\le \pi-\theta_0,
\end{equation}
with the boundary conditions
\begin{equation}
\label{BCLB-p}
\left. \psi\right|_{\theta=\theta_0}=\psi_1, \quad \left. \psi\right|_{\theta=\pi-\theta_0}=\psi_2.
\end{equation}

The problem (\ref{Laplace-Beltrami-p})-(\ref{BCLB-p}) admits an analytical solution, which is, after applying the boundary conditions,
$$
\psi=\frac{\psi_2-\psi_1}{2}\left\{1 - \frac{ \ln \frac{\sin\theta}{1+\cos\theta} }{\ln\frac{\sin\theta_0}{1+\cos\theta_0}} \right\} +\psi_1.
$$ 
One can now calculate the total flux 
$$
Q_T=-2\pi \delta_L\,   \frac{\kappa_m}{\mu} \sin\theta_0 \left. \frac{\partial \psi}{\partial \theta}\right|_{\theta=\theta_0}
$$
$$
=-\pi \delta_L\, \frac{\kappa_m}{\mu} \frac{ \psi_2-\psi_1 }{\ln \frac{1+\cos\theta_0}{\sin\theta_0}}. 
$$

Now, one can define the effective coefficient of permeability of a sphere $K_1$, which is approximately equivalent to $K$, by 
$$
Q_T=-4R\, (\psi_2-\psi_1)\frac{K_1}{\mu}
$$
so that 
$$K_1=\frac{\delta_L}{4 R} \frac{\pi \kappa_m}{ \ln \frac{1+\cos\theta_0}{\sin\theta_0}}.$$

One can see that the permeability coefficient $K_1$ is divergent at $\theta_0=\pi/2$ and tends to zero at $\theta_0=0$ as expected, that is  
$$
K_1\approx \frac{\delta_L}{4 R} \frac{\pi \kappa_m}{\frac{\pi}{2}-\theta_0 }, \quad \theta_0 \to \frac{\pi}{2}
$$
and
$$
K_1\approx \frac{\delta_L}{4 R} \frac{\pi \kappa_m}{|\ln\theta_0|}, \quad \theta_0 \to 0.
$$
In what follows, we approximate the coefficient of permeability $K$ by $K_1$ obtained in an azimuthally symmetric case. As we have already demonstrated, this approximation is very reasonable in particulate porous media even with non-spherical particle shapes involved~\cite{Penpark2018}.

To incorporate $K_1$ into the model, we should express it through the saturation $s$. Using an approximate relationship between the radius of curvature $R\sin\theta_0$ of the boundary contour $\partial \Omega_{1}$ and the pendular ring volume at $\theta_0\ll 1$ or $(s-s_0^e)\ll 1$, see details in ~\cite{Herminghaus-2005},
$$
R\sin\theta_0\approx R \left(\frac{V_B}{R^3}\right)^{1/4},
$$
one can get
$$
\sin\theta_0\approx \theta_0 \approx (s-s_0^e)^{1/4}.
$$
That is 
$$
K_1\approx \frac{\delta_L}{R} \frac{\pi \kappa_m}{|\ln(s-s_0^e)| }.  
$$

So, finally,  (\ref{Superfast-1}) becomes
\begin{equation}
\label{Superfast-2}
\frac{\partial s}{\partial t}= \nabla \cdot \left\{  \frac{D_0 \, \nabla s}{|\ln(s-s_0^e)|(s-s_0^e)^{3/2}}\right\}, 
\end{equation}
where
$$
D_0= \delta_L \frac{\bar{\delta}_R^2}{R^2}\frac{\pi \kappa_0(s)}{\mu}\frac{\gamma\, \cos\theta_c\, A_c }{\phi }. 
$$

The model now includes a logarithmic correction to the non-linear coefficient of diffusion due to the specific permeability of spherical particles and the local coefficient of permeability of the grooves $\kappa_0(s)$ due to the variable liquid content in the surface rough layer.

\subsection{Self-similarity and superfast diffusion}

The obtained non-linear partial differential equation is known in mathematical literature as the superfast non-linear diffusion equation, which has distinctive mathematical properties~\cite{Vazquez-2006}. In particular, it is well known that many non-linear diffusion models, such as the porous medium equation, exhibit the so called self-similar behaviour, which allows to obtain universal long-time limiting asymptotic solutions. For example, there are compactly supported Barenblatt self-similar distribution profiles satisfying a natural set of boundary conditions with finite velocity of the moving boundary~\cite{Aronson1986, Barenblatt2003, Vazquez2006-II, Vazquez2014-III}. These asymptotic distributions are very useful in practical applications, since solutions to the porous medium equations of different types are practically independent of initial conditions and ultimately tend to the asymptotic distributions with time~\cite{Barenblatt2003, Vazquez2006-II, Vazquez2014-III}.

But, this is not the case here. The super-fast diffusion model in our case does not demonstrate this universal behaviour. While initial distributions of saturation evolve with time to a distinctive saturation profile (as we will discuss later in detail), there was no true self-similar behaviour identified in our simulations so far. Indeed, consider a simplified non-dimensional version of (\ref{Superfast-2}) in a one-dimensional domain $\Omega\subset \mathbb{R}$ with the boundary $\partial\Omega$ moving with velocity $v$. Neglecting relatively slow variations of $\kappa_0$ with saturation, one has 
$$
\displaystyle \frac{\partial u}{\partial t}= \frac{\partial }{\partial x} \left\{ \frac{(u+u_0)^{-\frac{3}{2}}}{\ln (u+u_0)  } \, \frac{\partial u}{\partial x}  \right\}, \quad x\in \Omega, \,\,\, t>0
$$
with
\begin{equation}
\label{BC1-simpd}
\left. u\right|_{\partial \Omega}=0
\end{equation}
and
\begin{equation}
\label{BC2-simpd}
\left. {v}\right|_{\partial \Omega}=\left. -\frac{1}{s_f}\frac{ u_0^{-\frac{3}{2}}} {\ln u_0}\frac{\partial u}{\partial x}\right|_{\partial \Omega},
\end{equation}
where $u=s-s_f$ and $u_0=s_f-s_0^e$.
Consider a one-parameter group of transformations of the variables $t\to \epsilon t$, $u\to \epsilon^q u$ and $x\to \epsilon^m x$, which is used to obtain self-similar solutions, in particular the Barenblatt self-similar distribution profiles; $\epsilon>0$. One can immediately see that the moving boundary value problem is not invariant under the group of transformations, that is one can not determine such $q$ and $m$ at $u_0\neq 0$ so that to obtain an invariant equation with invariant boundary conditions. This conclusion is consistent with our numerical simulations, where no global self-similar behaviour of the distribution profiles $s({\bf x},t)$ with time has been identified so far.

\FloatBarrier

\section{Results and discussion}
In this part, we consider and discuss applications and comparisons of the developed macroscopic model with our experimental data. First, we analyse steady states of the spreading process, that is the final extends of the spreading volumes, in three-dimensional spherically symmetric geometry. We will discuss and demonstrate how steady state data can be used to obtain estimates of the model parameters $s_f$ and $s_f-s_0^e$. We will also evaluate the effects of evaporation. Next, we augment our super-fast diffusion model to extend its applicability domain to the entire funicular regime of spreading. We will discuss saturation profiles and their intimate connection with the universal power law of liquid spreading observed in the experiments. To verify the fidelity of our approach and the mathematical model, we consider spreading and compare with the experiments in one-dimensional geometry using parameter set obtained in the three-dimensional experiments and comparisons. Finally, we will analyse liquid spreading in pre-wet porous matrices.

\subsection{Steady state distributions}

\begin{figure}[ht!]
\begin{center}
\includegraphics[trim=-1.5cm 1.cm 1cm -0.5cm,width=0.9\columnwidth]{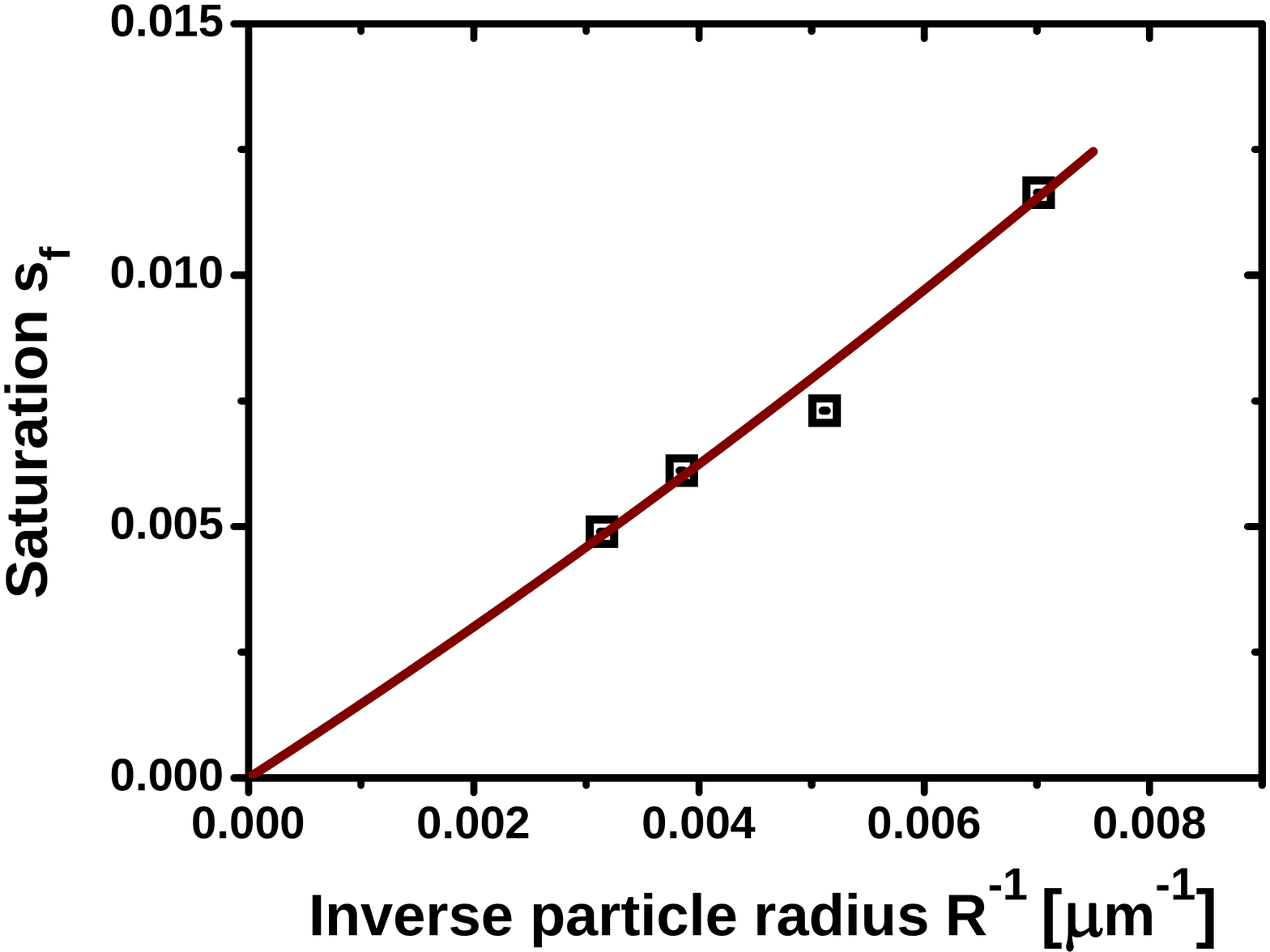}
\end{center}
\caption{Saturation $s_f$ as a function of $R^{-1}$ shown by symbols. The solid line is the fit $s_f=A_f R^{-1} + B_f R^{-2}$ at $A_f=1.5\, \mu\mbox{m}$ and $B_f=29\, \mu\mbox{m}^{2}$.} 
\label{Rough-saturation}
\end{figure}

Consider final extends of spreading obtained in a series of experiments with TCP liquid drops placed on sands with different grain sizes $R$, that is runs II, IV, V, and VI, see details in Table \ref{Table1}. The dependence of the equilibrium saturation in the end of the spreading process, $s_f$, on the inverse particle radius $R^{-1}$ is shown in Fig. \ref{Rough-saturation}. Assuming scaling (\ref{Scaling-P-2}), we fit the dependence by a function 
\begin{equation}
\label{scaling-sf-s0}
s_f=s_0^e+B_f R^{-2}=A_f R^{-1}+B_f R^{-2}
\end{equation}
with  $A_f=1.5\, \mu\mbox{m}\pm 0.2\,\mu\mbox{m}$ and $B_f=29\, \mu\mbox{m}^{2}\pm 24\, \mu\mbox{m}^{2}$. This implies that one can only determine one parameter with sufficient accuracy, and place upper and lower bounds for the other parameter.

Then, using obtained value of $A_f$ and the most probable value of $B_f$, from (\ref{Scaling-P-2}) and typical surface roughness parameters, one can estimate parameters $\delta_L$ and $\alpha_R$.  From $A_f=3\alpha_R  \, \frac{1-\phi}{\phi} \delta_L$ using $\phi=0.3$, one gets $ \alpha_R \delta_L \approx 0.2 \,\mu\mbox{m}$. Then, from $B_f = 4 \frac{A_c^2 \gamma^2 \cos^2\theta_c}{p_f^2}$, one can estimate assuming constant front pressure $p_f$ that at $\theta_c=\pi/6$ and $B_f=29\, \mu\mbox{m}^{2}$, $p_f\approx -3.6\times 10^{4}\,\mbox{Pa}$. That is, from $p_f=-\gamma/\delta_p$, $\delta_p\approx 1.2 \,\mu\mbox{m}$, and from (\ref{dl-scaling}), one gets $\delta_L\approx 0.7 \,\mu\mbox{m}$ and $\alpha_R\approx 0.3$ at $\bar{\delta}_R\approx 0.8\,\mu\mbox{m}$. 

The equilibrium value of saturation $s_f$ observed in the spreading of TEHP drops is consistent with the above estimates, while the observed value for TBP is slightly off. Indeed, the equilibrium level of TEHP in $R=0.25\,\mbox{mm}$ sand was found to be $s_f\approx 0.68\%$, which is consistent with estimates using (\ref{Surface-saturation}), if one presumes similar value of $\alpha_R=0.3$ and $\delta_L\approx 0.8\,\mu\mbox{m}$, Table \ref{Table1}, obtained scaling the front pressure $p_f$ with $\gamma$.

At the same time, the final saturation level of TBP in the sand with the same average grain radius, estimated assuming conservation of mass of the liquid, was found to be at much higher level $s_f=0.93\%$.  Since wetting properties of both liquids, TEHP and TBP, are very similar, Table \ref{Table1}, such deviation is likely to be due to much higher equilibrium vapour pressure of TBP, Table \ref{Table1}, and hence much higher evaporation rates involved in this case. Roughly, the observed level $s_f=0.93\%$  corresponds to evaporation of a quarter of a liquid drop $V_D=6\,\mbox{mm}^3$.

To obtain an estimate of the amount escaped from the surface of the TBP wet spot and compare with the observations, one can utilize evaporation rate calculated on the basis of the vapour pressure in quiescent conditions~\cite{Mackay2014}. That is, evaporation rate at $P_{ve}=0.15\,\mbox{Pa}$ in quiescent conditions (no air flow) for TBP (molar weight $266.32\,\mbox{g/mol}$) is $q_{ev}\approx 1.6\times10^{-8}\,\mbox{kg}/\mbox{m}^2\mbox{s}$~\cite{Mackay2014}. Then, the total mass evaporated during time $t$ is $Q_{ev}(t)=q_{ev}\pi \int_0^{t} R^2(t)\, dt$, where $R(t)$ is the radius of the observed wet spot. If we use experimentally observed dependence of $R(t)$ for TBP, Fig. \ref{Numerical-solution-3}, the amount of the liquid equivalent to a quarter of a liquid drop $V_D=6\,\mbox{mm}^3$ would evaporate in about four days ($t/t_0=5.4\times 10^{-2}$, Fig. \ref{Numerical-solution-3}) at this rate, which is comparable with the characteristic time to reach the steady state in the case of TBP. At the end of the power law phase, $t/t_0\approx 6\times 10^{-3}$ in Fig. \ref{Numerical-solution-3} (about $10$ hours), only $2\%$ of the drop volume would be lost at this evaporation rate. One can see, that the total amount of the TBP liquid evaporated at the end of the run is consistent with the observed level of the saturation calculated assuming conservation of mass (as if no evaporation occurred), that is $s_f=0.93\%$. On the other hand, apparently, evaporation plays no role during the power law phase of the TBP spreading. 

It is also instructive to compare the total evaporation rate $\frac{dQ_{ev}}{dt}$ with the total mass flux due to the dispersion processes, which is $Q_d=\rho s_f \phi \frac{dV}{dt}$, where $\rho$ is the liquid density. It turned out that at the beginning of the spreading process $Q_d\gg dQ_{ev}/dt$, but the two quantities are becoming comparable (in the case of TBP) $Q_d\approx dQ_{ev}/dt$ at $t/t_0\approx 1.5\times 10^{-2}$, Fig. \ref{Numerical-solution-3}, basically at the end of the spreading process when approaching the steady state, where some small distortion of the evolution curve can be observed.  So, one can conclude that even in the case of TBP (TCP and TEHP liquids have vapour pressure almost four orders of magnitude lower), the evaporation effects can be neglected during the power law spreading phase.

\subsection{The dynamics of spreading}
To understand the dynamics of liquid spreading and evolution of the moving front, that is the wetting volume, consider the superfast diffusion model (\ref{Superfast-2}). One can present (\ref{Superfast-2}) in non-dimensional form by normalizing distances $\tilde{\bf x}={\bf x}/L_0$ and time $\tilde{t}=t/t_0$. As the characteristic length scale, we use the wet spot radius $L_0$ at some moment of time, which will be initial time for simulations $\tilde{t}=0$, and $\displaystyle t_0=L_0^2/D_0^e$. Then, omitting tilde in the notations, equation (\ref{Superfast-2}) can be presented as  
\begin{equation}
\label{Superfast-nd}
\frac{\partial s}{\partial t}= \nabla \cdot \left\{  \frac{ \widehat{\kappa_0}(s) \, \nabla s}{|\ln(s-s_0^e)|(s-s_0^e)^{3/2}}\right\}, 
\end{equation}
with two boundary conditions
\begin{equation}
\label{BC1nd}
\left. s \right|_{\partial \Omega}=s_f
\end{equation}
and
\begin{equation}
\label{BC2nd}
\left. v_n \right|_{\partial \Omega}=-\frac{\widehat{\kappa_0}(s) \, \, ({\bf n}\cdot\nabla) s}{s_f|\ln(s_f-s_0^e)|(s_f-s_0^e)^{3/2}}.
\end{equation}
Here
\begin{equation}
\label{D0E}
D_0^e =  \delta_L \frac{\bar{\delta}_R^2}{R^2}\frac{\pi \kappa_0^{(1)}}{\mu}\frac{\gamma\, \cos\theta_c \, A_c }{\phi }
\end{equation}
and
$$
\widehat{\kappa_0}(s) = \frac{\kappa_0^{(2)}/\kappa_0^{(1)} - 1}{s_c-s_0^e} \left ( s-s_0^e \right) +1
$$
in
$$
s_0^e \le s \le s_c
$$
otherwise
$$
\widehat{\kappa_0}(s) = \kappa_0^{(2)}/\kappa_0^{(1)} 
.
$$
So, the problem has three essential non-dimensional parameters $s_f, s_f-s_0^e$ and $V_D/L_0^3$. The last parameter is only reflected  by the initial profile of saturation $s({\bf x},0)$ at $t=0$. We have already seen that variations of initial drop volume $V_D$ at $s_f=const$ and $s_f-s_0^e=const$ result in  collapse on a single master curve after re-normalizing time $t$ by a factor of $V_D^{2/3}$. This implies that one can further assume that $ L_0^3\propto V_D$, so that parameter $L_0$ can be solely defined by the initial drop volume $V_D$. This leaves us with just two non-dimensional parameters.

The role of parameter $s_f$ is clear, it defines the final level of saturation and the final size of the wetting zone in porous media after the spreading comes to standstill. To understand the role of the remaining parameter $s_f-s_0^e$, which represents the capillary action, that is the inverse of the reduced capillary pressure at the moving front, consider numerical solutions to the problem. The details of the numerical moving mesh method can be found in the appendix. 

\subsection{Augmented superfast diffusion model}

In the experiments, only the spot wetting area is measured giving the average value of saturation, while the accurate estimation of the liquid distribution within  the porous matrix is still unattainable. This implies that the initial saturation profile at the onset of the pendular regime of wetting is basically unknown and should be simulated starting from a liquid distribution at much higher saturation levels $s>10\%$, that is in the funicular regime of wetting, where the permeability is also a function of saturation~\cite{Koorevaar-1983}. 

To obtain realistic distributions of the liquid at the onset of the pendular regime of wetting, we augment the diffusion law (\ref{Superfast-nd}) using empirical permeability relationships found in sands~\cite{Koorevaar-1983}. In unsaturated porous media (in particular in sands) at high saturation values, permeability decreases very fast with liquid saturation $\log_{10} K\propto s$, as it could be anticipated, such that the augmented diffusion law takes the form  
\begin{figure}[ht!]
\begin{center}
\includegraphics[trim=-1.5cm 1.cm 1cm -0.5cm,width=\columnwidth]{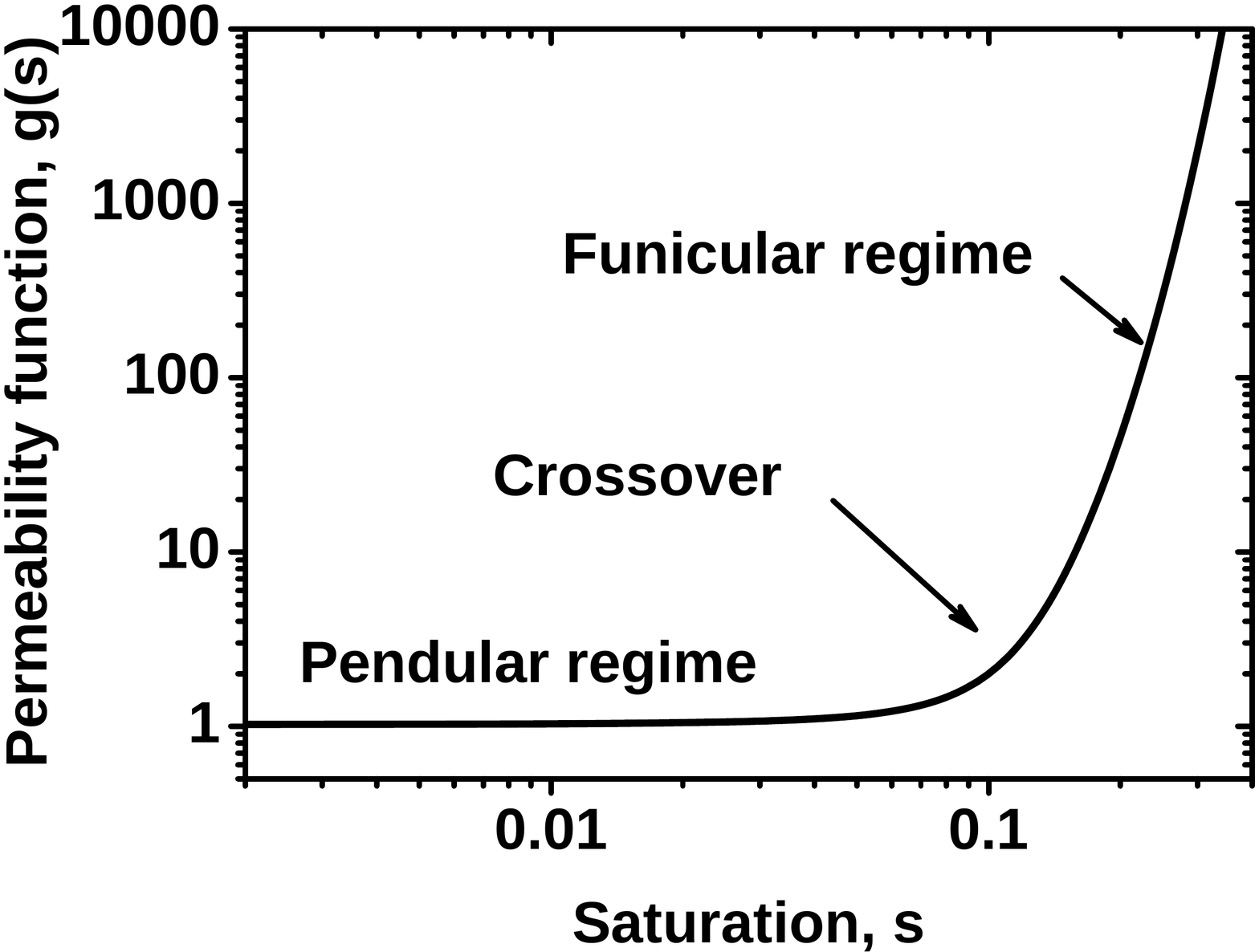}
\end{center}
\caption{Permeability function $g(s)$ versus saturation $s$ at $\alpha_g=16.5$, $\beta_g=1.65$ and $f_0=1$.} 
\label{Global-K}
\end{figure}
\begin{equation}
\label{Superfast-nd-am}
\frac{\partial s}{\partial t}= \nabla\cdot \left\{  \frac{ \widehat{\kappa_0}(s)\, g(s)\, \nabla s}{|\ln(s-s_0^e)|(s-s_0^e)^{3/2}}\right\}, 
\end{equation}
where augmenting permeability function $g(s)$, Fig. \ref{Global-K},
\begin{equation}
\label{Funcg}
g(s)=1+f_0 \, 10^{\alpha_g s -\beta_g}
\end{equation}
with
$$
 \,\alpha_g=16.5,\, \beta_g=1.65 
$$
and
\begin{equation}
\label{Pf0}
f_0=\left(\frac{R}{R_m}\right)^2 \frac{\gamma}{\mu} \frac{\mu_w}{\gamma_w},
\end{equation}
where $\gamma_w$ and $\mu_w$ are the surface tension and viscosity of water respectively.

The values of the coefficients in (\ref{Funcg}) have been chosen such that, according to~\cite{Koorevaar-1983}, in the medium fine sands ($R_m\approx 260\,\mu\mbox{m}$) and water
$$
g(s)\left. \right|_{s=0.1}=2,\quad g(s)\left. \right|_{s=0.3}=2000
$$
and $f_0=1$. As one can see, Fig. \ref{Global-K}, the augmenting function $g(s)$ due to the strong decline with the saturation has a very short crossover region quickly reaching a constant value $g(s)\approx 1$ at $s\approx 0.1$, where the pendular regime begins. We note that we still use pressure-saturation relationship (\ref{Pressure-saturation}), which provides a reasonable approximation considering strong variations of permeability. Alternatively, the model can be easily generalized by using a Leverett J-function~\cite{Leverett1941} or more complex, and general, porous media models for the retention curves and coefficients of permeability at higher saturation values~\cite{Mualem1976, Genuchten1980}.

\begin{figure}[ht!]
\begin{center}
\includegraphics[trim=-1.5cm 0.cm 1cm -0.5cm,width=\columnwidth]{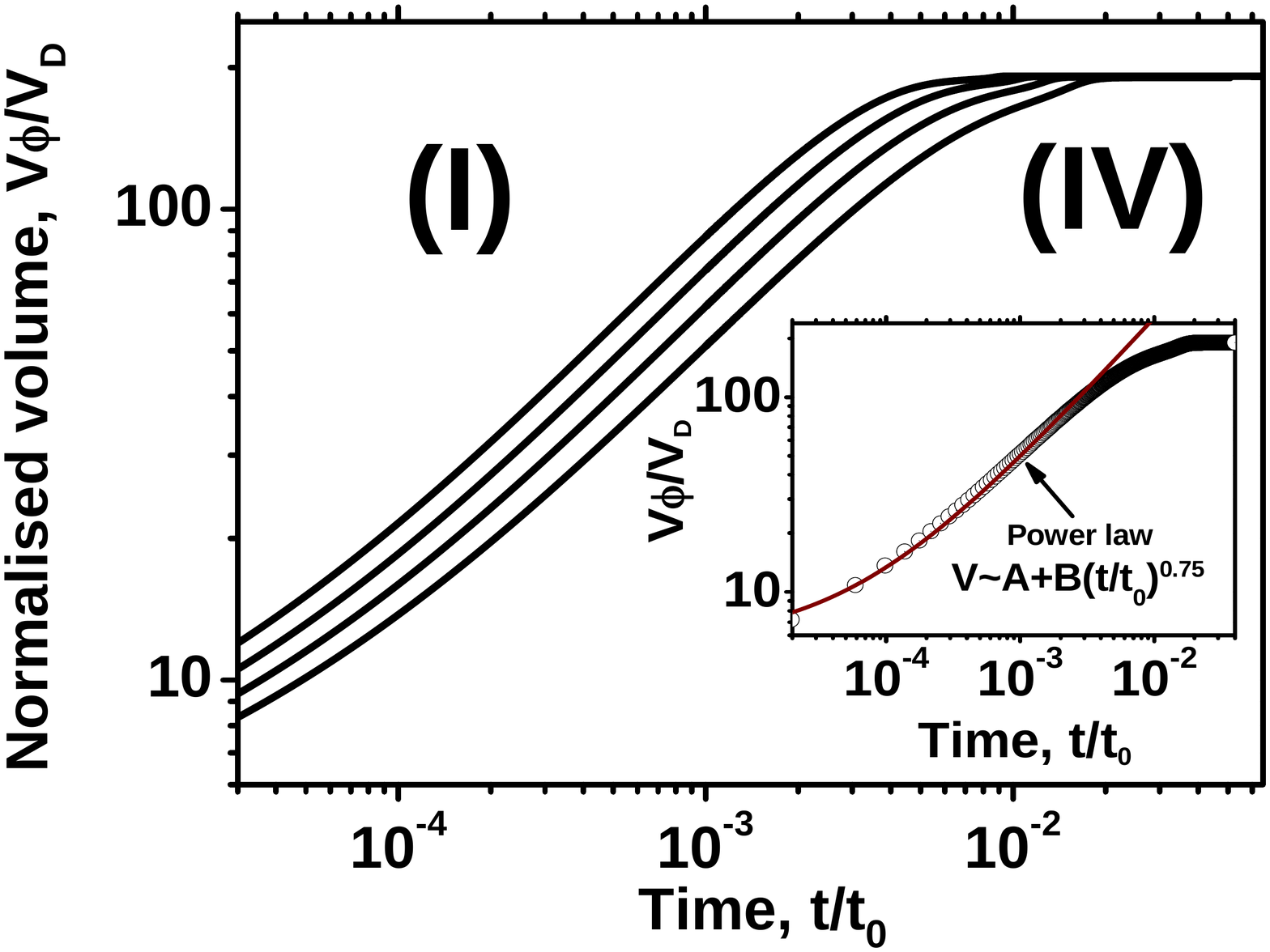}
\end{center}
\caption{Simulation of spreading in a three-dimensional spherically symmetric case using augmented superfast diffusion model (\ref{Superfast-nd-am}) with initial distribution (\ref{Initial}) at $\lambda_a=0.3$, $s_a=0.4$, $\alpha_g=16.5,\, \beta_g=1.65,\, f_0=1$ and $s_f=0.0052$, but at different values of parameter  $s_f-s_0^e$. Normalised wet volume $V\phi/V_D$ (inverse average saturation $\bar{s}^{-1}$, solid lines) as a function of the reduced time $t/t_0$, $t_0=L_0^2/D_0^e$. From left to right: (I) $s_f-s_0^e=0.0001$, (II) $s_f-s_0^e=0.0002$, (III) $s_f-s_0^e=0.0004$, (IV) $s_f-s_0^e=0.0008$. Insert shows the power law $V\phi/V_D =A+B(t/t_0)^{0.75}$ (solid line, brown) in comparison with the numerical data (symbols, black) at $s_f-s_0^e=0.0008$.} 
\label{Numerical-solution-1}
\end{figure}

\subsection{Numerical simulations and experimental results in three-dimensional spherically symmetric cases}

To compare numerical solutions of the superfast diffusion model (\ref{Superfast-nd-am}) with experimental observations, we first consider simulations in a three-dimensional spherically symmetric case, where saturation $s(r,t)$ is a function of time and the radius $r$ in a spherical coordinate system with its origin at the centre of the hemisphere representing the wet volume, Fig. \ref{Wet-spot}.  We have started our simulations in this case with 
\begin{equation}
\label{Initial}
s(r,t)\left.\right|_{t=0}=s_f + s_a\cos^{\lambda_a}(\pi r/2), \,\,\, 0\le r\le 1
\end{equation} 
at different values of parameters $0.2 \le s_a\le 1-s_f$ and $0.2 \le \lambda_a\le 0.4$. The value of $L_0$ then is defined by conservation of the liquid, neglecting the evaporation effects,
$$
2\pi \phi\, \int_{0}^1\, s(r,0)\, r^2 \, dr = V_D L_0^{-3}.
$$   
We note, that due to the use of a spherical coordinate system, we also require that at $r=0$ the first derivative $\partial s/\partial r=0$. 

The choice of parameter $s_a$ in the initial distribution and even its functional form is not obvious. We observed in the experiments that just in about ten minutes of spreading, the wetting spot volume shape becomes spherically symmetric, when the average saturation level $\bar{s}\approx 0.5$, Fig. \ref{Fig1}. But what is the liquid distribution at this stage? 

If we fix parameters of the initial distribution ($s_a$ and $\lambda_a$) and parameter $s_f$, then evolution of the moving front at different values of $s_f-s_0^e$ represents a family of curves shown in Fig. \ref{Numerical-solution-1}. One may notice that, first of all, the smaller is the parameter $s_f-s_0^e$ (that is the higher is the reduced capillary pressure at the moving front) the faster the spreading occurs. Secondly,  the power law found in the experiments $V\propto A+B(t/t_0)^{0.75}$ is very well observed in the simulations, see insert in Fig. \ref{Numerical-solution-1}.  

\begin{figure}[ht!]
\begin{center}
\includegraphics[trim=-1.5cm 0.cm 1cm -0.5cm,width=\columnwidth]{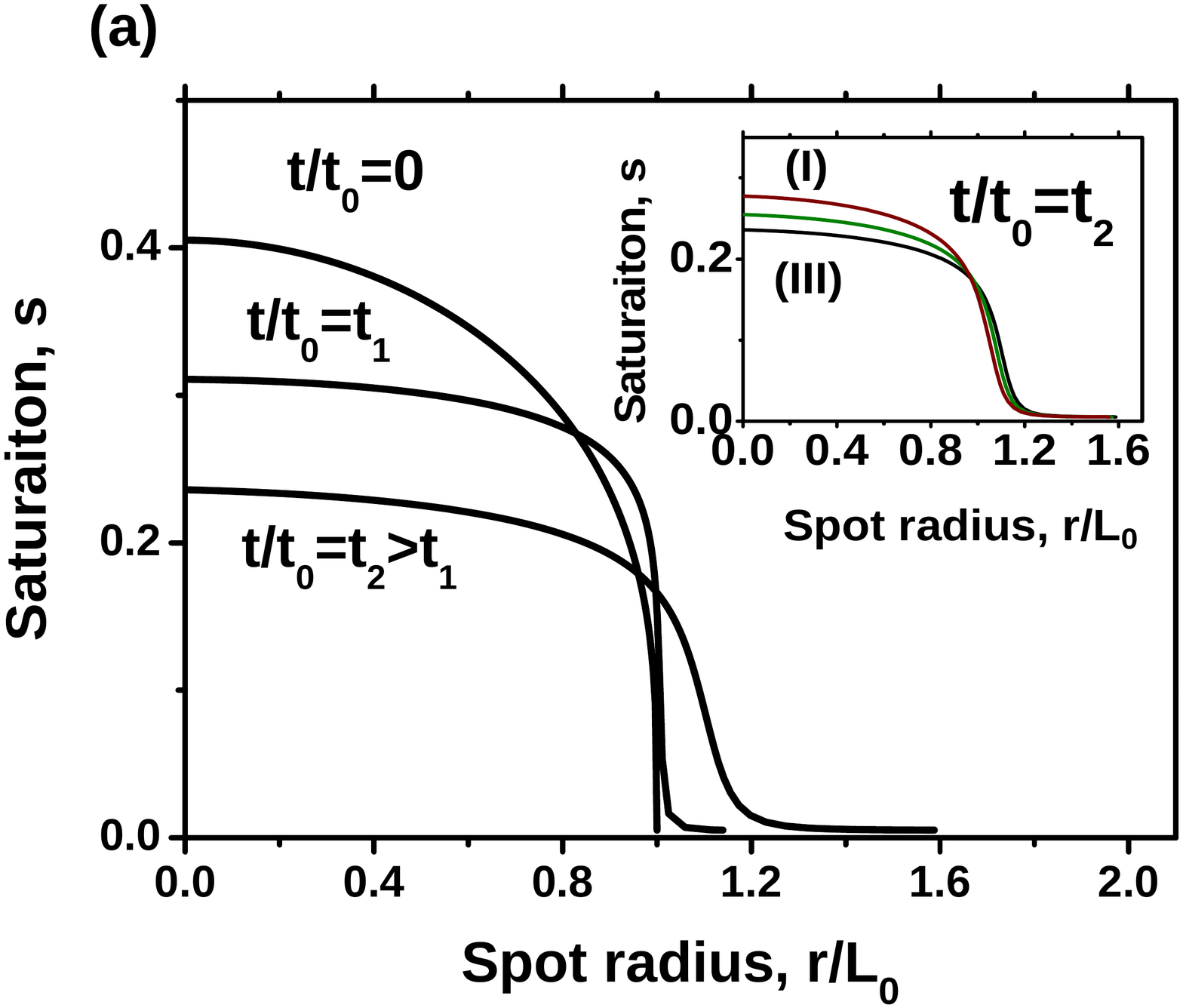}
\includegraphics[trim=-1.5cm 0.cm 1cm -0.5cm,width=\columnwidth]{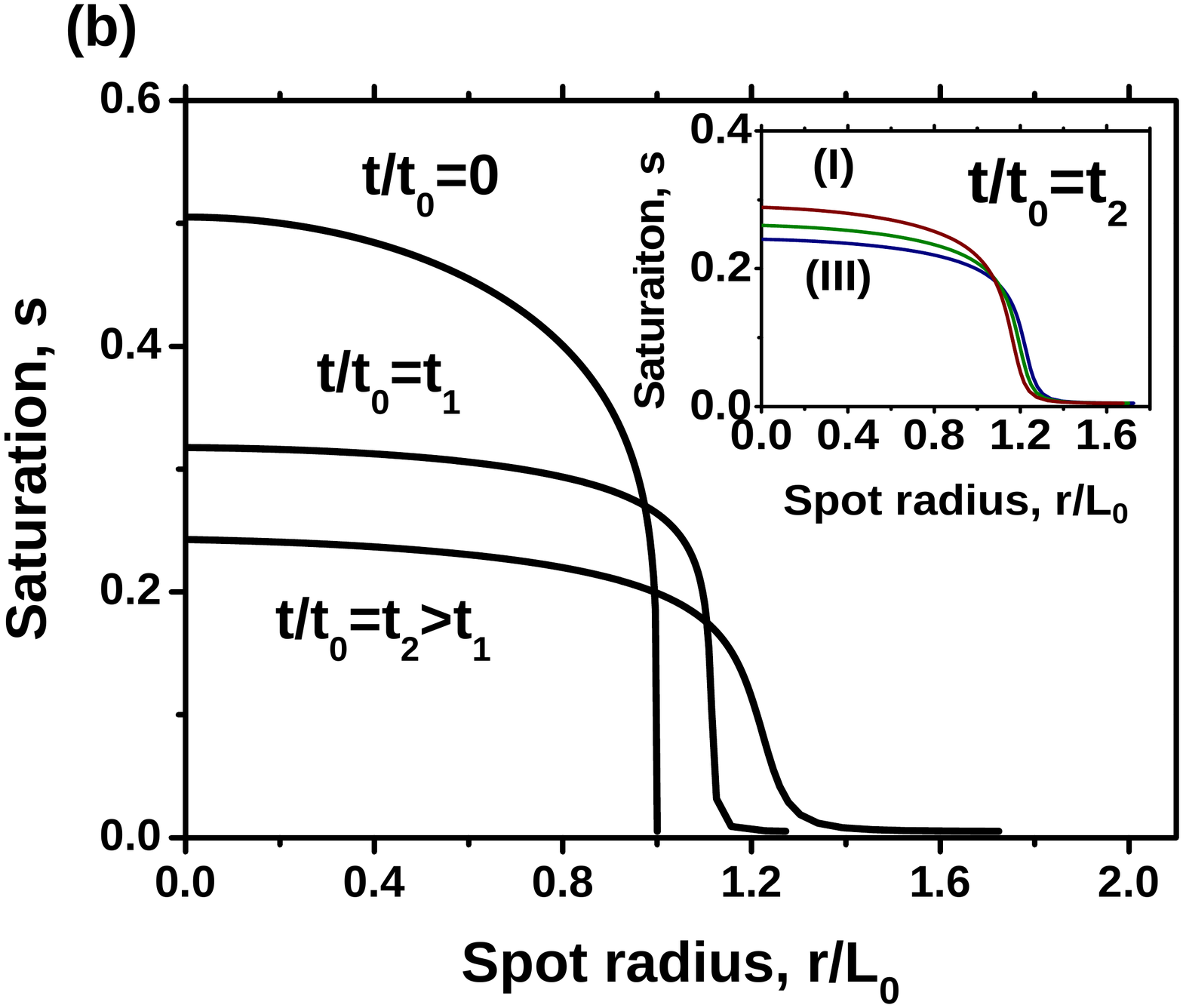}
\end{center}
\caption{Simulation of spreading in a three-dimensional spherically symmetric case using augmented superfast diffusion model (\ref{Superfast-nd-am}) at different initial distributions (\ref{Initial}). Saturation $s(r,t)$ as a function of the reduced spot radius $r/L_0$ at $t/t_0=0$, $t/t_0=t_1=3\times 10^{-6}$ and $t/t_0=t_2=6\times 10^{-5}$ at fixed values of $\alpha_g=16.5,\, \beta_g=1.65,\, f_0=1$, $s_f=0.0052$ and $s_f-s_0^e=0.0002$. (a) $\lambda_a=0.3$ and $s_a=0.4$; (b) $\lambda_a=0.2$ and $s_a=0.5$. The inserts show variation of the saturation profile at $t=t_2$ with the parameters of the augmenting function $g(s)$. Here (I) $\alpha_g=12.5$ and $\beta_g=1.25$, (II) $\alpha_g=14.5$ and $\beta_g=1.45$,  and (III) $\alpha_g=16.5$ and $\beta_g=1.65$.} 
\label{Numerical-solution-2}
\end{figure}

As one can see from the distribution of the liquid at $t>0$, Fig. \ref{Numerical-solution-2} (a)-(b), the saturation profile quickly relaxes to a universal distribution at fixed values of $s_f$, $s_f-s_0^e$ and $V_D$. The distribution $s(r,t)$ at $t=t_2=6\times10^{-5}$, when the average value of saturation is already $\bar{s}\approx 0.1$, does not depend much on the details of the initial conditions. This implies that we may not need to worry about the initial profile in the simulations as far as the spreading at low saturation levels is concerned. The profile shape is very distinctive and is in good qualitative agreement with direct nuclear magnetic resonance imaging of inflow in porous materials such as gypsum building plaster, Portland lime stone and Portland cement~\cite{Gummerson1979}.  It is flat in the central part, where the saturation levels are still in the funicular regime, and sharply declines to the boundary value $s=s_f$ through a zone with an accentuated tail, where the saturation levels are characteristic to the pendular regime of wetting. We note that the saturation profile with the value in the central part $s\approx 0.3$ already corresponds to an average saturation level $\bar{s}\approx 0.1$. This implies that, first of all, there is no purely pendular or funicular regimes of spreading in dry porous materials and both mechanisms are in operation simultaneously.  The overall dynamics of the wetting spot area seems to be defined to the large extent by the superfast diffusion processes in the tail region of the saturation distribution, while the role of the standard diffusion mechanisms inherent to the funicular regime is to level the liquid distribution  by smoothing the profile in the central part. This can be directly seen, if we change the values of the augmenting function parameters $\alpha_g, \beta_g$ keeping the other model parameters $f_0$, $s_f$ and $s_f-s_0^e$ at the same level. One can observe that such a change has almost no influence on the overall dynamics at $t=t_2$, see the inserts in Fig. \ref{Numerical-solution-2}. Indeed, while in the central part the permeability coefficients are almost two orders of magnitude different, the position of the front at $s=s_f$ is practically the same and the saturation level in the centre has only variations within approximately $15\%$. In what follows, we fix parameters of the augmented function at $\alpha_g=16.5$ and $\beta_g=1.65$ and scale parameter $f_0$ according to (\ref{Pf0}) using particular properties of the sand and the wetting liquid.

To understand the origin of the sharp transition observed in the saturation profiles, consider an intermediate asymptotic in the pendular regime of wetting, when $g(s)\approx 1$. Introducing new variable $\xi=(r-r_0)/\epsilon$, $\epsilon=const$ and $r_0=const$, $\epsilon\ll 1$, and neglecting terms of the order of $\epsilon$ and relatively slow variations in the logarithmic term and in $\widehat{\kappa}_0(s)$, from (\ref{Superfast-nd-am})

$$
\frac{\partial^2 }{\partial \xi^2}\frac{1}{(s-s_0^e)^{1/2}}\approx 0.
$$

\begin{figure}[ht!]
\begin{center}
\includegraphics[trim=-1.5cm 0.cm 1cm -0.5cm,width=\columnwidth]{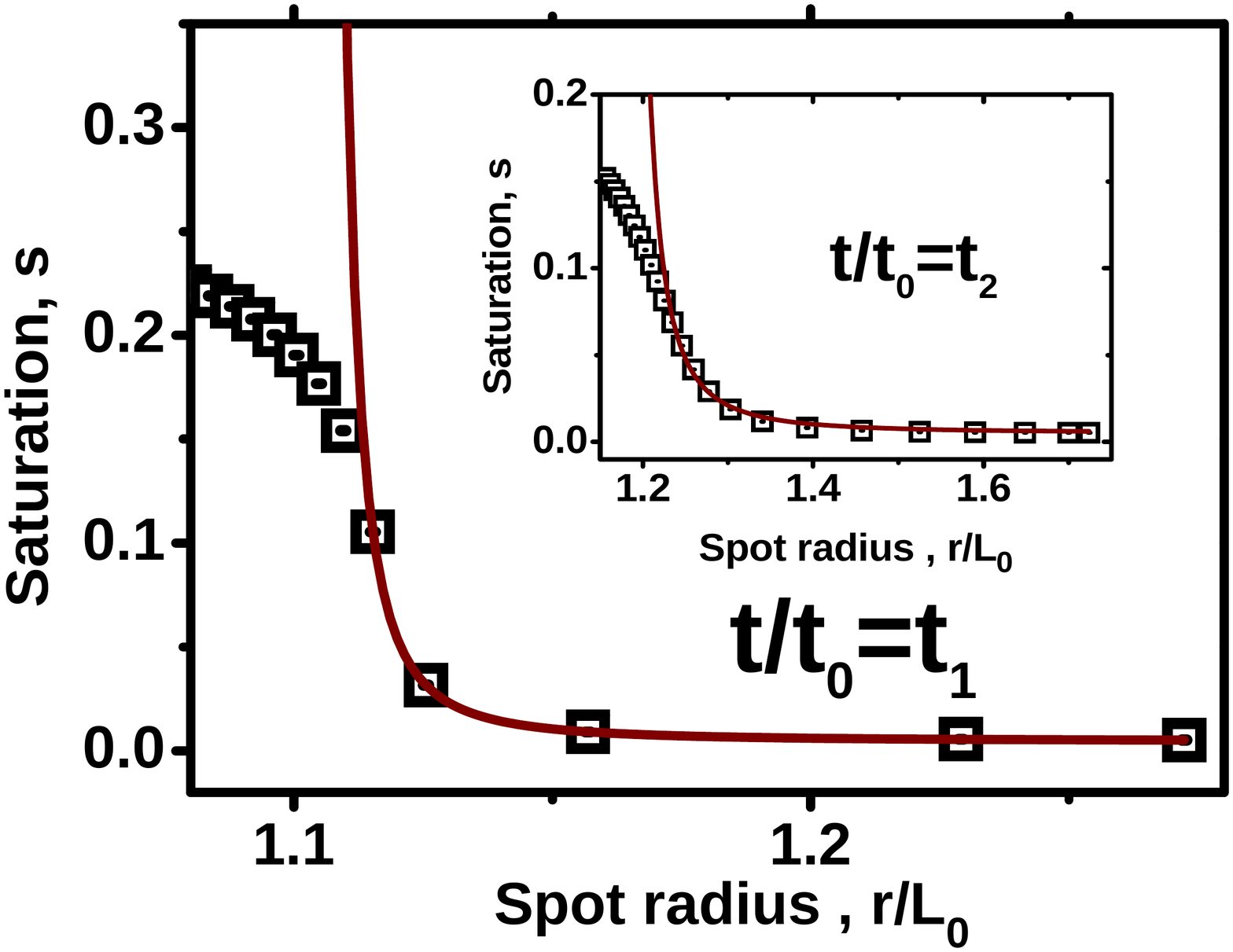}
\end{center}
\caption{Simulation of spreading in a three-dimensional spherically symmetric case using augmented superfast diffusion model (\ref{Superfast-nd-am}) with initial distributions given by (\ref{Initial}) at $\lambda_a=0.2$ and $s_a=0.5$. Saturation $s(r,t)$ as a function of the reduced spot radius $r/L_0$ at fixed values of $\alpha_g=16.5,\, \beta_g=1.65,\, f_0=1$, $s_f=0.0052$ and $s_f-s_0^e=0.0002$.  A comparison between the asymptotic solution (\ref{inter}) (solid line, brown) and the numerical solution at $t=t_1=3\times 10^{-6}$ shown by symbols, $W_0\approx 287$. The insert shows a similar comparison, but at $t/t_0=t_2=6\times 10^{-5}$, $W_0\approx 62$.} 
\label{Numerical-solution-2-asymp}
\end{figure}

Then 
\begin{equation}
\label{inter}
s \approx s_0 + \frac{1}{(W_0(r-r_0) + W_1)^2}.
\end{equation}
As one can see, Fig. \ref{Numerical-solution-2-asymp}, the asymptotic behaviour matches very well the simulated saturation profiles at the point of the sharp transition and even in the tail region. 

\subsection{Universal scaling laws of the moving front propagation and the super-fast diffusion model}

The distinctive shape of the saturation profiles suggests an explanation of the characteristic power laws of the front motion observed in the experiments. First of all, the total flux $\Pi_0(t)$ at the moving front $X_{n}(t)$ should be proportional to the moving front velocity, that is
$$
\Pi_0(t)\propto X_{n}^{n-1} \frac{dX_{n}}{dt},
$$
where index $n$ designates here the dimension of the diffusion problem. At the same time, the asymptotic behaviour (\ref{inter}) suggests that parameter $W_0(t)$, Fig. \ref{Numerical-solution-2-asymp} and the insert, should be inversely proportional to the length of the tail region $X_n(t)-r_0(t)$. Hence, when $X_n\gg r_0$, the total flux $\Pi_0(t)$ (since it is proportional to the gradient of saturation) should scale with $X_n(t)$ as $\Pi_0(t)\propto \frac{1}{X_n(t)}$. That is in the one-dimensional case  
$$ 
\frac{dX_1}{dt}\propto \frac{1}{X_1}.
$$  
This results in $X_1(t)\propto t^{1/2}$, which is the well-known Lucas-Washburn law for fluid motion in a circular capillary observed in our one-dimensional experiments. In a general case
$$
X_n(t)\propto t^{1/(n+1)},
$$
which in the three-dimensional case would give rise to 
$X_3\propto t^{1/4}$ or $V(t)\propto X_3^3\propto t^{0.75}$ - the power law observed in the three-dimensional experiments and simulations. We have also verified by numerical solution of the model that $X_2(t)\propto t^{1/3}$ in two-dimensional radially symmetric cases.

\subsection{A comparison between numerical simulations and experimental results}

\begin{figure}[ht!]
\begin{center}
\includegraphics[trim=-1.5cm 0.cm 1cm -0.5cm,width=\columnwidth]{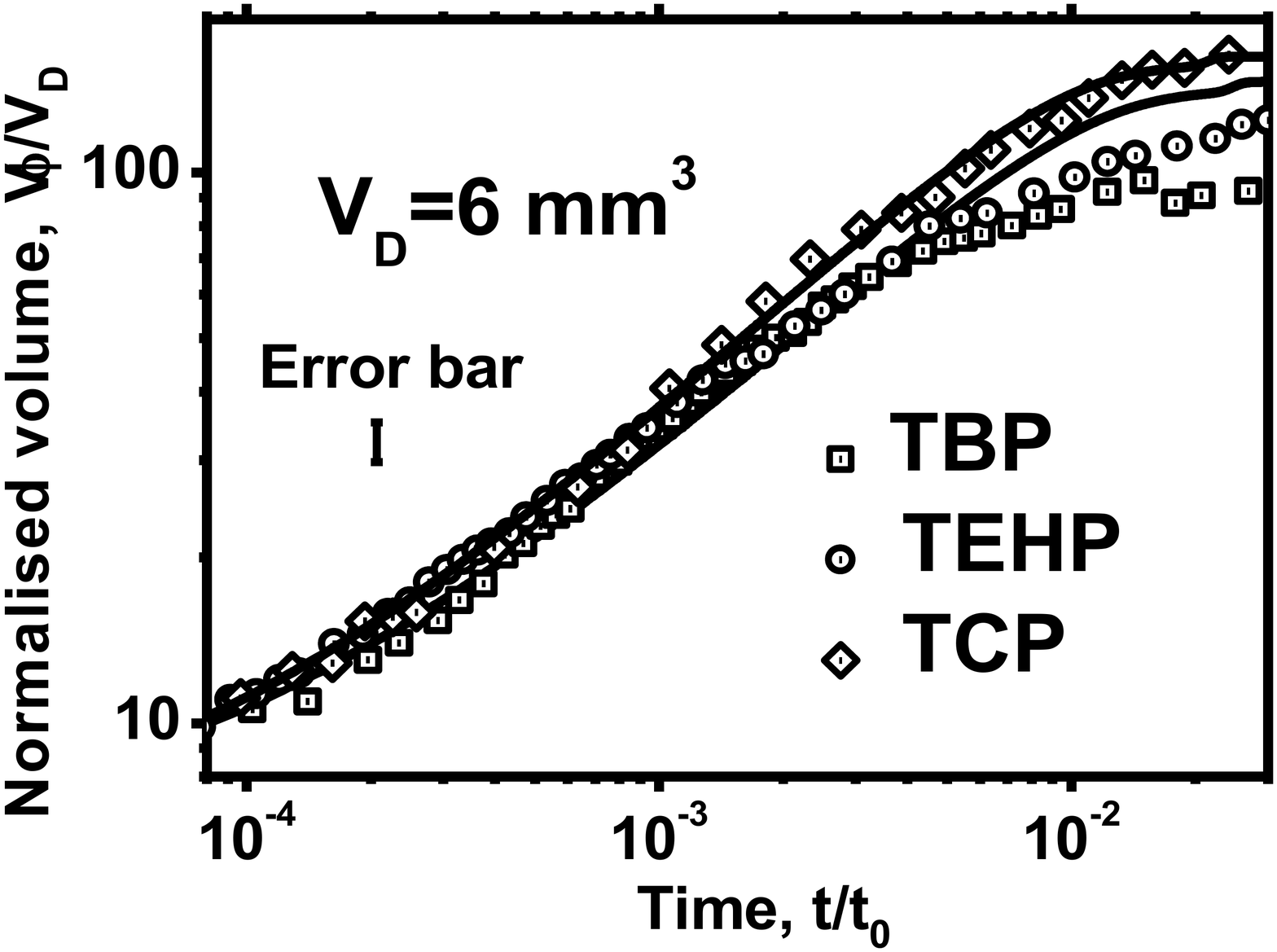}
\end{center}
\caption{Spreading of TCP, TEHP and TBP liquid drops ($V_D=6\,\mbox{mm}^3$) in $R=0.25\,\mbox{mm}$ sand ($R=0.26\,\mbox{mm}$ for TCP liquid). Comparison between experimental data and simulations using superfast diffusion model (\ref{Superfast-nd-am}) with initial distribution of saturation given by (\ref{Initial}). Normalised wet volume $V\phi/V_D$ (inverse average saturation $\bar{s}^{-1}$) as a function of the reduced time $t/t_0$, where $t_0=L_0^2/D_f$ for experimental data and the numerical results were scaled by $t_0=L_0^2/D^e_0$. Experimental data are shown by symbols and simulations are presented by the solid lines. Parameters of the simulations and the fitting are summarized in Table \ref{Table1}.} 
\label{Numerical-solution-3}
\end{figure}

Consider now a comparison between numerically found evolution curves of the moving front using (\ref{Superfast-nd-am}) and the experimental observations. In all simulations we start from a profile with $\lambda_a=0.3$ and $s_a=0.4$, such that $L_0=3.24\,\mbox{mm}$ at $V_D=6\,\mbox{mm}^3$. Consider spreading of TCP liquid drops ($V_D=6\,\mbox{mm}^3$) in $R=0.26\, \mbox{mm}$ sand, Fig. \ref{Numerical-solution-3}. In the simulations, we fixed the value of $s_f=0.0061$ according to the experimental observations, Table \ref{Table1}, and $s_f-s_0^e = 4.3\times 10^{-4}$ according to the scaling (\ref{Scaling-P-2}) at $B_f=29\, \mu\mbox{m}^{2}$. The experimentally observed evolution curves $V(t)$ have been shifted by renormalising time $(t-t_s)/t_0$, $t_0=L_0^2/D_f$, where an effective coefficient of diffusion $D_f$ was the fitting parameter. The time $t_s$ corresponds here to the actual time when the simulations started (about $30-90$ minutes of spreading), when the average saturation levels $\bar{s}$ observed in the experiments coincide with the initial average saturation levels in the simulations. As one can see the numerical solution is a good match to the observations. In the comparison, parameter $D_f$ was determined by the best match between experimental data and the numerical solution, then the value of the fitting parameter $\xi_f=0.038$ was obtained by achieving $D_0^e=D_f$. Considering that parameter $\alpha_R\approx 0.3$ and the effects of tortuosity can reduce permeability at least two-fold~\cite{Tortuosity-Review}, the connectivity factor contribution into $\xi_f$ can be estimated on the level of $0.25$ in comparison with $0.1$ found in the studies of surface flows~\cite{Tuller-2000}. This may imply that indeed surface grooves with sharp opening angles $\theta_R$ are poorly interconnected (that is serving mostly as liquid reservoirs) and could be neglected while considering surface flow permeability.  

Now, in a similar way, we compare evolution of the moving front for TEHP and TBP liquid spots with numerical solutions, but with already fixed value of $\xi_f=0.038$. To obtain parameter $s_f-s_0^e$ for those liquids, we scale the capillary front pressure $p_f$ with the liquid surface tension $\gamma$, Table \ref{Table1}. Those liquids have much smaller contact angle on a flat smooth/rough surface of quartz, $\theta_c\approx 10^{\circ}/0^{\circ}$ against $\theta_c\approx 30^{\circ}/20^{\circ}$ in the case of TCP liquids. Therefore, the surface grooves are expected to be fully filled in the range of capillary pressures in question, hence one can expect much higher permeability according to (\ref{Permeability-contact-angle}). In the comparison, we presumed that for both TEHP and TBP the equilibrium saturation level is $s_f=0.68\%$ ignoring the higher value of $s_f=0.93\%$ found for TBP. This implies that the formation (and the thickness) of the liquid film on the rough surfaces of the sand grains, given similar wetting properties of both liquids, should be the same. One can observe, Fig. \ref{Numerical-solution-3},  very good agreement between numerical solutions and the experimental data, demonstrating the scaling of the propagation rates with the surface tension $\gamma$, liquid viscosity $\mu$ and contact angle $\theta_c$ through the permeability of the surface layer $\kappa_0^{(1)}$, (\ref{Permeability-contact-angle}), suggested by the diffusion coefficient $D_0^e$. 

\FloatBarrier

\begin{figure}[ht!]
\begin{center}
\includegraphics[trim=-1.5cm 0.cm 1cm -0.5cm,width=\columnwidth]{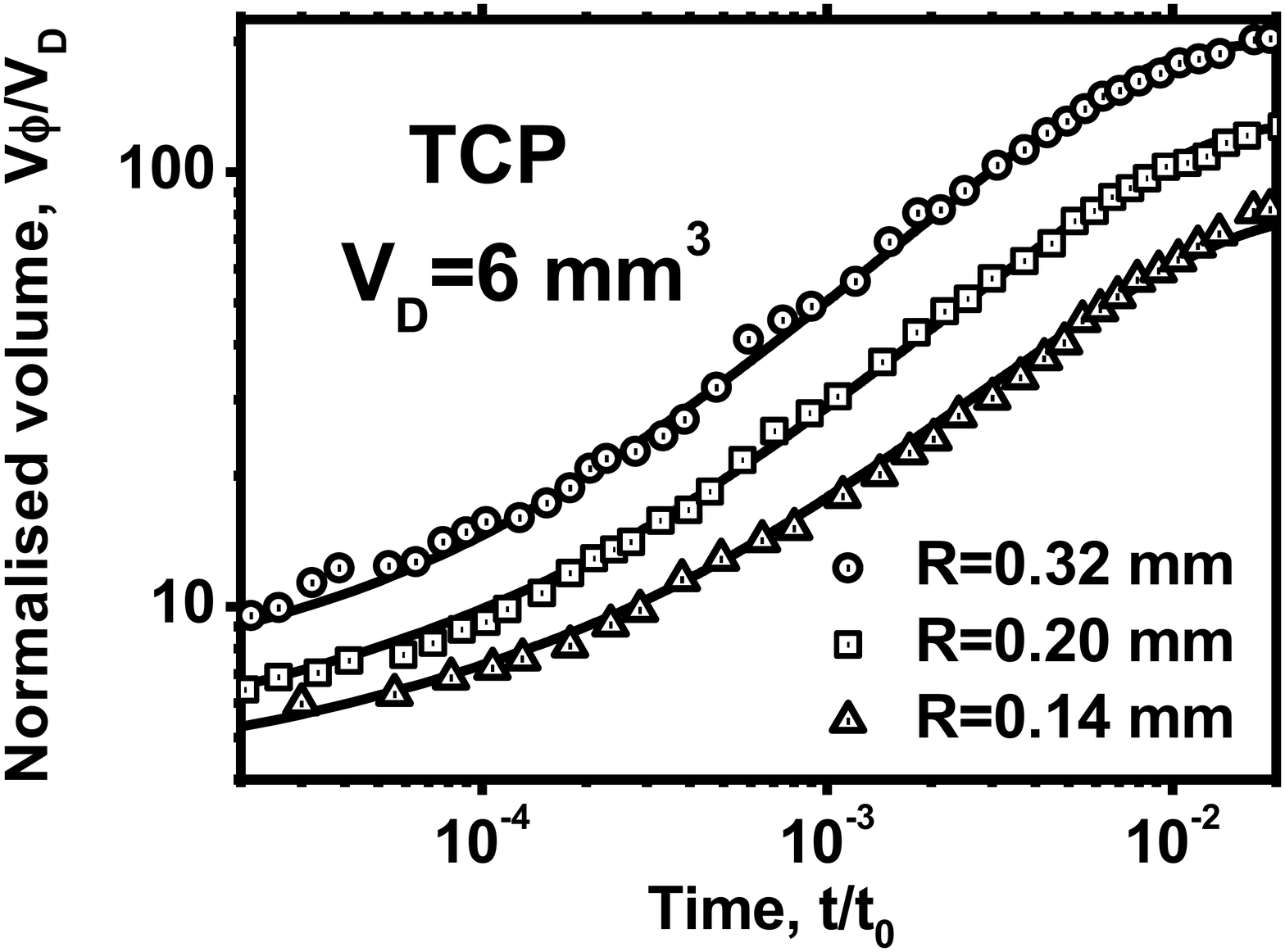}
\end{center}
\caption{Spreading TCP liquid drops ($V_D=6\,\mbox{mm}^3$) in sands with different grain radii $R=0.14, 0.20$ and $0.32\,\mbox{mm}$. Comparison between experimental data and simulations using superfast diffusion model model (\ref{Superfast-nd-am}) with initial distribution of saturation given by (\ref{Initial}). Normalised wet volume $V\phi/V_D$ (inverse average saturation $\bar{s}^{-1}$) as a function of the reduced time $t/t_0$, where $t_0=L_0^2/D_f$ for experimental data and the numerical results were scaled by $t_0=L_0^2/D^e_0$. Experimental data are shown by symbols and simulations are presented by solid lines. Parameters of the simulations and the fitting are summarized in Table \ref{Table1}.} 
\label{Numerical-solution-4}
\end{figure}

Consider now how the average grain size affects the spreading in the pendular regime. We have done a series of experiments using TCP liquid drops ($V_D=6\,\mbox{mm}^3$) placed on sand beds with different average radius $R$, runs II, IV, V and VI, Table \ref{Table1}. The results of a comparison between numerical solutions of the model and the data are shown in Figs. \ref{Numerical-solution-3} and \ref{Numerical-solution-4}. In the comparison, we used the fixed value of $\xi_f=0.038$ obtained previously and scaling (\ref{Scaling-P-2}) with $B_f=29\, \mu\mbox{m}^{2}$ to estimate parameter $s_f-s_0^e$ and $\kappa_0$. As one can see, the model demonstrates the same trend as it was observed in the experiments. That is that the evolution is slower for smaller grain sizes $R$. In the model, this is a manifestation of the scaling of the parameter $s_f-s_0^e\propto R^{-2}$. The obtained values of the fitting parameter $D_f$ were also in agreement with the values predicted by the theory $D_0^e$. The only exception is observed at the smallest value of $R$, which can be in principle mitigated by adjusting parameter $s_f-s_0^e$ within the uncertainty window. Alternatively, one can think that properties of the surface roughness, while not seen in the equilibrium distribution of the liquid, may be different for large and small grains. So that further improvement of the model would require, perhaps, more accurate characterizations of the sand particles and considerations of the flows within surface roughness, at the micro-scale.  

Consider now, how the dynamics observed in three-dimensional spherically symmetric cases can be translated into one-dimensional geometry. 

\subsection{Dynamics of spreading in one-dimensional geometries}

The liquid spreading was observed in the open channels, as is shown in Fig. \ref{Wet-channel}, by placing a $V_D=3\,\mbox{mm}^3$ liquid drops of TCP, TEHP and TBP at one end of the groove. The numerical solutions were obtained by solving the augmented model (\ref{Superfast-nd-am}) with $\alpha_g=16.5$ and $\beta_g=1.65$, as before, and with initial distributions given by 
\begin{equation}
\label{Initial-1D}
s(x,t)\left.\right|_{t=0}=s_f + s_a\cos^{\lambda_a}(\pi x/2), \,\,\, 0\le x\le 1
\end{equation}
at $\lambda_a=0.3$, $s_a=0.2$. We use the same set of boundary conditions, together with $\partial s /\partial x =0$ at $x=0$ to reflect the absence of the flux at the end of the channel.

\begin{figure}[ht!]
\begin{center}
\includegraphics[trim=-1.5cm 0.cm 1cm -0.5cm,width=\columnwidth]{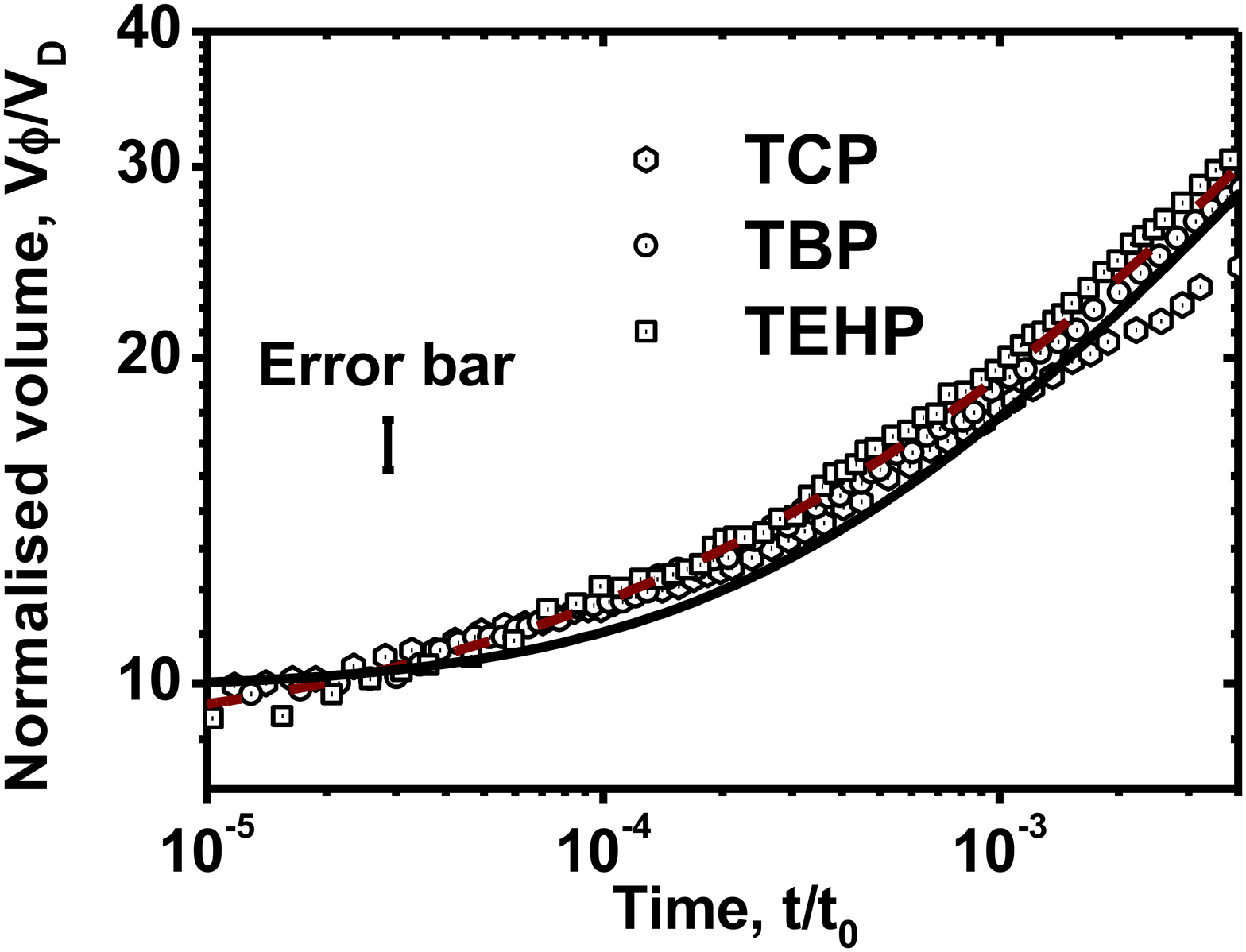}
\end{center}
\caption{Spreading TCP, TEHP and TBP liquid drops ($V_D=3\,\mbox{mm}^3$) in sands with $R=0.25\,\mbox{mm}$ in one-dimensional geometry. Comparison between experimental data and simulations using superfast diffusion model model (\ref{Superfast-nd-am}) with initial distribution of saturation given by (\ref{Initial-1D}). Normalised wet volume $V\phi/V_D$ (inverse average saturation $\bar{s}^{-1}$) as a function of the reduced time $t/t_0$, where $t_0=L_0^2/D_f$ for experimental data and the numerical results were scaled by $t_0=L_0^2/D^e_0$. Experimental data are shown by symbols and simulation is presented by the solid line. Parameters of the simulations and the fitting are summarized in Table \ref{Table1}. The dashed line (brown) is the fit $V\phi/V_D = A+B(t/t_0)^{0.5}$ at $A\approx 8.5$ and $B\approx 340$.} 
\label{1D}
\end{figure}

A comparison between the experimental data and the numerical solutions is shown in Fig. \ref{1D}. In the comparison, we have taken  all parameter values directly from the similar comparison in the three-dimensional geometry, Table \ref{Table1}, with parameter $L_0$ defined according to the initial distribution (\ref{Initial-1D})
$$
\pi d_c^2  \phi \int_0^1 s(x,0)\, dx = 8 V_D L_0^{-1}.
$$ 
We note that practically all parameter values in the comparison were fixed, we have only taken the liberty to vary $L_0$ within $1\,\mbox{mm}$ to take into account the fact that the shape of the groove is hemispherical rather than cylindrical at the ends, Fig. \ref{Wet-channel}, so that the one-dimensional model is an approximation.

As is seen, Fig. \ref{1D},  the numerical solutions follow the propagation law observed in the experiments $X_1(t)\propto t^{0.5}$. Secondly, one can observe that the scaling suggested by the diffusion coefficient, $D_0^e\propto \kappa_0^{(1)} \frac{\gamma}{\mu}$, is well observed. Indeed, after re-scaling the time $t/t_0$, $t_0=\frac{L_0^2}{D_f}$, the TCP, TBP and TEHP data collapsed into a single curve. The overall comparison is looking very good considering that there were practically no fitting parameters involved. 

\subsection{Spreading in pre-wetted porous media}

\begin{figure}[ht!]
\begin{center}
\includegraphics[trim=-1.5cm 0.cm 1cm -0.5cm,width=\columnwidth]{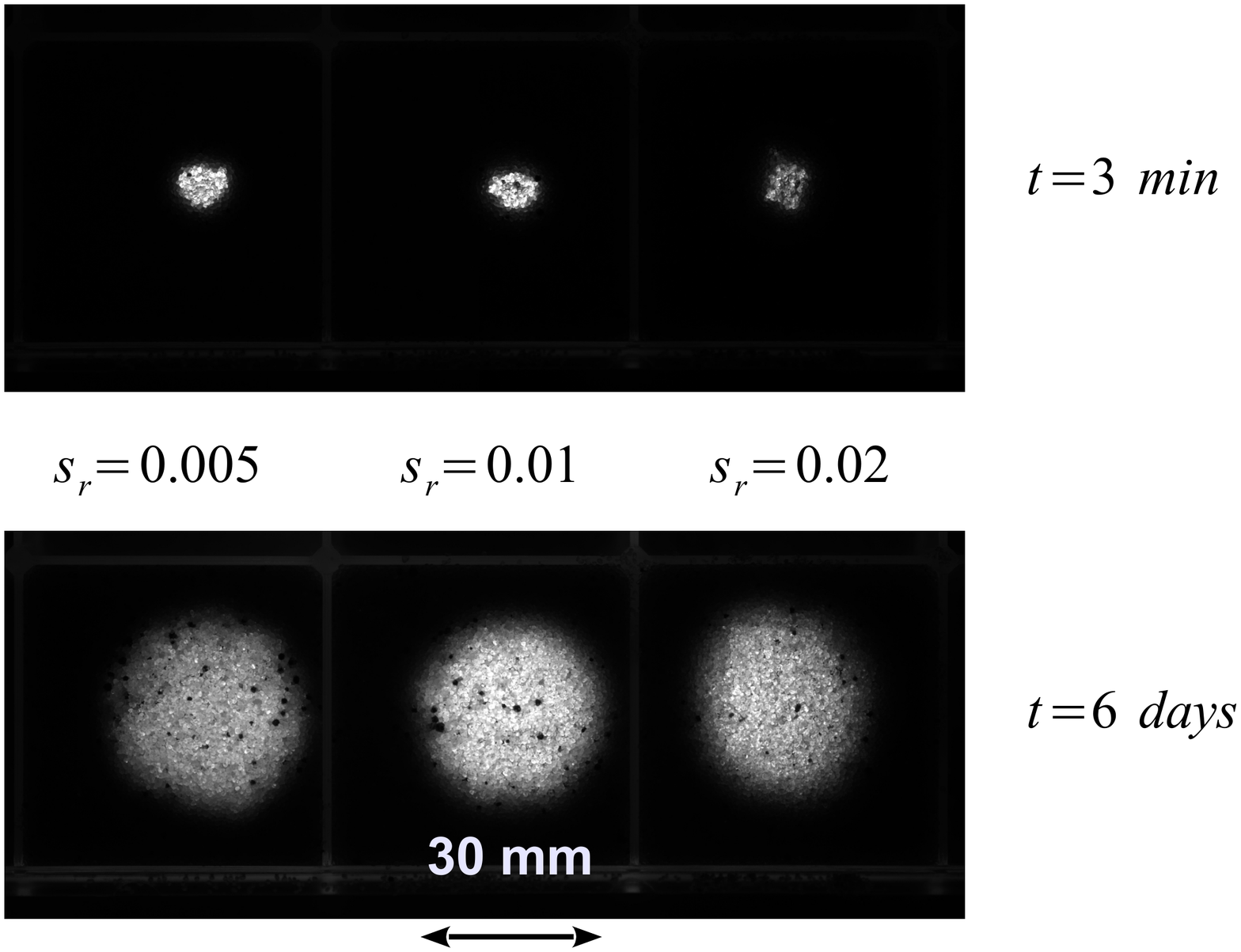}
\end{center}
\caption{Spreading of TEHP liquid drops ($V_D=6\,\mbox{mm}^3$) in pre-wetted sands with different background saturations levels $s_r=0.5, 1$ and $2\%$. UV fluorescence wet spot areas taken at $t=3\,\mbox{min}$ and at $t=6\,\mbox{days}$ after the deposition of the drops on the sand bed.} 
\label{Prewett1}
\end{figure}

\begin{figure}[ht!]
\begin{center}
\includegraphics[trim=-1.5cm 0.cm 1cm -0.5cm,width=\columnwidth]{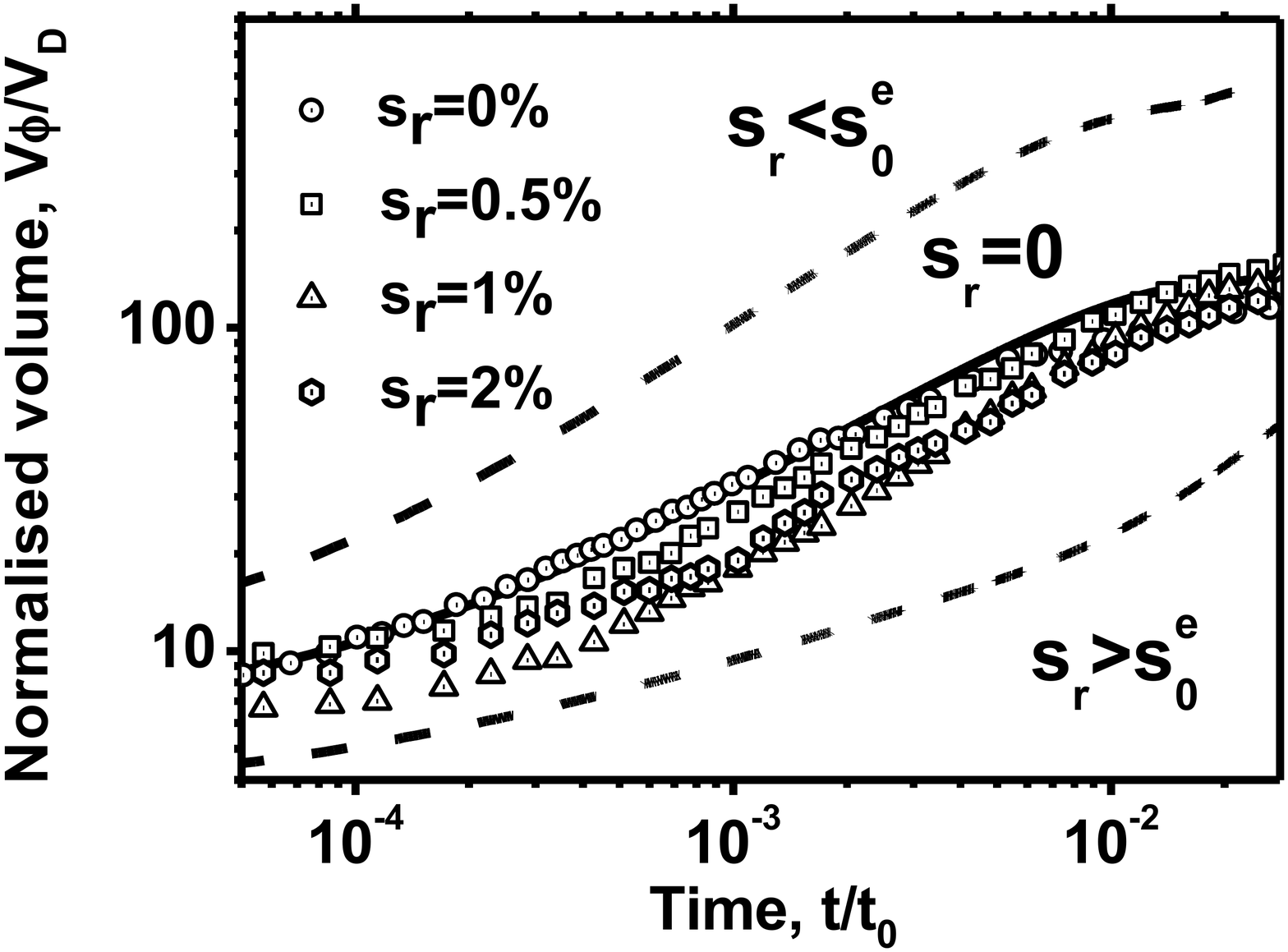}
\end{center}
\caption{Spreading of TEHP liquid drops ($V_D=6\,\mbox{mm}^3$) in pre-wetted sands with different background saturations levels $s_r=0, 0.5, 1$ and $2\%$. Normalised wet volume $V\phi/V_D$ (inverse average saturation $\bar{s}^{-1}$) as a function of the reduced time $t/t_0$, where $t_0=L_0^2/D_f$ for experimental data and the numerical results were scaled by $t_0=L_0^2/D^e_0$. The experimental results are shown by symbols. The results of numerical simulations are shown by solid lines ($s_r=0\%$) and by dashed lines $s_r=1\%$ and $s_r=0.5\%$.} 
\label{Prewett2}
\end{figure}

Even kiln-dried sands in open-chamber conditions would absorb some amount of the liquid present in the gas phase due to capillary condensation processes~\cite{Kierlik2001}. So we have conducted a series of spreading experiments in the presence of some background level $s_r$ of the wetting liquid in the porous matrix to understand how the spreading dynamics would be affected by the pre-wet conditions. The pre-wetted sand samples were prepared by shaking and mixing a certain amount of the TEHP liquid with the sand in a closed container over a long period of time to ensure that the liquid is equally distributed in the sample. The experimental results of spreading of $V_D=6\,\mbox{mm}^3$ TEHP liquid drops in $R\approx 0.25\, \mbox{mm}$ pre-wet sands are shown in Figs. \ref{Prewett1} and \ref{Prewett2} at different levels of $s_r$.  The main question here is to understand if the mixing and shaking of the pre-wetted sand samples would have produced a similar liquid distribution on the grain surfaces to that obtained during the natural liquid spreading at similar saturation levels. Apparently, one might expect that the distributions could be different due to the hysteresis effect commonly observed in porous media spreading processes~\cite{Leverett1941, Mualem1974, Kierlik2001}. For example, if some areas on the grain surfaces were inaccessible to the liquid flow at low saturation levels~\cite{Tuller-2000}, then during shaking and mixing those areas might be wet. The assumption is in agreement with the analysis presented in~\cite{Tuller-2000} and our observations that the equilibrium value of $\alpha_R\approx 0.3$ after natural spreading is small. That is, during the natural spreading, large surface areas of the grains were left dry. This implies that the liquid content in equilibrium would depend on the way this equilibrium was achieved, and this seemed to be observed in our experiments, Figs. \ref{Prewett1} and \ref{Prewett2}. Indeed, as is seen from the figures, the rate of the front evolution and the final size of the wet spot area were practically independent of the value of $s_r$, as if the sand was almost dry. One can observe some small effect of the background moisture presence, but as we will argue below, this was way too low.

\begin{figure}[ht!]
\begin{center}
\includegraphics[trim=-1.5cm 0.cm 1cm -0.5cm,width=\columnwidth]{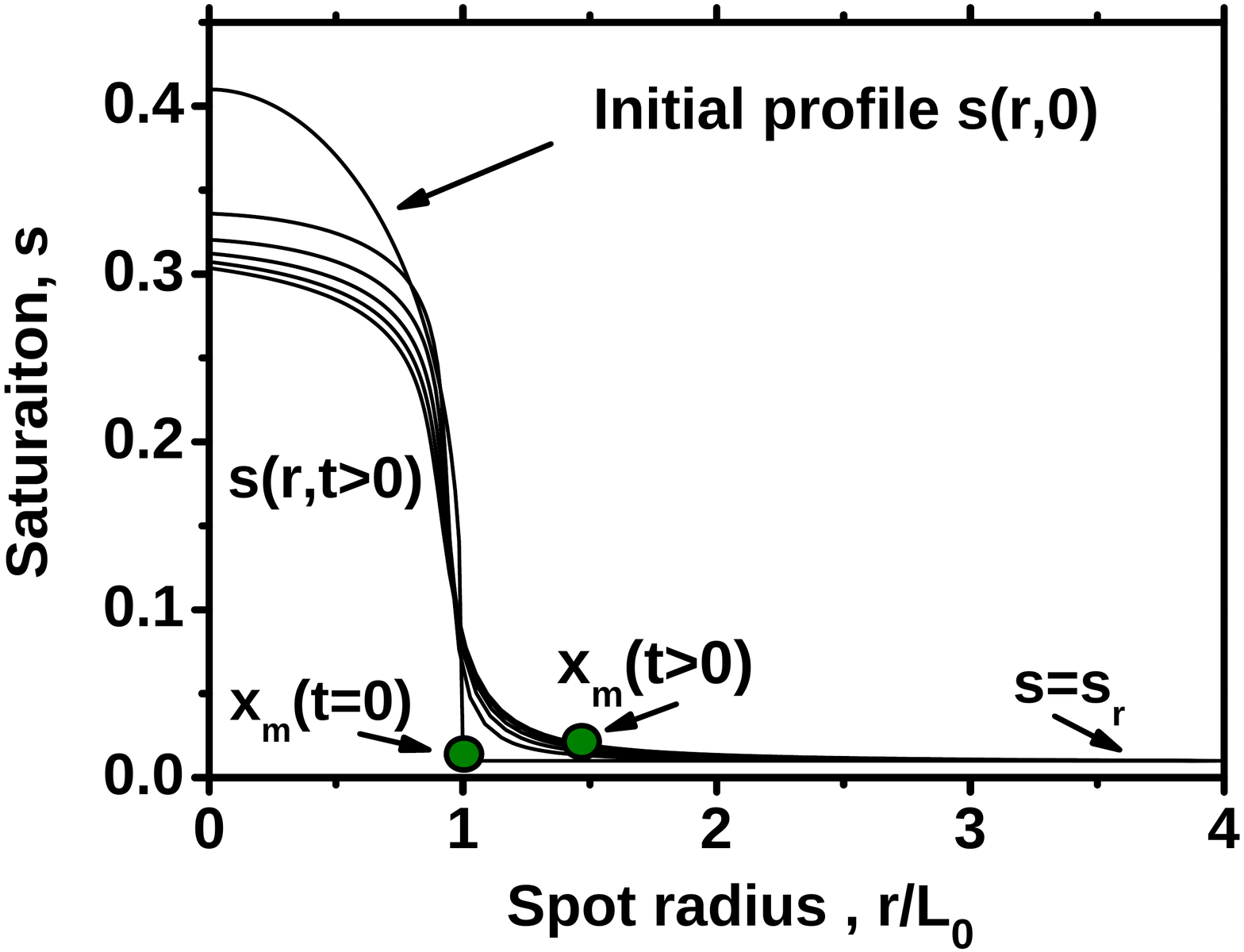}
\end{center}
\caption{Simulation of spreading of TEHP liquid drops ($V_D=6\,\mbox{mm}^3$) in pre-wetted sands with the background saturations level $s_r=1\%$ using the augmented model (\ref{Superfast-nd-am}) with initial conditions (\ref{Prewet-ini}) at $s_a=0.4$, $r_a=4$ and $\lambda_a=0.3$ with $s_0^e\approx 0.0068$.} 
\label{Prewett3}
\end{figure}

Theoretically, if we presume for a while that our pre-wetted sands with some background level of saturation $s_r$ have similar liquid morphology to that during the natural spreading, one should distinguish two cases. In the first case, when $s_r>s_0^e$, there should be liquid bridges present in the background porous material. In the second case, when $s_r<s_0^e$, the global network connection is broken. In the former case, the notion of the moving wetting front is absent as a matter of fact. Consider, as an example, again a spherically symmetric three-dimensional case, when initial liquid distribution at $t=0$ is given by
\begin{equation}
\label{Prewet-ini}
s(r,0)=s_r + s_a\cos^{\lambda_a}(\pi r/2), \,\,\, 0\le r\le 1
\end{equation}
$$
s(r,0)=s_r, \,\,\,  1 \le r \le r_a
$$ 
and there is no flux at the end of the simulation domain at $r=r_a$, Fig. \ref{Prewett3}. Due to the nature of our numerical method, which is using moving meshes, the amount of the liquid is conserved in between any moving mesh points. Hence, one can easily follow the evolution of a benchmark point $x_m(t)$, as is shown in Fig. \ref{Prewett3}. The result at $s_r=1\%$, shown in Fig. \ref{Prewett2} in terms of the evolution of the volume contained within $0\le r\le x_m$, indicates that while there is some initial plateau in the distribution of the saturation, as is observed in the experiments presented in the same figure, in general the evolution is much slower. One can conclude then that, while the initial plateau observed during the volume evolution at high average saturation values $\bar{s}\approx 20\%$ at both $s_r=2\%$ and $s_r=1\%$ indicates that the mechanism of spreading is sensitive to the background levels, see Fig. \ref{Prewett2}, to the large extent the spreading dynamics is still defined by the front capillary pressure generated on the scale of surface roughness. One can also conclude that the liquid morphology of that background liquid distribution seemed to be different from the liquid morphology observed at these saturation levels during the natural spreading.   

In the second case, $s_r<s_0^e$, one needs to modify the original model to include the presence of some background saturation level. Using conservation of the liquid in the domain $\Omega$ with a front $\partial \Omega(t)$ moving into the area with background saturation $s_r$ and the transport Reynolds theorem
$$
\frac{d}{dt}\int_{\Omega(t)}\, s\, d^3 x =\int_{\Omega(t)}  \left(\frac{\partial s}{\partial t} + \nabla\cdot (s {\bf v})\right) \, d^3 x = 
$$
$$
=\int_{\partial \Omega(t)} ({\bf v\cdot  n})\, s_r\, dS,
$$
where $\bf n$ is the normal vector to $\partial \Omega$.

Transforming the surface integral into the volume integral
$$
\int_{\Omega(t)}  \left(\frac{\partial s}{\partial t} + \nabla\cdot ((s-s_r) {\bf v})\right) \, d^3 x =0.
$$

This implies that an equivalent moving boundary-value non-linear diffusion problem of transport in pre-wetted sands can be formulated in terms of a function $\varphi=s-s_r$
\begin{equation}
\label{ndpw}
\frac{\partial \varphi}{\partial t} = \nabla \cdot \left\{ \frac{ \widehat{\kappa_0}(\varphi)\, \widehat{g}\,  (\varphi)\nabla \varphi}{|\ln(\varphi-\varphi_0)|(\varphi-\varphi_0)^{3/2}}\right\}, 
\end{equation}
$$
\varphi_0=s_0^e-s_r
$$
with the boundary conditions
$$
\left. \varphi \right|_{\partial \Omega}=s_f-s_r
$$
and
\begin{equation}
\label{BC2ndpw}
\left. v_n \right|_{\partial \Omega}=-\frac{\widehat{\kappa_0}(\varphi)\, \widehat{g} \, ({\bf n}\cdot\nabla) \varphi}{(s_f-s_r)|\ln(s_f-s_0^e)|(s_f-s_0^e)^{3/2}}.
\end{equation}

\bigskip
One can see that in general due to a smaller factor at the moving front $s_f-s_r$ (instead of just $s_f$), the front motion is expected to proceed with much higher velocity. This is understandable, since one requires lesser amount of the liquid to move the front by an infinitesimal value $\Delta x$ within a time interval $\Delta t$, and this is exactly what was observed in the numerical solutions of (\ref{ndpw})-(\ref{BC2ndpw}) at the parameters of set VIII, Table \ref{Table1}, and initial distribution (\ref{Initial}) at $\lambda_a=0.3$ and $s_a=0.4$, Fig. \ref{Prewett2}. As one can see, the propagation of the front is indeed  much faster than that at $s_r=0$ shown in the same figure. One might expect that the value of the parameter $s_f-s_0^e$ would be larger in this case, since in the pre-wetted sand the small length scales of the surface roughness may not be available. This might reduce the capillary pressure at the moving and slow down the propagation rate. But, we have checked that even increasing the value of $s_f-s_0^e$ by three times was insufficient to match the slower propagation observed in the experiment. This again indicates that the liquid morphology is different at $s_r=0.5\%$ than one would anticipate. Basically, the wetting process is unaffected by the presence of small background levels. In a way, this result is in accord with the characteristic values of the coefficient  $\alpha_R\approx 0.3$ obtained in the comparison with experimental data. This indicates, that only a limited part of the surface area of the grains is fully participating in the liquid transport in the system. We note, that given the length scale of the liquid films involved in the transportation in the pendular regime, it is unlikely to have stochastic enhancement of the dye transport in the wet porous matrix~\cite{Stoch-book}. These are very interesting results, which definitely require further, specific studies.

\section*{Conclusions}
In our previous study we established that: 

\begin{itemize}
\item The process of spreading can be described by a special type of non-linear diffusion process, where the driving force is the capillary pressure at the moving front generated by the particle surface roughness and the coefficient of diffusion has a characteristic singular form $D(s)\propto (s-s_0^e)^{-3/2}$. The resulting mathematical model belongs to a class known as super-fast diffusion equation, and the so-suggested scaling with viscosity and surface tension is as expected for capillary flows, $D\propto \gamma/\mu$.

\item Motion of the wetting front $X_3(t)$ in a three-dimensional spherically symmetric domain (when the wetted volume has a shape of the hemisphere) exhibits universal scaling behaviour with time $t$, such that $X_3(t)\propto t^{1/4}$, ultimately going to standstill at finite saturation levels $s_0\approx 0.6\%$. This behaviour led us to a conjecture, confirmed in numerical simulations of the superfast diffusion model, that in general, depending on the geometry, basically on its dimension $n$, $X_n(t)\propto t^{1/(n+1)}$, which may be used in practical applications to analyse such kind of spreading processes.  
\end{itemize}

In the work reported here, with the help of a new set of experiments, we have delved deeper into the theoretical formulation aiming to refine the modelling of relevant permeability and include the funicular regime, so as to supply improved initial conditions for the super-fast regime. The experiments were carried out with a set of low-dispersed (with small deviations of the grain radius $R$ from its average value), well characterized  sands, using different geometric set-ups and regimes of spreading (one- and three-dimensional symmetric regimes). The new results can be summarized as follows:

\begin{enumerate}  

\item The motion of the liquid wetting front $X_n(t)$ in geometrically different set-ups and regimes of liquid spreading indeed follows the universal scaling law $X_n(t)\propto t^{1/n+1}$, with $n$ being solely defined by the dimension of the moving front diffusion problem. As it was shown by the numerical analysis, the augmented superfast diffusion model (\ref{Superfast-nd-am}) clearly demonstrates this universal behaviour, which may be used in the practical applications for the analysis of spreading at low saturation levels. Analysis of the mathematical model has revealed that this behaviour is manifestation of the specific shape of the saturation profile (a Mexican hat), predicted by the model, with a distinctive tail at almost equilibrium saturation levels $s\approx s_f$. In the one-dimensional case, when spreading is confined within long, open channels, the advancing-front motion conforms to the well-known Lucas-Washburn law $X_1(t)\propto t^{1/2}$ for a single capillary.
   
\item The overall evolution of the wetted volume is predominantly defined by the diffusion rates in the tail region, that is by the processes described by the super-fast diffusion model. On the other hand, the standard diffusion mechanisms, commonly applied for the analysis of spreading in the funicular regime of wetting, only smooth out the distribution profile at higher levels of saturation, usually found in its central part. Thus, the funicular and the pendular regimes are found to simply operate simultaneously but in different locations.  

\item Experimental data obtained using liquids of different viscosities and wettabilities confirm our previous finding that the spreading dynamics of different liquids  obeys the scaling law when the driving force is the capillary pressure, and the coefficient of diffusion $D\propto \gamma/\mu$, as is depicted by our super-fast diffusion model. Further, we have been able to identify the scaling behaviour of diffusion with the wettability of the liquid-solid combinations involved, that is with the contact angle $\theta_c$. As it might be expected, the diffusion rate is found to be smaller for larger contact angles. This effect is directly related with the available amount of the surface roughness groove filling, which diminishes as the contact angle increases.

\item A set of experiments using low-dispersed sand samples with different distributions of the grain sizes has allowed to obtain more accurate estimates of the main non-dimensional parameters of the model, such that only one adjusting parameter $\xi_f$ was left incorporating nothing but specific microscopic properties of the surface roughness. Spreading dynamics observed in sands with different grain size distributions was found to be slightly counter-intuitive. The spreading was slower when the grain size reduces, while the effective surface area per unit volume $S_T\propto 1/R$ (and hence the effective free surface energy) increases. This behaviour is in accord with the mathematical model and is manifestation of the scaling of the main non-dimensional model parameter $s_f-s_0^e \propto 1/R^2$, which is in fact the inverse of the capillary front pressure, the main driving force of the process. 

\item Analysis of spreading in pre-wet sands with a small background level of saturation $s_r \approx  1-2\%$ have shown, that the distributions of the same amount of a liquid are different after natural spreading and mechanical mixing procedures. If a small background saturation level was achieved by a mechanical mixing process, it does not change dramatically the dynamics of spreading predicted by the superfast diffusion model.  

\item While the dynamics of liquid spreading was found to depend on the liquid and porous media properties, the equilibrium thickness of the liquid film on the surface of grains was solely defined by the surface roughness, at least for the well-wetting liquid-solid combinations used in our study. Such universal behaviour allows to estimate one of the main parameters of the model $s_f\approx s_0$ with sufficient accuracy only on the basis of the effective surface area $S_T\propto 1/R$, porosity $\phi$ and the average amplitude of the surface roughness $\bar{\delta}_R$.  
\end{enumerate}

One can then finally conclude that on the basis of comparison with experimental data the augmented superfast non-linear diffusion model (\ref{Superfast-nd-am}) provides an adequate description of liquid transport at low saturation levels, which therefore can be used in practical applications.

\section*{Appendix: Numerical moving mesh method.}

The numerical technique used to solve the partial differential equations in this study is a moving mesh method driven by conservation,
similar to that presented in \cite{Lee-2015} and described in \cite{Baines-2017}. A nodal velocity $v$ is constructed from a combination of 
a non-linear diffusion equation, for example the 3-D radially symmetric nonlinear diffusion equation 
\[
\frac{\partial s}{\partial t}=\frac{1}{r^2}\frac{\partial}{\partial r}\left(r^2D(s)\frac{\partial s}{\partial r}\right),
\]
and the conservation law
\begin{equation}
\label{E}
\frac{\partial s}{\partial t}+\frac{1}{r^2}\frac{\partial}{\partial r}\left(r^2sv\right)=0,
\end{equation}
yielding the velocity formula
\begin{equation}
v(r,t)=-\frac{D(s)}{s}\frac{\partial s}{\partial r}\label{v}
\end{equation}
where $v(0,t)=0$.  
An equation for $\mathrm{d} s/\mathrm{d} t$ following the motion is then
\[
\frac{\mathrm{d} s}{\mathrm{d} t}=\frac{\partial s}{\partial t}+v(r,t)\frac{\partial s}{\partial r}
=-\frac{1}{r^2}\frac{\partial}{\partial r}\left(r^2sv\right)+v(r,t)\frac{\partial s}{\partial r}
\]
\begin{equation}
\label{1}
=-s\frac{1}{r^2}\frac{\partial}{\partial r}\left(r^2v\right)
=s\frac{1}{r^2}\frac{\partial}{\partial r}\left(r^2\frac{D(s)}{s}\frac{\partial s}{\partial r}\right)
\end{equation}

Introducing
moving nodes $\widehat r_i(t)$  
and corresponding saturation values $\widehat s_i(t),\,(i=1,\ldots,N)$, 
an approximation to (\ref{v}) is
\begin{equation}
v_{i+1/2}^n=-\frac{D(\widehat s_{i+1/2}^n)}{\widehat s_{i+1/2}^n}\frac{\widehat s_{i+1}^n-\widehat s_i^n}{\widehat r_{i+1}^n-\widehat r_i^n}\label{vi}\end{equation}

The system (\ref{1}) is approximated by the first-order-in-time semi-implicit scheme
\[
\frac{\widehat s_i^{n+1}-\widehat s_i^n}{\Delta t}=\frac{\widehat s_i^n}{(\widehat r_{i+1/2}^n-\widehat r_{i-1/2}^n)(\widehat r_i^n)^2}
\]
\[
\left\{ \left. ({\widehat r}_{i+1/2}^n)^2\frac{D(\widehat s^n)}{\widehat s^n}\right|_{i+1/2}  \frac{(\widehat s_{i+1}^{n+1}-\widehat s_{i}^{n+1})}
{(\widehat r_{i+1}^n-\widehat r_{i}^n)} \right.
\]
\begin{equation}
-\left. \left. ({\widehat r}_{i-1/2}^n)^2\frac{D(\widehat s^n)}{\widehat s^n}\right|_{i-1/2}
\frac{(\widehat s_{i}^{n+1}-\widehat s_{i-1}^{n+1})
}{(\widehat r_{i}^n-\widehat r_{i-1}^n)}
 \right\}
\label{si}
\end{equation}
$(i=1,\ldots,N-1)$
where $\Delta t$ is the time step,
which has the property that no new local extrema in $\widehat s_i$
are created in the 
interior of the domain in a time step, thereby preserving positivity of $\widehat s_i$
and avoiding oscillations. This allows arbitrarily large numbers of nodes without $\Delta t$ being
restricted by stability conditions.

The scheme (\ref{si}) can be written in the matrix form
\begin{equation}B\underline{\widehat s}^{n+1}=\underline {\widehat s}^n\label{B}\end{equation}
where $\underline{\widehat s}^{n+1}=\{{\widehat s}_i^{n+1}\}$,  
$\underline{\widehat s}^n=\{{\widehat s}_i^n\}$, and
$B$ is a tridiagonal matrix
modified to take into account the boundary condition 
$\widehat s_N=s_f$ and the continuity condition $\partial s/\partial r=0$ at $r=0$.

Once the $\widehat s_i^{n+1}$ have been obtained the mesh nodes $\widehat r_i^{n+1}$ can be found from
the Lagrangian form of the conservation principle (\ref{E}), i.e.
\begin{equation}
\int \widehat s(r,t)r^2\mathrm{d} r\mbox{ is constant in time,}\label{L}
\end{equation}
valid when $s(r,t)>0$. 

A discretisation of (\ref{L}) is
\begin{equation}\{(\widehat r_{i+1}^{n+1})^3\widehat s_{i+1}^{n+1}-(\widehat r_i^{n+1})^3\widehat s^{n+1}_i\}
=\mbox{ its initial value}\label{ri}\end{equation}
$(i=2,\ldots,N)$, yielding $\widehat r_i^{n+1}$
by recursion over $i$, given $\widehat r_0^{n+1}=0$.
Since the $s_i^{n+1}$ are positive the recursion process ensures that the nodes remain ordered.

To summarise the algorithm, 
given the $r_i^n$ and $s_i^n$ values at time step $n$,
\begin{itemize}
\item approximate the $v_{i+1/2}^n$ from (\ref{vi})
\item determine the $\widehat s_i^{n+1}$ from (\ref{si}), equivalently (\ref{B})
\item recover the $r_i^{n+1}$ from (\ref{ri}).
\end{itemize}

\end{document}